\documentclass{aastex61}

\accepted{December 19, 2017 for publication in The Astrophysical Journal}

\shorttitle{Pulses of Short Gamma-Ray Bursts}
\shortauthors{Hakkila et al.}

\begin{document}

\title{Properties of Short Gamma-Ray Burst Pulses from A BATSE TTE GRB Pulse Catalog}

\correspondingauthor{Jon Hakkila}
\email{hakkilaj@cofc.edu}

\author{Jon Hakkila}
\affiliation{The Graduate School, University of Charleston, SC at the College of Charleston, 66 George St., Charleston, SC 29424-0001, USA}
\affiliation{Department of Physics and Astronomy, College of Charleston, 66 George St. Charleston, SC 29424-0001, USA}

\author{Istv{\'a}n Horv{\'a}th}
\affiliation{National University of Public Service, 1441, Budapest, Hungary}

\author{Eric Hofesmann}
\affiliation{Department of Physics and Astronomy, College of Charleston, 66 George St. Charleston, SC 29424-0001, USA}

\author{Stephen Lesage}
\affiliation{Department of Physics and Astronomy, College of Charleston, 66 George St. Charleston, SC 29424-0001, USA}




\begin{abstract}

We analyze pulse properties of Short gamma-ray bursts (GRBs)
from a new catalog containing 434 pulses from 387 BATSE 
Time-Tagged Event (TTE) GRBs.
Short GRB pulses exhibit correlated properties of duration, fluence,
hardness, and amplitude, and they evolve hard-to-soft
while undergoing similar triple-peaked light curves 
similar to those found in Long/Intermediate bursts. 
We classify pulse light curves using their temporal complexities,
demonstrating that Short GRB pulses exhibit a range of
complexities from smooth to highly variable.
Most of the bright, hard, chaotic emission seen
in complex pulses seems to represent a separate
highly-variable emission component.
Unlike Long/Intermediate bursts, as many as 90\% of Short GRBs are single-pulsed.
However, emission in Short multi-pulsed bursts is coupled such that
the first pulse's duration is a predictor
of both the interpulse separation and subsequent pulse durations.
These results strongly support the idea that external shocks
produce the prompt emission seen in Short GRBs. The similarities between 
the triple-peaked structures and spectral evolution of 
Long, Short, and Intermediate GRBs then suggests that
external shocks are responsible for the prompt emission
observed in all GRB classes.
In addition to these findings, we
identify a new type of gamma-ray transient 
in which peak amplitudes occur at the end of the burst rather than at earlier times.
Some of these ``Crescendo'' bursts are preceded by rapid-fire ``Staccato'' pulses, whereas 
the remaining are preceded by a variable episode
that could be unresolved staccato pulses.

\end{abstract}




\section{Introduction} \label{sec:intro}

Gamma-ray bursts (GRBs) radiate at such large rates
over tens of milliseconds to hundreds of seconds
that they must by necessity extract their energies 
($E_{\rm tot} \approx 10^{51}$ ergs; \cite{fra01,pan01,pir01}) from the violent gravitational 
collapse that accompanies black hole formation.
Production of these energy rates could require a variety of progenitors: in the 1990s
the broad logarithmic distribution of GRB durations (spanning six decades)
showed evidence of bimodality, with Long and Short bursts separated at roughly 
$T_{90}=2$ s \citep{kou93}.
Bursts in the Long class were shown to have softer average spectral hardnesses than
those in the Short class. Theoretical models favoring accretion scenarios
involving stellar core have difficulty explaining GRB timescales
shorter than $2-3$ seconds ({\em e.g., }\cite{woo93}), supporting the idea of a second 
GRB population arising from merging neutron stars or other compact massive objects.

Significant evidence has been presented indicating that Long and Short GRB classes 
represent different source populations \citep{nor01,bal03,pir04,zha09,luli10,li16}. 
These two burst classes appear to originate in different types of host galaxies,
belong to different redshift distributions, and produce different types of afterglows 
({\em e.g., }\cite{hog99,hjo06,ber14}).
Some low-luminosity Long GRBs have been associated with Type Ic 
supernovae (SN) \citep{hjo03,cam06,pian06,blan16}, supporting
the idea that the Long GRBs in general are related to deaths of 
massive stars \citep{woo93,pac98,wb06,blan16}.
In contrast, Short GRBs are found in metal-poor regions and are less luminous
than Long GRBs, suggesting origins from compact binary mergers 
\citep{pac86,usov92,ber14}.

Despite this supportive evidence, the apparent clarity of the 
simple $T_{90}$-based classification scheme used by many 
is a stark over-simplification. 
Application of statistical clustering techniques and 
machine learning algorithms to prompt emission properties indicate
that GRBs fall into three or more separate classes
\citep{muk98,hor98,hak00,bal01,rm02,hor02,hak03,bor04,hor06,chat07,dup11,zit15,zha16,cha17}. 
The favored solution involves Long, Short, and Intermediate classes identified
on the basis of duration, hardness and fluence. This result has been 
repeatedly found from observations by many GRB experiments including 
BATSE (the Burst And Transient Source Experiment on NASA's Compton 
Gamma Ray Observatory), BeppoSAX, Swift, and Fermi GBM
({\em e.g., }\cite{hor08,hor09,huja09,hor10,ht16}).
The Intermediate GRB class is composed of bursts with durations 
overlapping Short and Long bursts, and is characterized 
by GRBs having the softest spectra. However, only weak evidence 
has been provided arguing that the Intermediate class might be a distinctly different source population 
(the angular distribution of Intermediate GRBs is anisotropic at around the $2\sigma$ 
significance level), and theoretical models have been unable to account for them. 
Few studies have tried to explain the existence of the Intermediate
class, although \cite{hak03} proposed that this class could result
from instrumental biases: Intermediate GRBs are 
faint, soft, Long bursts that appear to be a separate class because they
are found close to the trigger threshold. 

With little evidence that the Intermediate class makes up a separate source population, 
one is forced to reassign each Intermediate GRB to either the Long or the Short class.
The Long/Short GRB duration boundary is therefore even less clear
than implied by the simple $T_{90} \approx 2$ s dividing line, and additional parameters
such as spectral hardness and fluence are likely needed
before assigning bursts near this line to a class. This is problematic, 
as Short and Long GRB class characteristics 
depend on the instrument that observes them, the classification 
techniques used, and the specific set of bursts being classified.  In other words, each
gamma-ray instrument has its own spectral and intensity response which can lead to 
redefinitions of the burst class properties and their associated dividing lines. 
Modern statistical and machine learning classification techniques are powerful
tools that are sensitive to the aforementioned data characteristics, if the instrumental
characteristics are also accounted for.
In the reassignment of Intermediate BATSE bursts, we have found
that short hard GRBs generally belong to the Short class, while short soft GRBs
generally belong to the Long class. 

{\em Pulses}, the basic units of GRB prompt emission, have the potential of 
delineating Long and Intermediate GRBs from Short GRBs. Pulses are
pervasive and have well-defined light curves as opposed to representing stochastic or 
chaotic emission. Isolated pulses observed by BATSE, Swift, and Fermi exhibit 
hard-to-soft evolution, longer durations at lower
energies, near-simultaneous initiation across the range of observed
energies, asymmetric shapes, and triple-peaked structures with re-hardening 
occurring around the time of each peak. Most of these pulse behaviors have been found
in Long, Short, and Intermediate bursts, but the triple-peaked pulse structure
has not been systematically studied in Short bursts. In BATSE archival data, 
this is because most observations are limited to data having 64-ms resolution, which is often
longer than the durations of the expected triple-peaked substructures. In Swift data, it is because
Swift's spectral response favors detection of soft GRBs over hard ones, making the 
division between Long and Short GRBs less clear. An analysis of BATSE Short bursts can be
performed using the instrument's TTE (Time Tagged Event) data type, and is 
described in this paper. An extensive high time resolution
study of Short Fermi GRBs is also possible, and will be examined in a separate paper.

BATSE ({\em e.g., }\cite{hor91,fish13}) was composed of eight large 
sodium iodide detectors located on the outside of the Compton 
Gamma-Ray Observatory; the faces of these detectors were
arranged to describe the shape of a regular octahedron. 
Gamma- and x-ray photons absorbed by the sodium iodide detectors
produced visible photons of roughly proportional energy that were 
detected by photomultipliers; these counts were parsed
into four energy channels (channel 1 energies of 20 - 50 keV, channel 2
energies of 50 - 100 keV, channel 3 energies of 100 - 300 keV,
and channel 4 energies of 300 keV - 1 MeV). Photon counts were collected 
in the form of a changing instrumental background throughout the 
mission on one-second timescales; the format of this data stream
changed at the moment an onboard trigger occurred, at which time the instrument 
switched to 64-ms resolution data for the duration of an event. 
A trigger occurred when the counts rose above a specified 
statistical threshold (usually $5.5\sigma$) of the signal (measured on three different 
trigger timescales: 64 ms, 256 ms, and 1024 ms) relative to the
background (based on a 17-second running average count rate) in a predefined 
set of energy channels (generally channels 2 plus 3 spanning the 
50 - 300 keV energy range) in each of the two
BATSE detectors most nearly facing the source (to eliminate single-detector
particle events). 

In addition to the data having 64-ms resolution, BATSE also collected a limited 
amount of high time resolution data referred to as TTE (Time Tagged Event) data. 
TTE data, containing specific
information on each photon collected, were stored in a ring buffer
starting around the time of the instrumental trigger. Photons were included
in the buffer until it was filled, which generally spanned a time interval of no more 
than two seconds.  If the count
rate was too high, the buffer contained fewer than two seconds worth of photon counts. 
The energy of each photon was independently measured and stored in the buffer.

Because of BATSE's large surface area and energy response, 
BATSE TTE data have sufficient temporal resolution and counts to permit useful
and unique analyses of some Short and Intermediate GRB light curves.
Short and Intermediate GRB light curves can be fully contained 
within the TTE ring buffer, whereas
Long bursts cannot. The high time resolution 
of TTE data allow for the the study of temporal structures within these bursts
that cannot be performed with the lower-resolution 64-ms data.
Additionally, count rates are often high enough for detected photons to be 
parsed into different energy bins, and the four-channel properties
of the 64-ms data can be reproduced on shorter timescales.

BATSE TTE data allow several important, yet unanswered questions
about GRB pulse structure to be addressed: 
Do Short GRB pulse light curves contain the same triple-peaked, hard-to-soft evolutionary
structures exhibited by Long and Intermediate GRB pulses? Do Short GRB pulse spectra
re-harden at the time of each of the three pulse peaks as they do for Long and 
Intermediate GRB pulses? How do the spectrotemporal characteristics of Short 
GRB pulses contrast with those of Long and Intermediate pulses? Can pulse 
characteristics be used to differentiate between Short and Intermediate 
or Long GRB pulses? To answer these questions we have undertaken a 
systematic study of short duration BATSE gamma-ray bursts using TTE data.

\section{TTE Pulse-Fitting} \label{sec:pulse}

BATSE obtained TTE data for 532 GRBs 
(2702 GRBs appear in the online BATSE Burst Catalog (Briggs et al., in preparation,
at http://gammaray.msfc.nasa.gov/batse/grb/catalog/current/).
Some of these are long GRBs with
durations extending far beyond the 2 s maximum boundary of the TTE window,
leaving a smaller number of shorter bursts 
available for high resolution pulse analysis.
Our initial sample consists of 392 of these BATSE TTE GRBs 
obtained from \cite{hor05}; these GRBs are all short enough to potentially
fit completely within their respective TTE windows.
The photon counts of these bursts
have been subdivided into 4 ms bins, as well as 
into the four standard BATSE energy channels.  

The values we are fitting are the 4 ms-binned
counts summed over the four BATSE energy 
channels.
These are short duration GRBs having
spectrotemporal resolutions similar to those of
the 64 ms Long and Intermediate BATSE and Swift
bursts used in prior pulse analyses \citep{hak14, hak15}.
Upon removal of five bursts with data problems, the sample
available for GRB pulse-fitting is reduced to 387 bursts.

\subsection{The Pulse-Fitting Model}

The pulse-fitting model consists of two parts. The first
is the general four-parameter empirical pulse model of \cite{nor05}.
The hypothesis is that a
GRB emission episode can be modeled by the following asymmetric, 
monotonically increasing
and decreasing intensity function:

\begin{equation}\label{eqn:function} 
I(t) = A \lambda e^{[-\tau_1/(t - t_s) - (t - t_s)/\tau_2]},
\end{equation}
where $t$ is time since trigger, $A$ is the pulse amplitude, $t_s$ is the pulse start time, $\tau_1$ is the pulse rise parameter, $\tau_2$ is the pulse decay parameter, and the normalization constant $\lambda$ is given as $\lambda = \exp{[2 (\tau_1/\tau_2)^{1/2}]}$. Poisson statistics and a two-parameter background counts model of the form $B=B_0+BS \times t$ are assumed (where $B$ is the background counts in each bin and $B_0$ and $BS$ are constants denoting the mean background (counts) and the rate of change of this mean background (counts/s)). 
Observable pulse parameters obtained from this model include the pulse peak time $\tau_{\rm peak}$ where
\begin{equation}\label{eqn:taupeak}
\tau_{\rm peak} = t_s + \sqrt{\tau_1 \tau_2},
\end{equation}
along with the pulse duration $w$ and the pulse asymmetry $\kappa$. As a result of the rapid smooth rise and fall of the pulse model, $w$ and $\kappa$ are measured relative to some fraction of the peak intensity. Using the fraction previously described \citep{hak11} as $I_{\rm meas}/I_{\rm peak}=e^{-3}$ (corresponding to $4.98\% I_{\rm peak}$),

\begin{equation}\label{eqn:duration}
w = \tau_2 [9 + 12\mu]^{1/2},
\end{equation}
where $\mu = \sqrt{\tau_1/\tau_2}$, and

\begin{equation}\label{eqn:asymmetry}
\kappa \equiv [1 + 4\mu/3]^{-1/2};
\end{equation}
Asymmetries range from symmetric (characterized by $\kappa=0$)
to asymmetric having longer decay times than rise times ($0 < \kappa \le 1$). 
The \cite{nor05} pulse model cannot physically describe pulses 
in which asymmetries are characterized by longer rise than decay times,
but it provides a good first-order fit to BATSE, Fermi GBM, and Swift pulses.

Residuals to the \cite{nor05} model can be produced by subtracting each best-fit
model from an observed pulse light curve. 
Small yet distinct deviations in the residuals are found to be systematically
in phase with the light curve \citep{hak14}, and these deviations are needed to accurately describe GRB pulse shapes. 
Although the deviations are closely aligned with the pulse duration, they are not always contained within it.
Thus we have defined the larger {\em fiducial time interval} $w_{\rm fid}$ as
\begin{equation}\label{eqn:res}
w_{\rm fid} = \tau_{\rm end} - \tau_{\rm start} = 4.4 \tau_2 (\sqrt{1+\mu/2}+1) + \sqrt{\tau_1 \tau_2},
\end{equation}
with the fiducial end time $t_{\rm end}$ given by 
\begin{equation}\label{eqn:tend}
t_{\rm end} = \frac{w}{2} (1+\kappa)+t_s+\tau_{\rm peak}
\end{equation}
and the fiducial start time $t_{\rm start}$ given by
\begin{equation}\label{eqn:tstart}
t_{\rm start} = t_s - 0.1 [\frac{w}{2} (1+\kappa) - \tau_{\rm peak} ].
\end{equation}
The strange, wavelike pattern of the residual variations can be fitted with an empirical function \citep{hak14}:
\begin{equation}\label{eqn:residual}
\textrm{res($t$)} = \left\{ \begin{array}{ll}
a  J_0( \sqrt{\Omega [t_0 - t - 0.005]}) & \textrm{if $t < t_0 - 0.005$} \\
a & \textrm{if $t_0 - 0.005 \le t \le t_0 + 0.005$} \\
a  J_0(\sqrt{s \Omega [t - t_0 - 0.005]}) & \textrm{if $t > t_0 + 0.005$.}
\end{array} \right. 
\end{equation}
Here, $J_0(x)$ is an integer Bessel function of the first kind, $t_0$ is the time of the residual peak (measured from the trigger time), $a$ is the amplitude of the residual peak, $\Omega$ is the Bessel function's angular frequency that defines the timescales of the residual wave (a large $\Omega$ corresponds to a rapid rise and fall), and $s$ is a scaling factor that relates the fraction of time which the function before $t_0$ has been compressed relative to its time-inverted form after $t_0$. The time during which the pulse intensity is a maximum is required to be a plateau instead of a peak, with a duration of $w_{\rm plateau} \approx 0.010 w_{\rm fid}$. Since there is no evidence in the pulse shape that the Bessel function continues beyond the third zero (following the second half-wave), the function is truncated at the third zeros $J_0 (x = \pm 8.654)$.

The fiducial values can be converted back to values in the measured time interval using:
\begin{equation}
a_{\rm meas}=a
\end{equation}
\begin{equation}
s_{\rm meas}=s
\end{equation}
\begin{equation}
t_{0; \rm meas}=t_0 (t_{\rm end}-t_{\rm start}) +t_{\rm start}
\end{equation}
and
\begin{equation}
\Omega_{\rm meas}=\Omega/(t_{\rm end}-t_{\rm start}),
\end{equation}
where $t_{\rm start}$ and $t_{\rm end}$ are the real time values corresponding to the start and end of the fiducial duration.

A convenient way to describe the pulse residual amplitudes is to normalize them to the pulse fit amplitudes, producing an intensity quantity that is independent of the instrument's signal-to-noise. The relative amplitude $R$ is given by
\begin{equation}\label{eqn:R}
R=a/A.
\end{equation} 

The 4 ms signal-to-noise ($S/N$) is a measure of the peak brightness of each TTE GRB,
relative to its background count measured with 4 ms temporal resolution. The $S/N$ is
\begin{equation}\label{eqn:SN}
S/N = (P_4-B)/\sqrt P_4
\end{equation}
where $P_4$ is the 4 ms peak counts and $B$ is the mean background count.
This signal-to-noise ratio primarily appropriate when analyzing
GRB pulses fit on the 4 ms time scale.

\subsection{TTE Pulse-Fitting Methodology} \label{sec:method}

The technique we use for extracting pulses from TTE light curves is a modification of that described previously (e.g., \cite{hak14} and references contained therein). This is because we have changed our expectations about the monotonic nature of GRB pulses based on our previous analyses. The 4 ms TTE light curves we are using generally exhibit clearly defined, isolated {\em emission episodes}, and many of these episodes appear to have shapes that are consistent with those identified for Long/Intermediate GRB pulses. Our {\em a priori} expectation of multiple peaks, rather than of strict monotonicity, allows us to hypothesize that every emission episode contains a potential pulse, with light curves that might be improved with the addition of the \cite{hak14} residual function. Overlapping pulses will be problematic, but these would have been problematic even if we had assumed that every bump in the light curve represented a monotonic, overlapping pulse. However, we can check our results for Short GRB pulses against our prior results for Long/Intermediate GRB pulses as our analysis proceeds.

This GRB pulse analysis approach is more comprehensive and systematic than any of our previous studies involving BATSE, Swift, and Fermi GBM 64-ms data. Here we attempt to analyze all bursts that entirely or mostly fit in the TTE temporal window, in contrast to the prior selection criterion of only bursts that appear to be composed of single, isolated pulses (from the 64-ms studies). This TTE study thus depends primarily on the duration of the fitted pulse, and not on how easy it is to fit the light curve. This allows us to estimate completeness for our results.

\section{The BATSE TTE Pulse Catalog} \label{sec:data}

The final BATSE TTE pulse catalog consists of 434 pulses found in 387 GRBs. Of the 387 TTE bursts for which TTE data are available, 206 completely fit within the TTE temporal window and 181 partially fit within the TTE temporal window. GRBs that completely fit within the window have been analyzed using 4 ms resolution data ({\em TTE Complete} pulses) while those that do not have been analyzed using 64-ms resolution ({\em TTE Partial} pulses). 

\begin{figure}[ht!]
\plotone{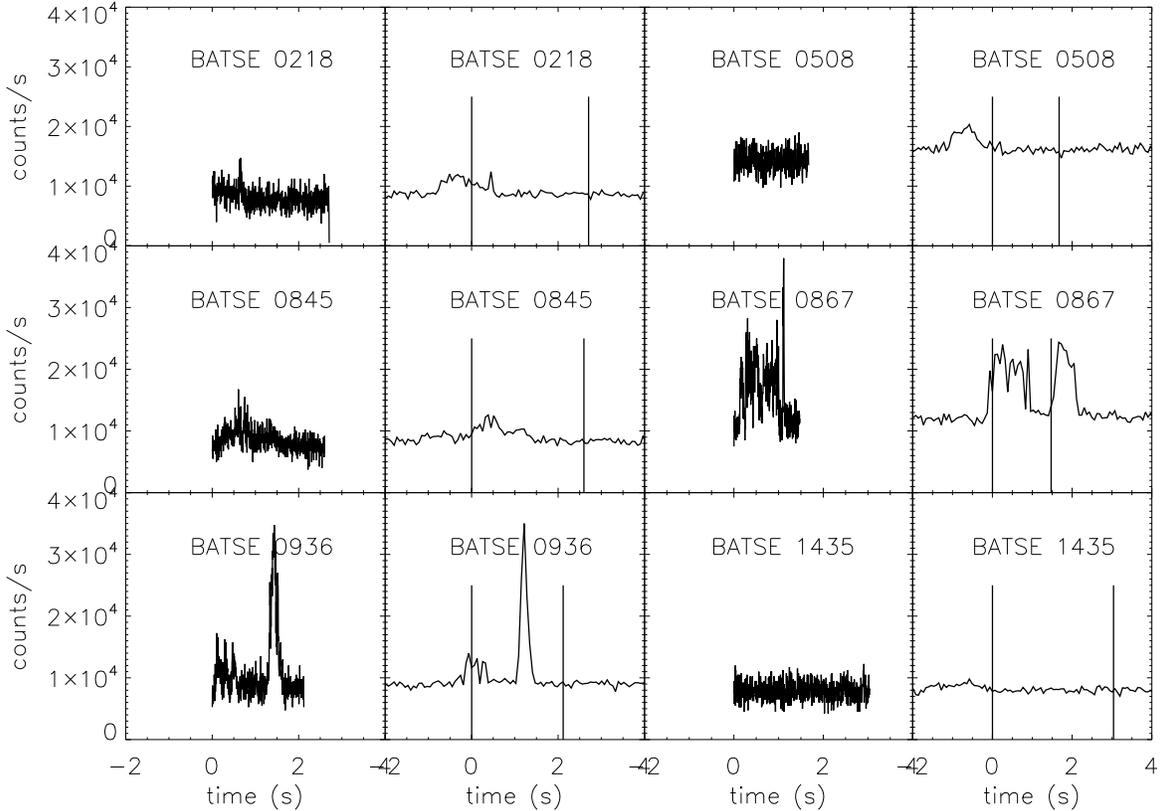}
\caption{Examples of the windowing bias responsible for causing some BATSE TTE bursts to be labeled as TTE Partial GRBs. For each of these six sample bursts, the left (noisy) panel shows the 4 ms TTE light curve. The right panel shows the corresponding 64-ms light curve, with vertical lines indicating the start and end times of the TTE window. For each of these bursts, too much flux lies outside the window for 4 ms structure to potentially be observed. Thus, the pulses in these bursts are fit using 64-ms resolution. \label{fig:f1}}
\end{figure}

The BATSE trigger is responsible for the fact that some Short bursts do not completely fit within the TTE temporal window. For BATSE to trigger, at least two of the LAD (Large Area Detectors) need to accumulate a necessary number of photon counts (typically $5.5\sigma$ above the background) in a predefined set of energy channels (generally channels $2+3$ spanning $50-300$ keV) on one of three trigger timescales (64 ms, 256 ms, and 1024 ms). Because GRB light curves have different peak intensities, spectral hardnesses, and variability, the times at which TTE windows start can be misaligned with the trigger times. The TTE window is so short that some of the TTE accumulation can occur prior to the trigger; when this happens the light curve found in the TTE window is incomplete. Additionally, the TTE window is not really a temporal window, but rather one based on photon count accumulation. If a burst is bright, then photon counts can fill the buffer quickly, and the TTE window will be shorter than 2 s. Examples of these effects are demonstrated in Figure \ref{fig:f1} for six TTE Partial pulses. 

Light curves for all fitted GRB pulses and their residuals are available from The Astrophysical Journal in the form of online electronic figures.

\subsection{Extended Burst Classification based on Duration, Hardness, and Fluence}\label{sec:class}

The 2 s temporal limit of the BATSE TTE window suggests that most TTE bursts belong to the Short GRB class. Rather than accepting this at face value using the ``$T_{90} \le 2$ s'' assertion, we prefer to classify GRBs using a formal statistical clustering or data mining technique. As discussed previously (see Section \ref{sec:intro}), formal techniques commonly prefer three GRB classes over two. Since the assignment of any GRB to a specific class depends on the classification technique being used, the burst characteristics being assessed, and the dataset being analyzed, we must choose a methodology for classifying the bursts in our sample. The three-class model obtained by \cite{hor06}, obtained through application of Principal Component Analysis (PCA) to GRB duration, fluence, and hardness data \citep{bag98,bal03}, provides a thoughtful systematic basis for classification. Unfortunately, \cite{hor06} do not provide classifications for many of the BATSE GRBs in our sample). 

We can extend the \cite{hor06} results to our TTE dataset using rules developed from supervised classification techniques. Our choice of a supervised classification algorithm is the decision tree C4.5 \citep{qui93}, found in the WEKA freeware suite of data mining tools under the J48 implementation \citep{fra16}. We use the classification results of \cite{hor06}, combined with the durations, fluences, and spectral hardnesses of all bursts found in the BATSE Final Catalog ({\tt http://gammaray.msfc.nasa.gov/batse/grb/catalog/ current/}). 
J48 uses previously-classified bursts (a training set) to identify simple {\tt IF THEN ELSE} classification branch rules. Each {\tt IF THEN ELSE} statement is a branch on this classification tree, and the terminal branches are referred to as {\em leaves}. A measure of entropy determines whether or not information is gained through the branching process. As it is being developed, a classification tree can be pruned if desired to eliminate sparsely-filled leaves.  Once a classification tree exists, the {\tt IF THEN ELSE} rules can be used to classify unknown objects while assigning to each a probability that each has been placed in the appropriate class. 

For this application of J48, we use the default parameter settings, which are known to generally provide a good performance \citep{fra16}. Pruning has been disabled because our goal is to extend rather than generalize the \cite{hor06} classification scheme. However, in order to avoid having rules that apply to individual bursts, the minimum number of objects per leaf has been set to two.

The resulting J48 tree has 14 leaves developed from 1557 GRBs using the attributes of logarithmic duration, hardness (channel 3/channel 2) and fluence, with a $97.88\%$ accuracy. The confusion matrix demonstrates the effectiveness of the resulting rule set: diagonal matrix elements indicate agreement between J48 and the original classification for 1524 Long (L), Short (S), and Intermediate (I) GRBs, while off-diagonal elements show how the remaining 33 GRBs are misclassified. \small{\tt{
\begin{tabbing}
1255 \= 10035 \= 3965 \= \kill
I  \>  L   \>  S   \>   <-- classified as \\
125 \>   5   \>  16  \>  |    I \\
6  \>   1003  \>  0  \>   |    L \\
5   \>   1  \>  396 \>  |    S \\
\end{tabbing}
}}
The final J48 tree shows the rules for each leaf, with parentheses indicating the number of originally- and re-classified GRBs placed in the leaf: \\ \small{\tt{
log(dur) <= 0.73 \\
|   log(dur) <= 0.382  \\
|   |   log(dur) <= -0.049: S (279.0) \\
|   |   log(dur) > -0.049 \\
|   |   |   log(hr) <= 0.307: I (15.0/5.0) \\
|   |   |   log(hr) > 0.307: S (111.0/8.0) \\
|   log(dur) > 0.382 \\
|   |   log(hr) <= 0.689: I (63.0) \\
|   |   log(hr) > 0.689 \\
|   |   |   log(dur) <= 0.561: S (22.0/8.0) \\
|   |   |   log(dur) > 0.561: I (21.0) \\
log(dur) > 0.73 \\
|   log(dur) <= 0.933 \\
|   |   log(hr) <= -0.143: I (11.0) \\
|   |   log(hr) > -0.143 \\
|   |   |   log(hr) <= 0.775 \\
|   |   |   |   log(hr) <= 0.1 \\
|   |   |   |   |   log(S) <= -6.005: I (4.0) \\
|   |   |   |   |   log(S) > -6.005 \\
|   |   |   |   |   |   log(S) <= -5.75: L (3.0) \\
|   |   |   |   |   |   log(S) > -5.75: I (2.0) \\
|   |   |   |   log(hr) > 0.1 \\
|   |   |   |   |   log(S) <= -6.31: I (3.0/1.0) \\
|   |   |   |   |   log(S) > -6.31: L (28.0/1.0) \\
|   |   |   log(hr) > 0.775: I (17.0/5.0) \\
|   |   log(dur) > 0.933: L (978.0/5.0) \\
}}

The Long and Short GRB classes are clearly identified at the ends of the duration distribution, as 97\% (983/1009) of the Long bursts have $T_{90} > 8.57$ s and 68\% (279/412) of the Short bursts have $T_{90} \le 0.89$ s. The rules for classifying Short GRBs are more successful ($97\%$ of the time for 398/412) when spectral hardness is also included (119 additional Short GRBs are characterized by 0.89 s $ < T_{90} \le 2.41$ s  and hr $> 2.03$). However, both Short and Intermediate bursts are found in time interval spanned by the TTE window (0.89 s $< T_{90} < 2$ s).

We can assign probable class membership to each previously unclassified GRB using these J48 rules. Some bursts are more difficult to classify because they lack one or more of the classification attributes, so J48 assigns greater uncertainty to these classifications. However, bursts lacking one or more classification attribute are rare, so most of the probabilities that a burst belongs to a preferred class exceed $90\%$. As a result, we consider burst classification probabilities of less than $90\%$ to represent questionable (denoted with a ``?'' in our catalog) class assignments. 

The classification results, shown in Table \ref{tab:complete}, verify that most ($\approx 90\%$) of the TTE GRBs belong to the Short GRB class. Use of these classification values allows us to examine the characteristics of Short GRB pulses with greater confidence than can be found by defining Short GRBs only as those having $T_{90} < 2$ s.

Ninety-one percent of the TTE Complete bursts and eighty-five percent of the TTE Partial bursts are Short. This difference results because TTE Partial bursts (which do not necessarily fit within the TTE window) are typically longer than TTE Complete bursts (which must fit within the TTE window). 
The large percentage of TTE Partial bursts belonging to the Short class suggests that our Short GRB sample is incomplete: Short bursts can be found with durations longer than the TTE window because at least some Short GRBs are longer than can be accommodated by the 2 s cutoff.

\begin{deluxetable*}{ccc}
\tablenum{1}
\tablecaption{Classification of TTE GRBs\label{tab:complete}}
\tablewidth{0pt}
\tablehead{
\colhead{Class} & \colhead{TTE Complete} & \colhead{TTE Partial}  \\
& 4 ms resolution & 64-ms resolution \\
}
\startdata
Short & 183 (+4) & 145 (+9) \\
Intermediate & 2 (+4) & 12 (+7) \\
Long & 5 (+8) & 3 (+5) \\
\enddata
\tablecomments{Parentheses () identify additional GRBs having less than certain classifications ($p < 90\%$).}
\end{deluxetable*}

\subsection{Pulse Classification based on Complexity}\label{sec:complex}

We use an iterative, heuristic approach to GRB pulse classification in order to recognize our partial yet still incomplete understanding of GRB pulse behaviors. This approach allows us to find and account for pulse behaviors that we expect while also allowing us to search for behaviors that we may not anticipate. An approach of this type (part statistical, part data mining) is needed because GRB pulses do not act like pulses in the standard sense of the term.

The Oxford English Dictionary defines a {\em pulse} as ``a single vibration or short burst of sound, electric current, light, or other wave." The standard assumption has been that that GRB pulses are monotonic structures overlaying stochastically varying backgrounds. Although this assumption has been applied almost universally in GRB pulse fitting, the residuals of isolated GRB pulses demonstrate that it is not always valid \citep{hak14, hak15}. There are negative repercussions to the measurement of GRB pulse properties if pulse monotonicity is assumed but not present. The assumption of pulse monotonicity serves to fragment larger structures and replace them with separate monotonic pieces. Using a monotonic pulse model instead of one that accounts for possible structure thus results in the recovery of more pulses, having shorter durations, and separated by smaller separations. These pieces cannot themselves have structure because any non-stochastic structure will be fragmented into smaller pulses by the monotonicity assumption. Important temporal correlations, such as the hard-to-soft spectral evolution characteristic of triple-peaked pulses, will go unrecognized and will be replaced by the less-pronounced spectral characteristics of individual monotonic pulse pieces.

The Oxford English Dictionary defines {\em complexity} as ``the state or quality of being intricate or complicated." Complexity is used rather vaguely in GRB analysis to define structure in GRB light curves. We have demonstrated \citep{hak14, hak15} that triple-peaked residual functions are responsible for one component of GRB pulse complexity. Because of this, we are open to the idea that other, more structured variations might also be present and extractable from GRB pulses. A useful technique for classifying GRB pulses is thus one that recognizes known pulse complexities while presuming that other unknown complexities might also exist.

GRB pulse light curves of Long and Intermediate bursts exhibit various degrees of structure and complexity. The smooth \cite{nor05} pulse model works best at fitting faint Long and Intermediate GRB pulses; brighter pulses show evidence of more complex structures such as the triple-peaked residual function \citep{hak14, hak15}. The existence of this non-monotonically increasing and decreasing complexity is problematic, as pulse-fitting techniques sometimes have difficulty determining whether complexity in a GRB emission episode results from identifiable substructures or if it represents embedded fainter pulses. To complicate matters, bright emission episodes often exhibit additional chaotic variations not present in faint ones. 

We note that some TTE bursts are characterized by what appear to be closely overlapping emission episodes. These are the most difficult events for us to fit because they could either represent two or more overlapping pulses or a single pulse that has an exceedingly complex temporal structure. We recognize that there will always be ambiguity in separating multi-pulsed emission episodes from multi-peaked pulses, and we have made efforts to adequately document these ambiguous cases. It is fortunate for our analysis that the emission episodes of most TTE GRBs are clearly defined.

We approach the identification of GRB pulse complexity by assuming to first order that each isolated emission episode represents a single GRB pulse that can be fitted by the \cite{nor05} model. We further assume that the simplest form of complexity, representing a second-order variation in the monotonic pulse structure, is the smoothly-varying triple-peaked \cite{hak14} residual function. We use $\chi^2_\nu$ as our goodness-of-fit statistic to determine the effectiveness of these models, where $\chi^2$ indicates the normalized deviation between the data and the model relative to the degrees of freedom $\nu$. The value of $\chi^2_\nu$ is dependent on the amount of non-stochastic emission (signal) relative to the amount of stochastic emission (noise), and thus to the temporal interval selected for the analysis. The number of bins (and thus the value of $\nu$) is sensitive to the bin size as well as to the temporal interval. We choose the fiducial timescale for pulse analysis because it has been defined to contain most of the pulse emission.

Our previous analyses of Long/Intermediate BATSE, Fermi GBM, and Swift GRB pulses have found that this approach generally results in fits having $\chi^2_\nu \approx 1$ in low signal-to-noise environments consistent with stochastic variations to the model. The model is less successful for pulses showing some complexity. Many of these less well-fit pulses exhibit larger precursor and/or decay peaks or additional faint residual structures that occur in addition to the residual fit.  Because the \cite{hak14} function still contributes to these pulse residuals, these findings suggest that the model is not incorrect so much as it is {\em incomplete} in accounting for augmented pulse structures.

We choose to define pulse complexity in terms of the $\chi^2_\nu$ $p-$values. Here, $\chi^2$ is defined over the fiducial timescale and $\nu$ from the \cite{nor05} model is the number of temporal bins minus the number of pulse- and background-fit parameters -- two for the background and four for a single pulse. The $p-$value associated with $\chi^2_\nu$ has the conventional meaning: is the probability that a $\chi^2$  statistic having $\nu$ degrees of freedom is more extreme than the measured value. As will be seen, the binning most appropriate for fitting a GRB pulse depends on the pulse's flux distribution: an optimal fit has sufficient signal-to-noise over the duration of the pulse to identify and match available structures. The burst sample is limited to 64-ms resolution outside the TTE window, which makes it difficult to fit shorter TTE Partial pulses and limits the number of bins being fitted. Inside the TTE window we have chosen to use 4ms resolution. The number of bins available for a pulse fit depends on a pulse's duration and shape, as these quantities define the fiducial time interval. Thus, the optimal choice of a bin size cannot is better made after a fit has been performed.

By limiting our comparisons to this fiducial interval we minimize the chance that a good fit is obtained simply because it uses a large number of background bins, and this approach assures that the $\chi^2$ fits are generally independent of pulse duration (exceptions can occur for fiducial timescales that extend beyond the TTE window; this can have the unfortunate effect of artificially increasing $\chi^2_\nu$ by decreasing the number of available background bins). We consider good fits (indicating relatively smooth light curves) to be those having best-fit $p-$values of $p_{\rm best} \ge 5 \times 10^{-3}$ (this is a standard choice for a ``good" fit criterion). A $\Delta \chi^2$ test is used to indicate the residual function needs to be included in the fit: $\Delta \chi^2$ is the difference in $\chi^2$ obtained from the \cite{nor05} model minus that obtained from the \cite{nor05} model combined with the \cite{hak14} residual model. The difference in the number of degrees of freedom between these fits is four per pulse. We require a $\Delta \chi^2$ $p-$value of $p_{\Delta} \le 10^{-3}$ for the model to be improved. This is more stringent than the criterion we use for a ``good'' fit because we want to ensure that the residual function, rather than some other structure, is most likely to be responsible for the improvement in the fit.

We have classified the TTE pulses into four groups using complexity as our classification parameter. We identify {\em Simple} pulses as those best characterized by the \cite{nor05} function alone. {\em Blended} pulses are explained by the \cite{nor05} function but also require the \cite{hak14} residual function to obtain a best-fit value of $p_{\rm best} \ge 5 \times 10^{-3}$. Many pulse fits are significantly improved by the \cite{hak14} residual function, but exhibit additional un-modeled structures. This may in part result from the inability of our empirical models to explain the true evolution of the light curve. We define {\em Structured} pulses as those having best-fit $p-$values of $10^{-5} \le p_{\rm best} < 5 \times 10^{-3}$; these pulses have many characteristics that can be explained by the pulse and residual models, but also have statistically significant variations from these structures. The value $p=10^{-5}$ appears somewhat arbitrary but has been selected for its potential use as a data mining attribute. This $p-$value subdivides complex pulses into two groups, such that 1) Structured pulses are expected to have characteristics common with Blended pulses, but also (perhaps to a lesser extent) with more complex pulses, and 2) the number of Structured pulses in our sample (50) is similar to the number of Blended pulses (38). The remaining pulses have complicated light curves as defined by $p_{\rm best} < 10^{-5}$. Some of these {\em Complex} pulses represent emission episodes having pronounced structures that might represent complex substructures, overlapping inseparable pulses, or a different physical phenomenon altogether. Although we have classified the TTE pulses according to these $p_{\rm best}$ values, the BATSE TTE GRB pulse catalog includes all $p-$values so that users may adjust these classification parameters as they wish.

In addition to finding some TTE emission episodes that upon re-examination appear to be overlapping pulses, we have found a small number that do not fit the existing pulse paradigm. The final pulses in these bursts are characterized by intensities that increase with time, producing asymmetric pulse shapes that are contrary to the intensity distribution function of \cite{nor05}. We call bursts containing these pulse structures {\em Crescendo} bursts. Some, but not all, Crescendo bursts are preceded by a series of short, symmetric {\em Staccato} pulses. Crescendo GRBs are discussed in greater depth in Section \ref{sec:crescendo}.

Two GRBs in this sample (BATSE triggers 1626 and 7427) have been identified by \cite{nor06} as being Short GRBs with extended emission. The residuals of both of these bursts show faint emission indicative of a brightening followed by a gradual decline, which is suggestive of a faint pulse structure rather than chaotic emission. However, the signal-to-noise of this extended emission is too faint to attempt to fit with the pulse model. An additional eight Short GRBs (four TTE Complete and four TTE Partial) are listed in \cite{bos13} as having extended emission (TTE Complete triggers 575, 1719, 5592, and 5634; TTE Partial triggers 3611, 3940, 7063, and 7599). Of these bursts, only trigger 575 exhibits extended emission during the TTE readout. We note that the double-pulsed nature of BATSE trigger 575 also makes it unique among the sample of Short GRBs with extended emission.

\begin{figure}
\plottwo{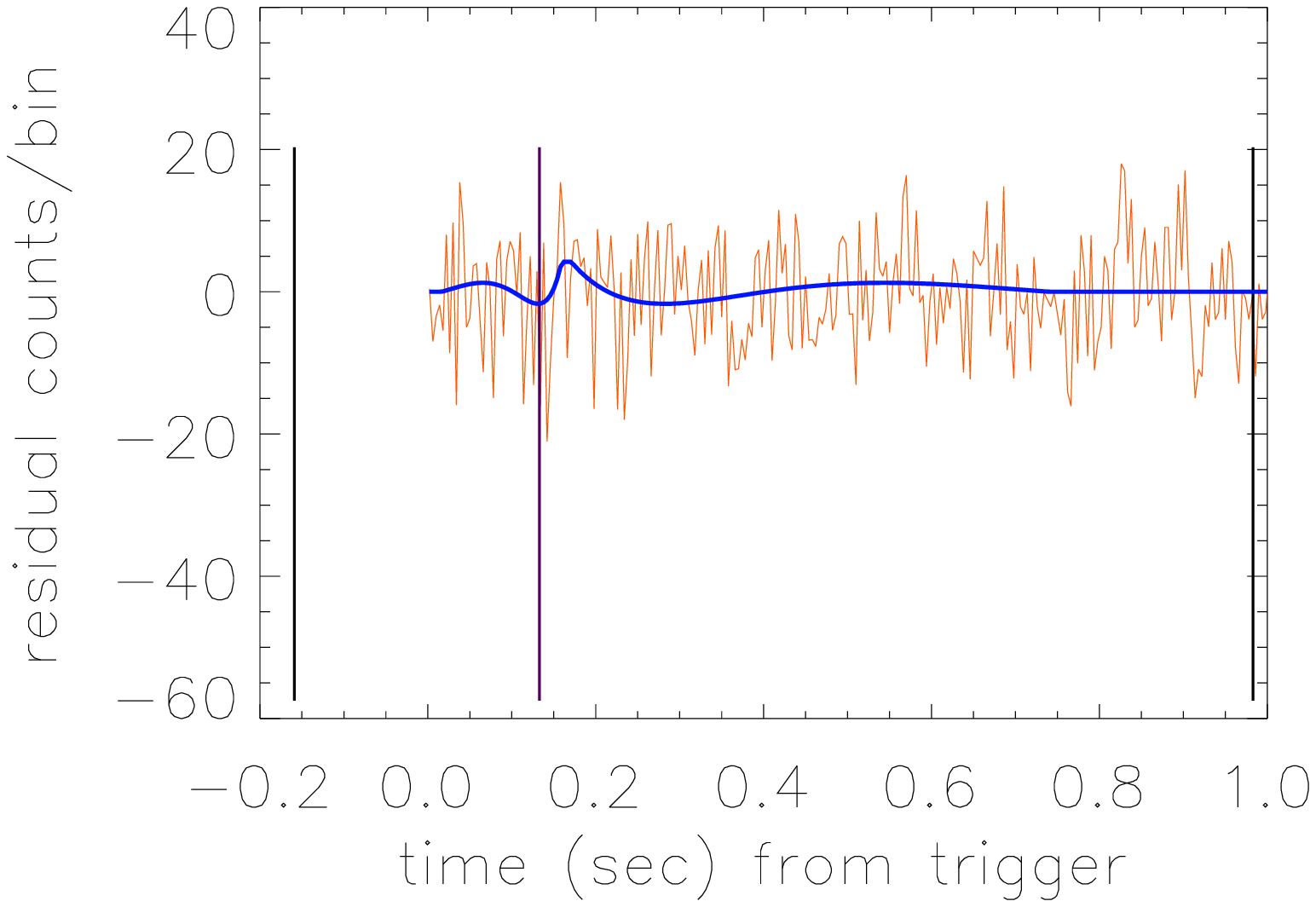}{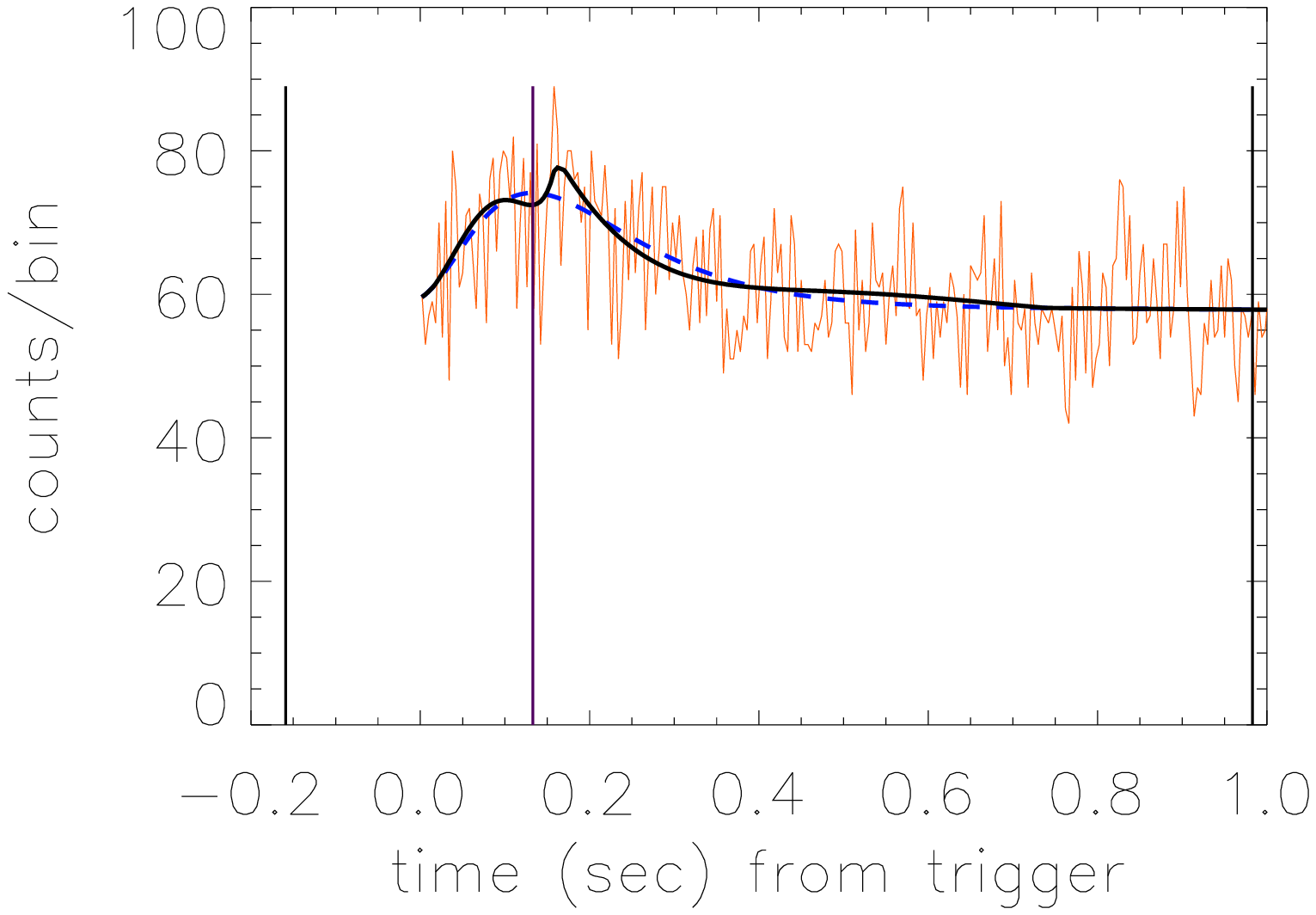}
\caption{Example of a Simple Short GRB pulse (Trigger 373). The left panel shows the \cite{hak14} residual structure, while the right panel shows both the \cite{nor05} pulse fit (dotted line) and the combined \cite{nor05} pulse plus \cite{hak14} residual fit (solid line). For Simple pulses, the residual structure is insignificant and is not used in the final fit.\label{fig:complexex1}}\end{figure}
\begin{figure}
\plottwo{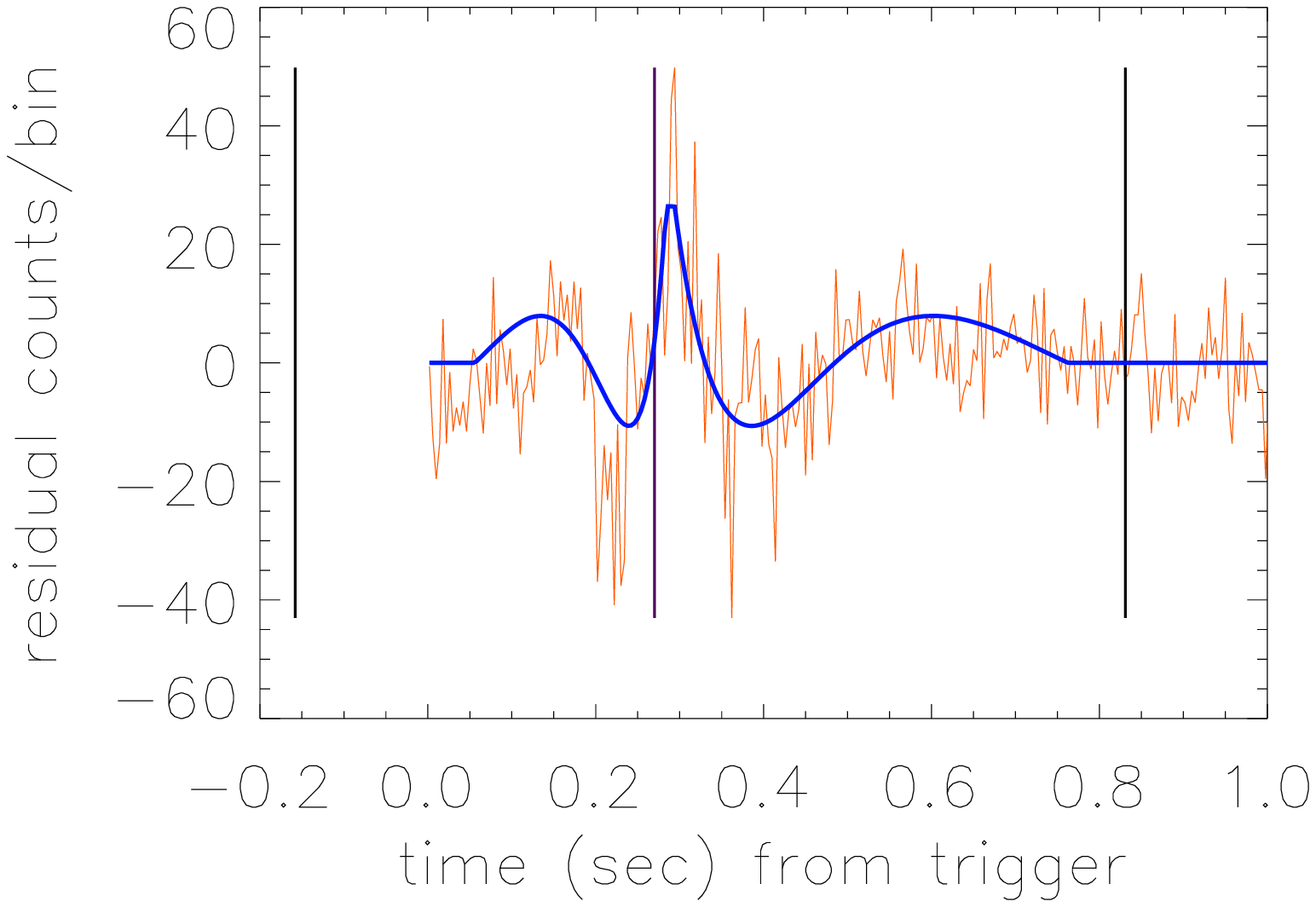}{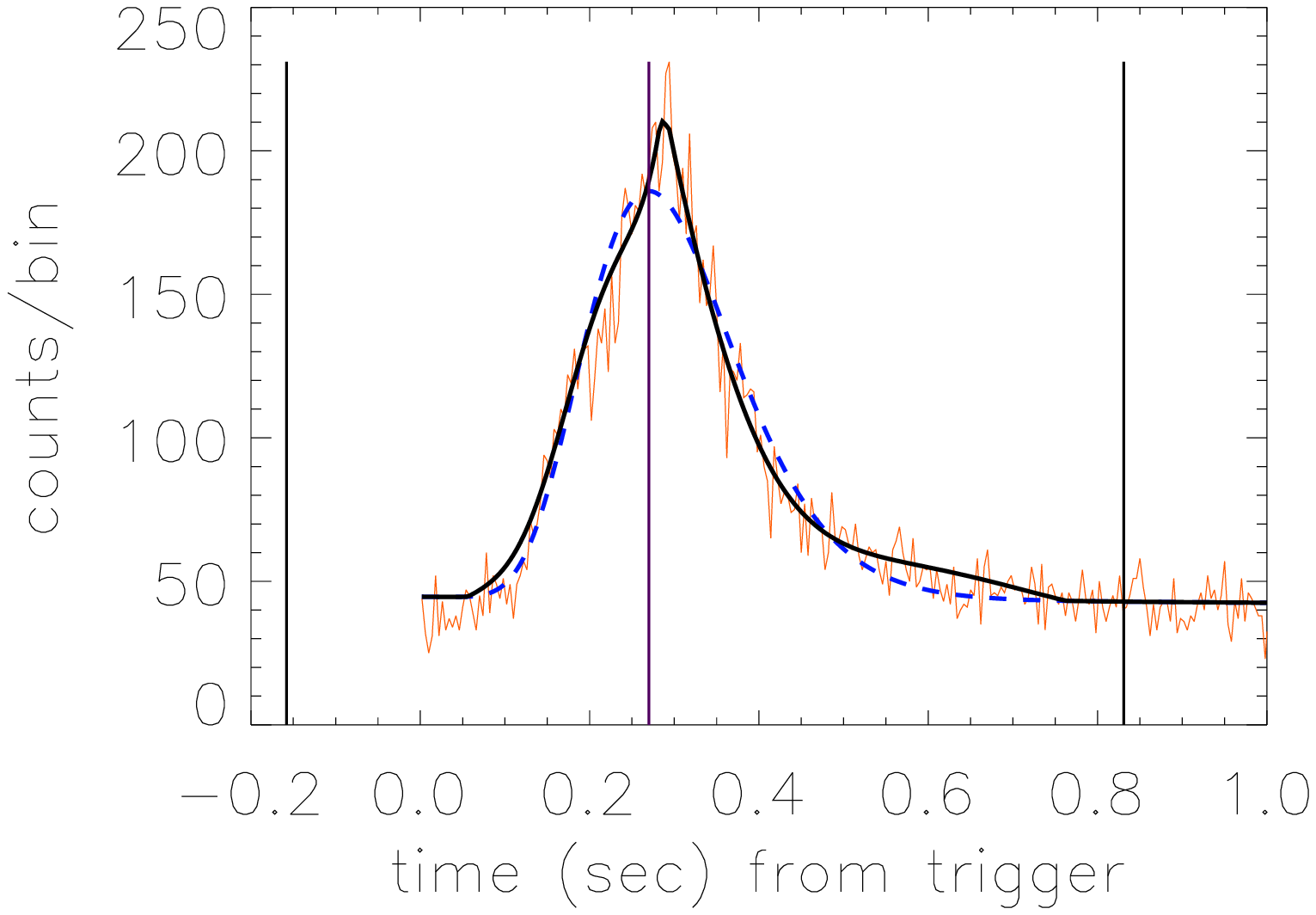}
\caption{Example of a Blended Short GRB pulse (Trigger 2896). The left panel shows the \cite{hak14} residual structure, while the right panel shows both the \cite{nor05} pulse fit (dotted line) and the combined \cite{nor05} pulse plus \cite{hak14} residual fit (solid line). The residual structure of Blended pulses is significant. 
\label{fig:complexex2}}\end{figure}
\begin{figure}
\plottwo{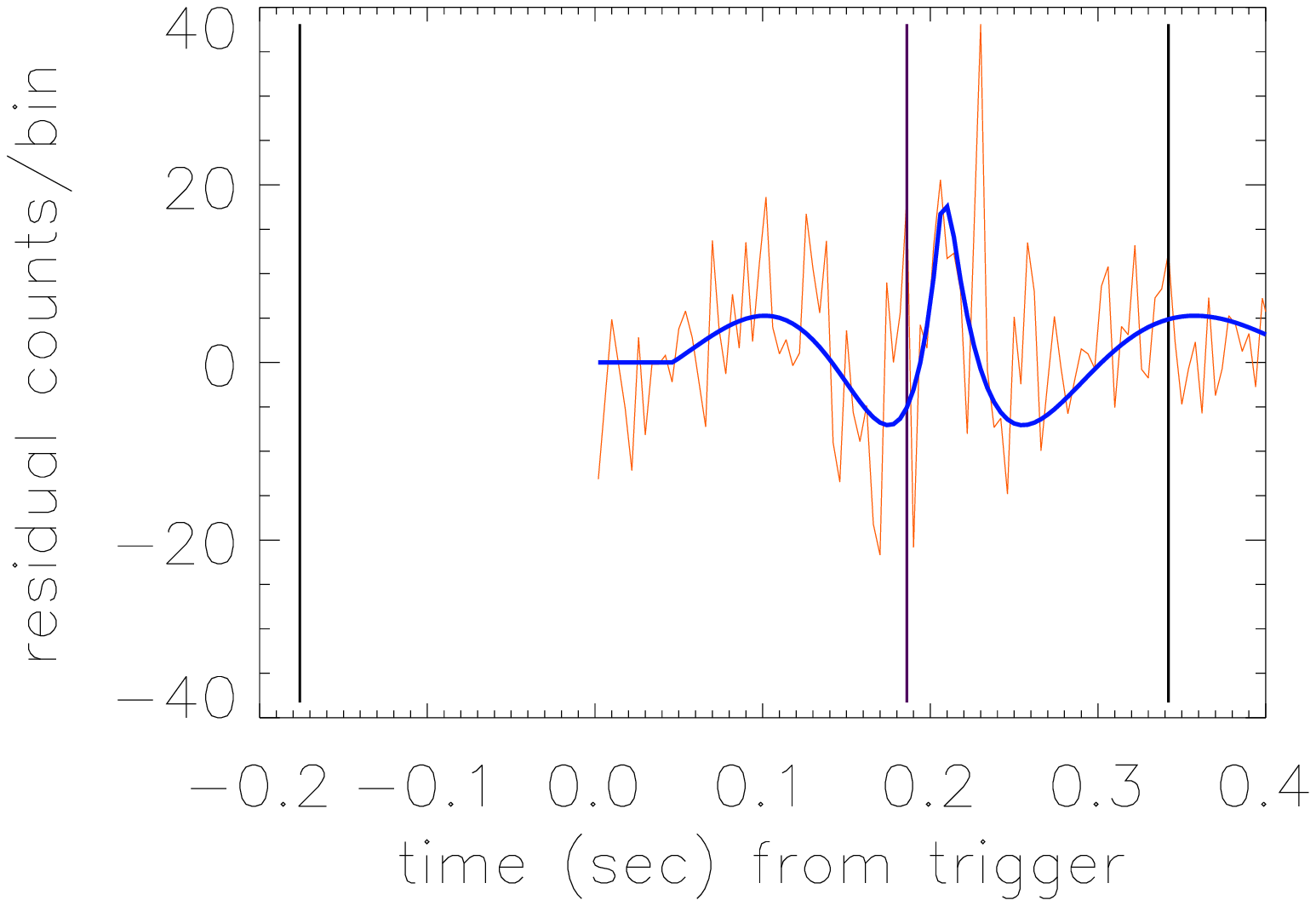}{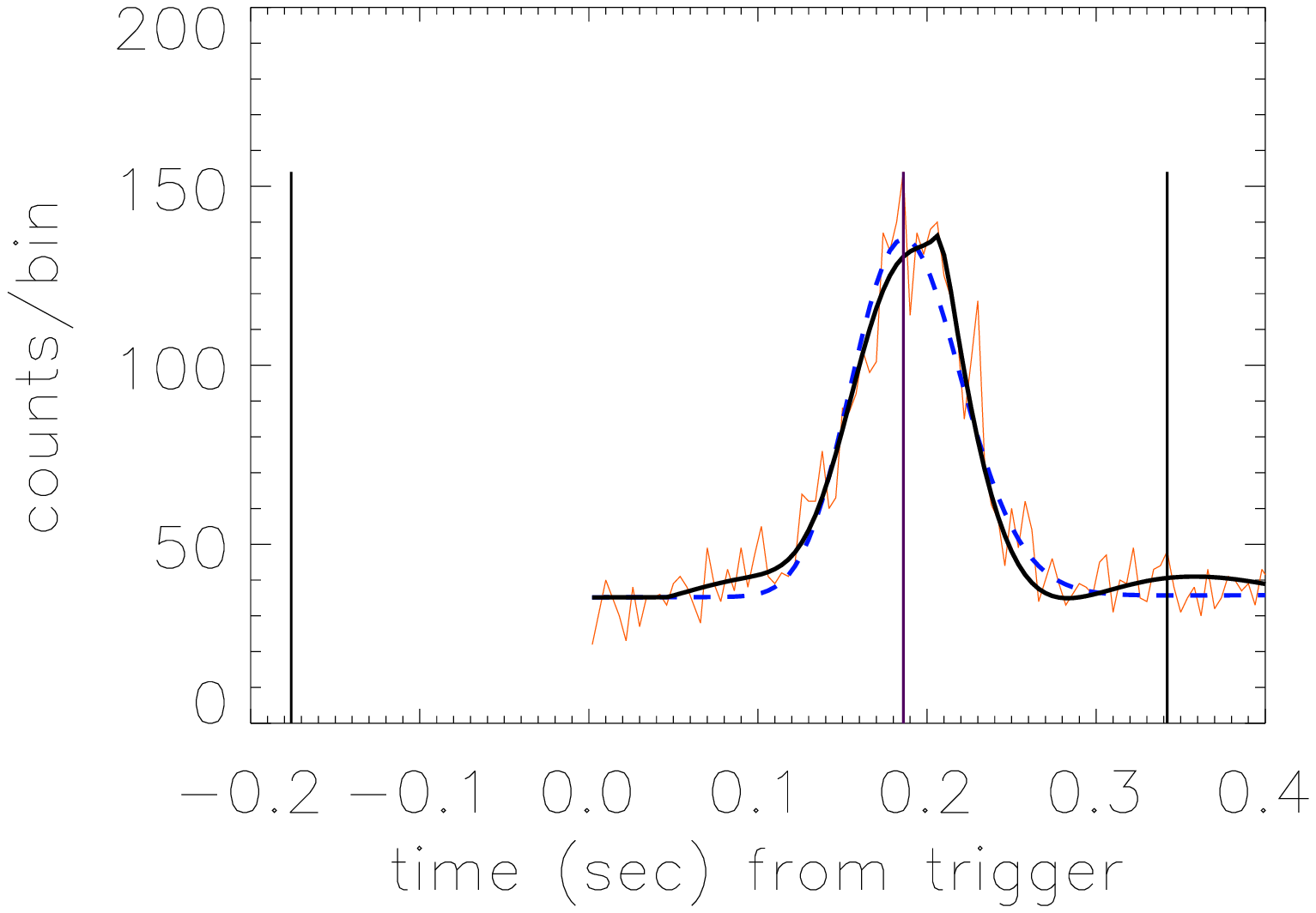}
\caption{Example of a Structured Short GRB pulse (Trigger 5564). The left panel shows the \cite{hak14} residual structure, while the right panel shows both the \cite{nor05} pulse fit (dotted line) and the combined \cite{nor05} pulse plus \cite{hak14} residual fit (solid line). Structured pulses exhibit residual structures so significant that they cannot be entirely explained by the \cite{hak14} model. 
\label{fig:complexex3}}\end{figure}
\begin{figure}
\plottwo{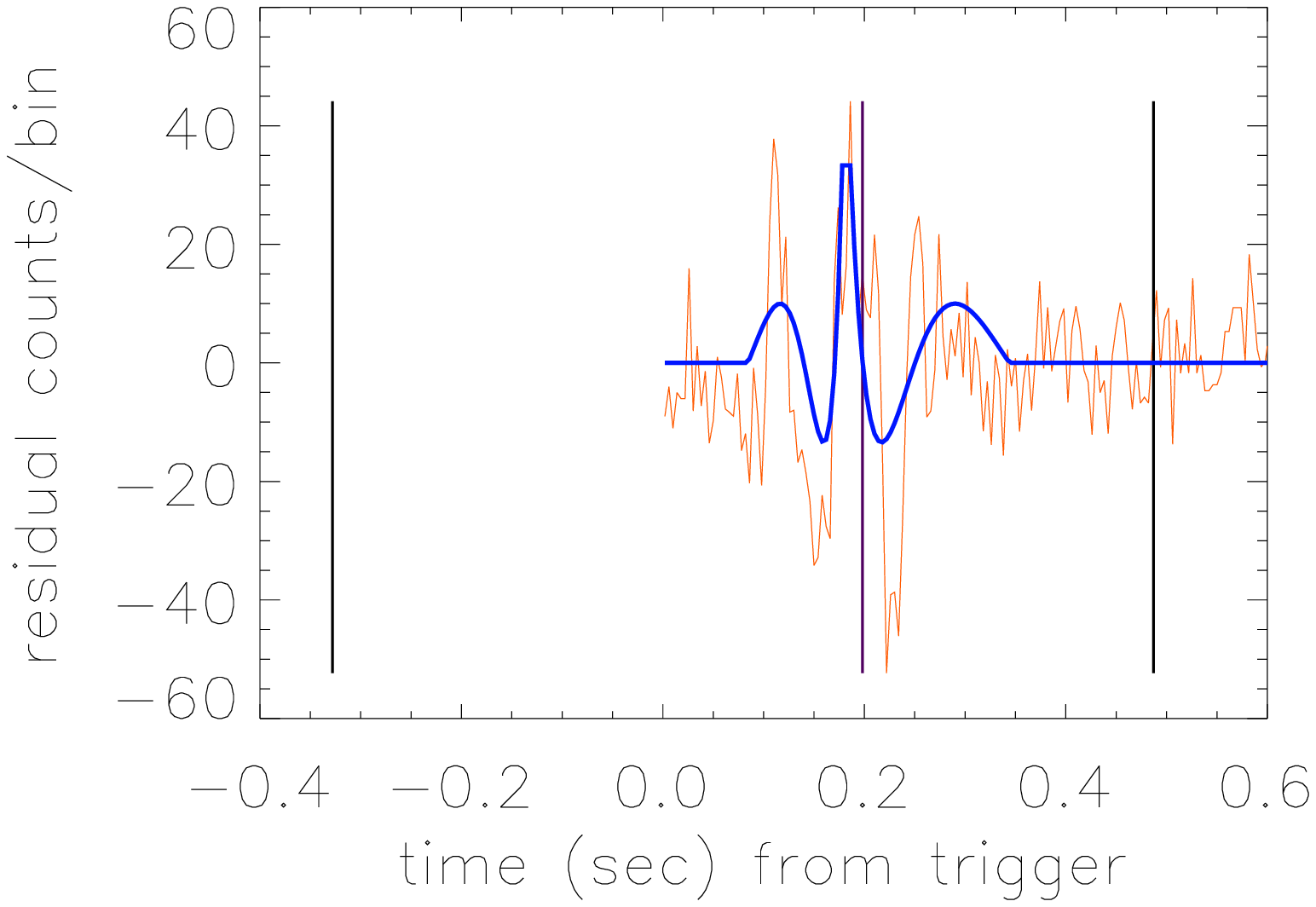}{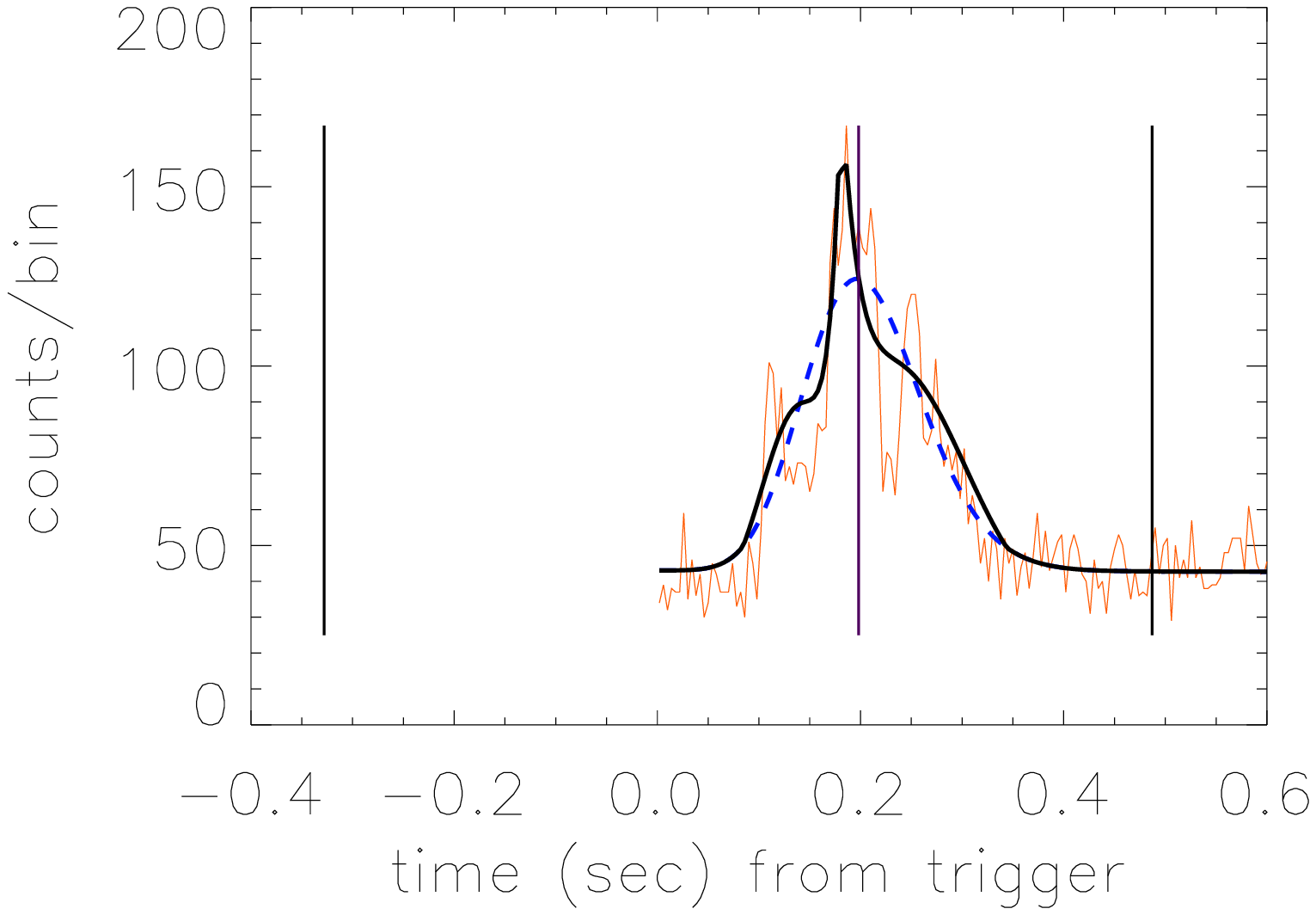}
\caption{Example of a Complex Short GRB pulse (Trigger 4955). The left panel shows the \cite{hak14} residual structure, while the right panel shows both the \cite{nor05} pulse fit (dotted line) and the combined \cite{nor05} pulse plus \cite{hak14} residual fit (solid line). Complex pulses exhibit significant residual structures that cannot be explained by the \cite{hak14} model. 
\label{fig:complexex4}}\end{figure}

Examples of representative pulse fits are shown in Figures \ref{fig:complexex1} $-$ \ref{fig:complexex4}. Figure \ref{fig:complexex1} shows an example of a Simple pulse (Trigger 373); the left panel shows the residual structure while the right panel shows both the \cite{nor05} fit (dotted line) and the combined fit (solid line). 
Figure \ref{fig:complexex2} similarly shows a Blended pulse (Trigger 2896), 
Figure \ref{fig:complexex3} shows a Structured pulse (Trigger 5564), 
and Figure \ref{fig:complexex4} shows a Complex pulse (Trigger 4955).

The complexity classifications of all 434 pulses in our catalog are summarized in Table \ref{tab:complex}. Although the total numbers of TTE Complete and TTE Partial pulses are similar, the distributions of Simple, Blended, Structured, and Complex pulses are noticeably different. This is a direct result of the better temporal resolution of the 4 ms data relative to the 64-ms data. Whereas the 4 ms binning allows substructures (including separable pulses and the residual function) to be identified and fitted, the 64-ms binning merges these together an prevents them from being properly delineated for fitting.  Thus, the lower-resolution Partial TTE group contains more Structured and Complex pulses and fewer Simple and Blended pulses. Examination of the TTE data for the bursts in this group shows that these structures and separable pulses are present, but just not resolved in the 64-ms data (for example, see the pulses in Figure \ref{fig:f1}). 

\begin{deluxetable*}{ccc}
\tablenum{2}
\tablecaption{Complexity of TTE GRB Pulses\label{tab:complex}}
\tablewidth{0pt}
\tablehead{
\colhead{Pulse Complexity} & \colhead{TTE Complete} & \colhead{TTE Partial}  \\
& 4 ms resolution & 64-ms resolution \\
}
\startdata
Simple & 133 & 89 \\
Blended & 31 & 7 \\
Structured & 20 & 30 \\
Complex & 43 & 71 \\
Staccato & 6 & 4 \\
\bf{Total} & 234 & 201 \\
\enddata
\end{deluxetable*}

\subsection{Multi-Pulsed Bursts}\label{sec:multi_intro}

The vast majority of the Catalog bursts ($345/387=89\%$) are single-pulsed. Most of the remainder are double-pulsed ($33/387=9\%$) and a few ($3/387=1\%$) are triple-pulsed. Six ($6/387=2\%$) are Crescendo bursts, having pulse structures similar to one another but different than that described by the standard pulse paradigm. We have likely underestimated the number of very short multi-pulsed GRBs having overlapping pulses because temporal resolution and pulse complexity make it hard to disentangle overlapping TTE pulses.  On the other hand, our attempt to fit all sampled TTE pulses gives us confidence that we have identified most of the multi-pulsed TTE GRBs. Table \ref{tab:multiplicity} summarizes the multiplicity of TTE GRBs in this sample. A list of the multi-pulsed GRBs is provided in Table \ref{tab:double}. 

Although $90\%$ of the Catalog bursts are single-pulsed, we cannot unequivocally state that $90\%$ of Short GRBs are single-pulsed. Because our sample contains only Short GRBs that entirely or mostly fit into the TTE window, we are missing the long-duration tail of the Short GRB distribution ({\em e.g.,} those with $T_{90} > 2$ s), and these missing bursts might well contain multiple pulses. However, we also believe that this long duration tail is small because Short bursts are separated from the other GRB classes with the most clearly delineated boundary (see \cite{muk98,hor98,hak03}), and many bursts with $T_{90} > 2$ s  have soft the spectra of Intermediate GRBs. Thus, we feel confident in stating that Short GRBs are overwhelmingly single-pulsed.


\begin{deluxetable*}{lcc}
\tablenum{3}
\tablecaption{TTE GRB Multiplicity \label{tab:multiplicity}}
\tablewidth{0pt}
\tablehead{
\colhead{Burst Type} & \colhead{TTE Complete} & \colhead{TTE Partial}  \\
& 4 ms resolution & 64-ms resolution \\
}
\startdata
Single-pulsed & 180 & 163 \\
+ extended emission & $2^\dagger$ & $0^\dagger$ \\
Double-pulsed & 18 & 15 \\
Triple-pulsed & 2 & 1 \\
Crescendo & 4 & 2 \\
\bf{Total} & 206 & 181 \\ 
\enddata
\tablecomments{The extended emission GRBs identified by \cite{nor06} show excess flux in the TTE interval.  Additionally, three single-pulsed TTE Complete and four TTE Partial GRBs have been described by \cite{bos13} as exhibiting extended emission. along with one double-pulsed TTE Complete GRB ($\dagger$). Since these do not exhibit prompt extended emission, they have not been identified as extended emission bursts.}
\end{deluxetable*}

\begin{figure}
\plottwo{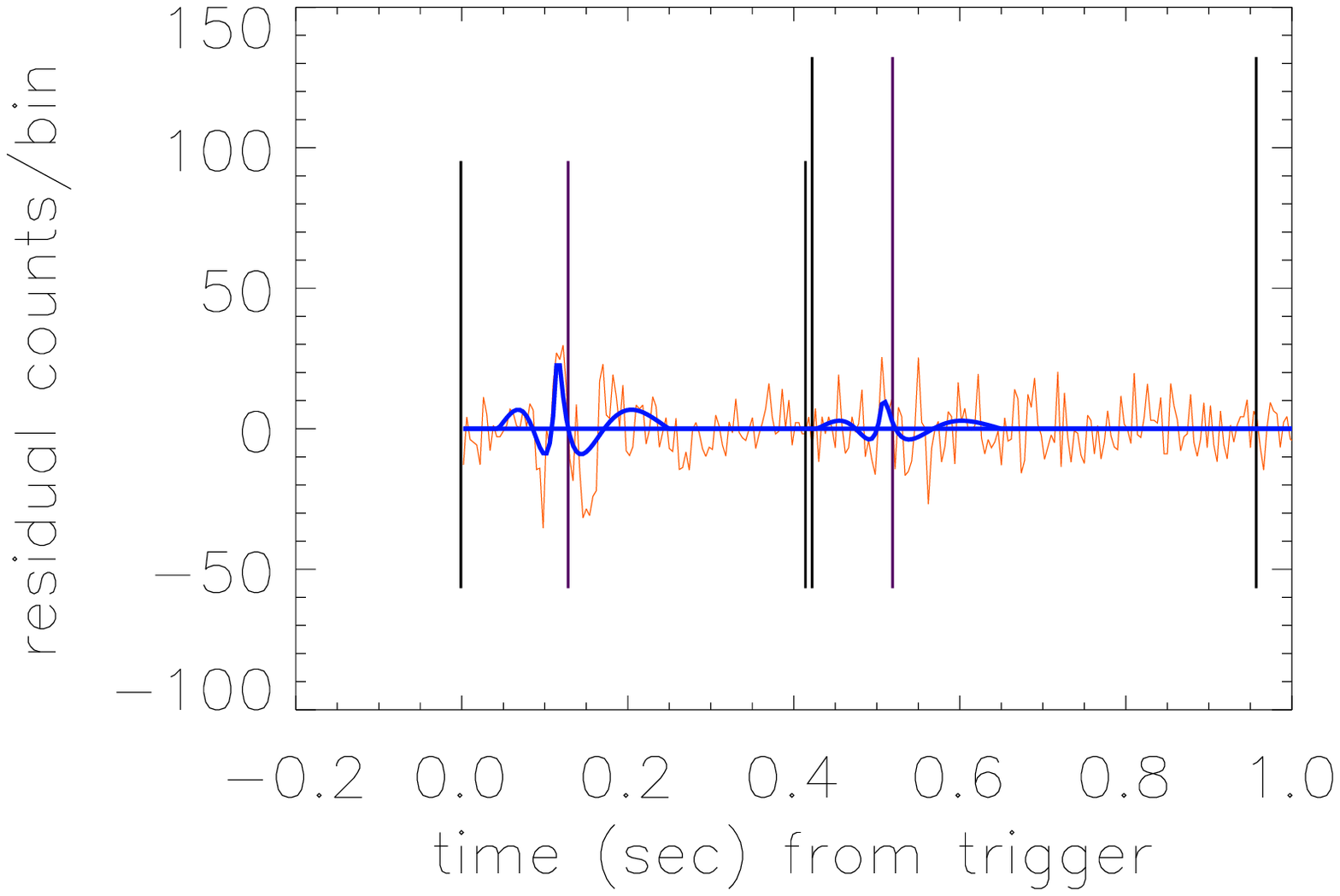}{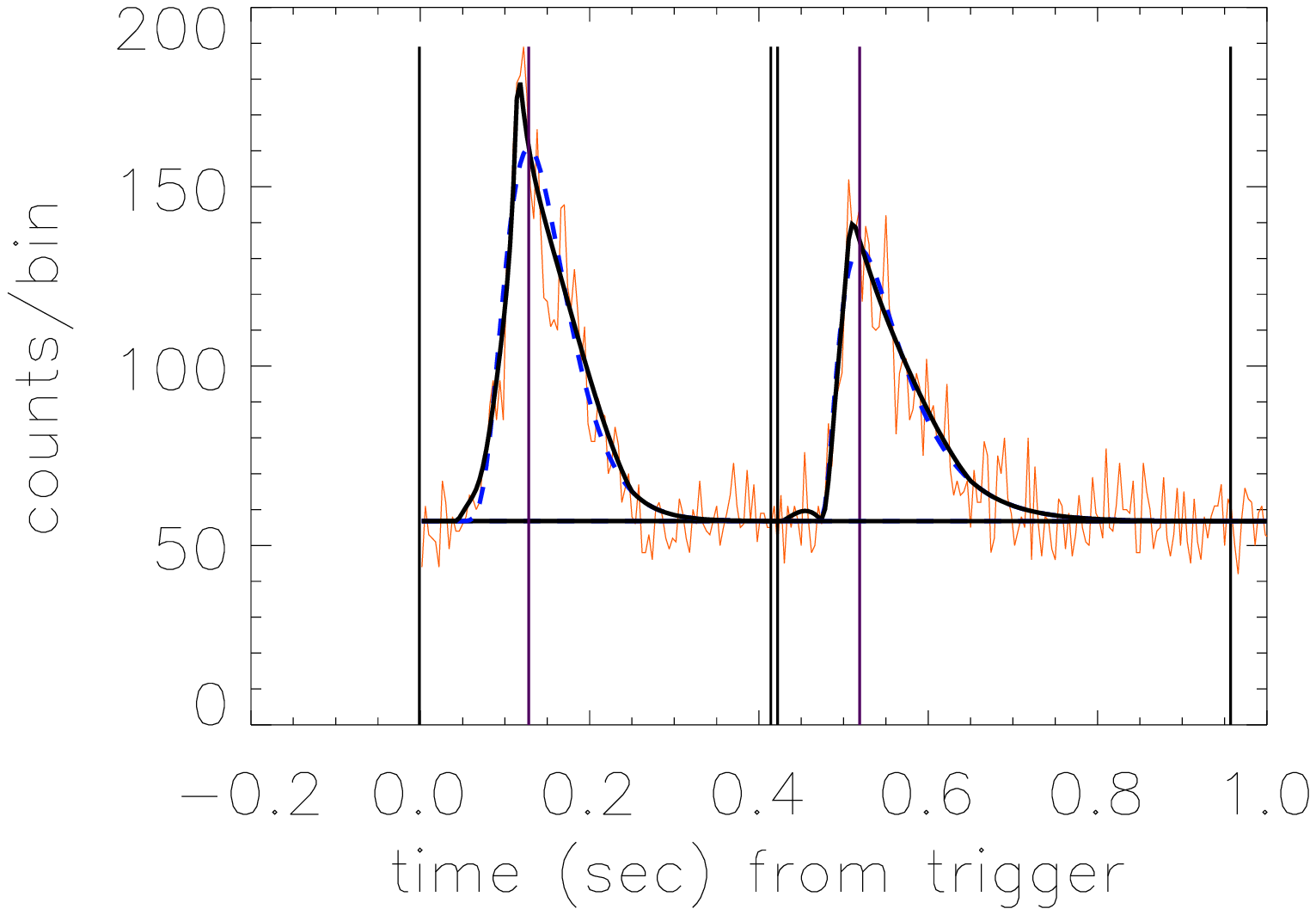}
\caption{Complex structures of the two pulses in BATSE Trigger 3770. The left panel shows the \cite{hak14} residual structure, while the right panel shows both the \cite{nor05} pulse fit (dotted line) and the combined \cite{nor05} pulse plus \cite{hak14} residual fit (solid line). Pulse 3770a is a Blended pulse, while 3770b is a Simple pulse.  \label{fig:complexdouble}}
\end{figure}

\begin{deluxetable*}{llll}
\tablenum{4}
\tablecaption{List of multi-pulsed GRBs. All GRBs have been classified as Short (S) except Triggers 5439, 7514, and 7559 which are Intermediate (I). \label{tab:double}}
\tablewidth{0pt}
\tabletypesize{\scriptsize}
\tablehead{
\colhead{BATSE ID} & \colhead{Class} & \colhead{No. Pulses} & \colhead{Resolution} 
}
\startdata
298 & S & 2 & 64 ms \\
551 & S & 2 & 4 ms \\
575 & S & 2 & 4 ms \\
867 & S? & 2 & 64 ms \\
936 & S & 2 & 64 ms \\
1453* & S & 2 & 4 ms \\
1694 & S & 2 & 4 ms \\
1747	 & S? & 2 & 4 ms\\
2217 & S & 2 & 64 ms \\
2330	 & S & 2 & 4 ms \\
2485	 & S & 2 & 4 ms \\
2715 & S & 2 & 4 ms \\
2776 & S & 2 & 64 ms \\
2834 & S & 2 & 64 ms \\
2860	 & S & 2 & 64 ms \\
2861	 & S & 2 & 64 ms \\
2918 & S & 2 & 4 ms \\
2952	 & S & 3 & 4 ms \\
2975 & S & 2 & 4 ms \\
3173* & S & 2 & 4 ms \\
3735* & S & 4 & 64 ms \\
3736	 & S & 2 & 64 ms \\
3770 & S & 2 & 4 ms \\
3791 & S & 2 & 4 ms \\
5212 & S & 2 & 4 ms \\
5439* & I & 3 & 4 ms \\
5529 & S & 3 & 4 ms \\
5633 & S & 2 & 4 ms \\
7273	 & S & 2 & 4 ms \\
7281 & S & 3 & 64 ms \\
7305 & S? & 2 & 64 ms \\
7375* & S & 2 & 4 ms \\
7378 & S & 2 & 64 ms \\
7514 & I? & 2 & 4 ms \\
7559 & I & 2 & 64 ms \\ 
7830	 & S & 2 & 64 ms \\
7912	 & S & 2 & 64 ms \\
7943	 & S? & 2 & 64 ms \\
8072	 & S & 2 & 4 ms \\
8079	 & S & 2 & 4 ms \\
8120	 & S & 2 & 64 ms \\
\enddata
\tablecomments{*Crescendo GRB}
\end{deluxetable*}

\subsection{Catalog Description} 

The BATSE TTE GRB pulse catalog is contained in three separate online files. Part I (Tables \ref{tab:cat1head} and  \ref{tab:cat1}) contains information related to the \cite{nor05} model fit. Part II (Tables \ref{tab:cat2head} and \ref{tab:cat2}) contains information pertaining to the \cite{hak14} residual fit (if available), descriptions of the overall pulse fit, and ancillary information such as pulse fluence and energy hardness. Part III (Table \ref{tab:cat3}) contains the names of the files containing both the residual fits and the total fits to the pulse light curves, as well as comments about the pulses.  These three files may be merged by the user to create a larger table, if desired. 

As indicated previously, this catalog strives to systematically create a characterization of Short GRB prompt emission by attempting to fit all Short GRB pulses that fit entirely or partially within BATSE's TTE window. We want to understand and characterize as many of our selection biases as we can. We believe that the selection of TTE bursts is random during the time of BATSE's operation. We have attempted to describe uncertainties in the pulse-fitting process, as well as characterizing uncertainties in the measured pulse parameters. We have created four descriptors based on our definitions of good fits: the temporal resolution (Table \ref{tab:cat1}, column 2), the pulse complexity (Table \ref{tab:cat2}, column 22), the GRB class (Table \ref{tab:cat2}, column 20), and the decision whether or not to include the residual fit in the final best fit (Table \ref{tab:cat2}, column 16).

\begin{deluxetable*}{cllll}
\tablenum{5}
\tablecaption{Information contained in the BATSE TTE GRB Pulse Catalog (part I) \label{tab:cat1head}}
\tablewidth{0pt}
\tabletypesize{\scriptsize}
\tablehead{
\colhead{Table Column} & \colhead{Header} & \colhead{Variable} & \colhead{Units} & \colhead{Description} 
}
\startdata
1 & pulse\_id & BATSE trigger & $-$ & number (+ letter) \\
2 & resolution & resolution & $-$ & 4ms for TTE Complete or 64ms for TTE Partial \\
3 & B & $B_0$ & cts & mean background per bin \\
4 & B\_err & $\sigma_{B0}$ & cts & mean background uncertainty per bin \\
5 & BS & $BS$ & cts/s & background rate change per bin \\
6 & BS\_err & $\sigma_{BS}$ & cts/s & background rate change uncertainty per bin\\
7 & ts & $t_s$ & s & pulse start time, from Equation \ref{eqn:function} \\
8 & ts\_err & $\sigma_{ts} $ & s & pulse start time uncertainty \\
9 & A & $A$ & cts & pulse amplitude, from Equation \ref{eqn:function}  \\
10 & A\_err & $\sigma_A$ & cts & pulse amplitude uncertainty \\
11 & tau1 & $\tau_1$ & s & pulse rise parameter, from Equation \ref{eqn:function} \\
12 & tau1\_err & $\sigma_{\tau1}$ & s & pulse rise parameter uncertainty \\
13 & tau2 & $\tau_2$ & s & pulse decay parameter, from Equation \ref{eqn:function}  \\
14 & tau2\_err & $\sigma_{\tau2}$ & s & pulse decay parameter uncertainty \\
15 & w & $w$ & s & pulse duration, from Equation \ref{eqn:duration} \\
16 & w\_err & $\sigma_w$ & s & pulse duration uncertainty \\
17 & kappa & $\kappa$ & $-$ & pulse asymmetry, from Equation \ref{eqn:asymmetry} \\
18 & kappa\_err & $\sigma_\kappa$ & $-$ & pulse asymmetry uncertainty \\
19 & tau\_pk & $\tau_{\rm peak}$ & s & pulse peak time, from Equation \ref{eqn:taupeak} \\
20 & tau\_pk\_err & $\sigma_{\tau \rm{pk}}$ & s & pulse peak pulse time uncertainty \\
21 & t\_start & $t_{\rm start}$ & s & fiducial start time, from Equation \ref{eqn:tstart} \\
22 & t\_end & $t_{\rm end}$ & s & fiducial end time, from Equation \ref{eqn:tend} \\
23 & chi\string^2 & $\chi^2$ & $-$ & goodness of fit for pulse + background model \\
24 & nu & $\nu$ & $-$ & degrees of freedom for pulse + background model \\
25 & chi\string^2\_nu & $\chi^2_\nu$ & $-$ & reduced goodness of fit for pulse + background model \\
\enddata
\end{deluxetable*}

\begin{splitdeluxetable}{ccccccccccBccccccBccccccccc}
\tablewidth{0pt} 
\tablenum{6}
\tablecaption{BATSE TTE GRB Pulse Catalog (Part I). Characteristics of the \cite{nor05} model pulse fit. \label{tab:cat1}}
\tabletypesize{\scriptsize}
\tablehead{
\colhead{pulse\_id} & \colhead{resolution} & \colhead{B} & \colhead{B\_err} & \colhead{BS} & \colhead{BS\_err} 
& \colhead{ts} & \colhead{ts\_err} & \colhead{A} & \colhead{A\_err} & \colhead{tau1} 
& \colhead{tau1\_err} & \colhead{tau2}
& \colhead{tau2\_err} & \colhead{w} & \colhead{w\_err}  & \colhead{kappa} & \colhead{kappa\_err}
& \colhead{tau\_pk} & \colhead{tau\_pk\_err} & \colhead{t\_start} 
& \colhead{t\_end} & \colhead{chi\string^2} 
& \colhead{nu} & \colhead{chi\string^2\_nu} 
} 
\colnumbers
\startdata 
138 & 4ms & 25.869881 & 25.869881 & 0.78208389 & 0.35183217 & 0.10942063 & 0.010948803 & 13.613831 & 1.2549311 
& 0.017597556 & 0.015627207 & 0.30104644 & 0.060408634 &1.0385573 & 0.19659315 & 0.869609534 & 0.094727124 
& 0.182205804 & 0.034894776 & -0.138 & 2.661 & 739 & 659 & 1.12  \\
185 & 4ms & 30.609108 & 0.47000447 & -0.37561279 & 0.26646661 & 0.055455937 & 0.01968871 & 28.178768 & 28.178768
& 0.13324944 & 0.13521754 & 0.031008664 & 0.009787145 & 0.009787145 & 0.047837702 & 0.515440072 & 0.194353043 
& 0.119735694 & 0.039424117 & 0.025 & 0.421 & 84.1 & 93 & 0.9 \\
206 & 4ms & 31.781707 & 0.65658421 & -0.41093742 & 0.37725601 & 0.11285123 & 0.001835358 & 34.510712 & 2.8177553 
& 0.001190187 & 0.0018917173 & 0.14323237 & 0.016561009 & 0.45506158 & 0.051605311 & 0.944261468 & 0.081376818 
& 0.12590777 & 0.010564292 & -0.003 & 1.285 & 355.5 & 315 & 1.13 \\
218 & 64ms & 552.31368 & 0.35900721 & -0.091601428 & 0.001941129 & -1.0981916 & 0.14571934 & 193.78792 & 10.559236
& 1.3918351 & 1.3918351 & 0.34425246 & 0.072217753 & 1.9814368 & 0.34996376 & 0.521216408 & 0.132481698 
& -0.405990287 & 0.289038132 & -1.431 & 2.921 & 144.8 & 62 & 2.33 \\
289 & 64ms & 0.289038132 & 0.324598 & 0.0941294 & 0.00179762 & -1.802 & 716.078 & 335.279 & 301709 
& 0.00179762 & 0 & 0.000711897 & 0.661024 & 0.661024 & 81.3093 & 0.018293982 & 4.245246788
& -0.207161978 & 1030.053863 & -1.812 & -0.109 & 22.7 & 21 & 1.08 \\
\enddata
\tablecomments{Table \ref{tab:cat1} is published in its entirety in the machine-readable format.
      A portion is shown here for guidance regarding its form and content.}
\end{splitdeluxetable}

\begin{deluxetable*}{cllll}
\tablewidth{0pt}
\tablenum{7}
\tablecaption{Information contained in the BATSE TTE GRB Pulse Catalog (Part II) \label{tab:cat2head}}
\tabletypesize{\scriptsize}
\tablehead{
\colhead{Table Column} & \colhead{Header} & \colhead{Variable} & \colhead{Units} & \colhead{Description} 
}
\startdata
1 & pulse\_id & BATSE trigger & $-$ & number (+ letter) \\
2 & t0 & $t_0$ & s & residual peak time, from Equation \ref{eqn:res} \\
3 & t0\_err & $\sigma_{\rm t0}$ & s & residual peak time uncertainty \\
4 & a & $a$ & cts & residual amplitude, from Equation \ref{eqn:res} \\
5 & a\_err & $\sigma_a$ & cts & residual amplitude uncertainty \\
6 & omega & $\omega$ & s$^{-1}$ & residual Bessel frequency, from Equation \ref{eqn:res} \\
7 & omega\_err & $\sigma_\omega$ & s$^{-1}$ & residual Bessel frequency uncertainty \\
8 & s & $s$ & $-$ & Bessel function stretching parameter, from Equation \ref{eqn:res} \\
9 & s\_err & $\sigma_s$ & $-$ & Bessel function stretching parameter uncertainty \\
10 & chi\string^2 & $\chi^2$ & $-$ & goodness of fit for pulse + residual + background model \\
11& nu & $\nu$ & $-$ & degrees of freedom for pulse + residual + background model \\
12 & chi\string^2\_nu & $\chi^2_\nu$ & $-$ & reduced goodness of fit for pulse + residual + background model \\
13 & delta\_chi\string^2 & $\Delta \chi^2$ & $-$ & goodness of fit improvement from residual model \\
14 & delta\_nu & $\Delta \nu$ & $-$ & difference in degrees of freedom \\
15 & p\_delta & $p_\Delta$ & $-$ & $p-$value of model improvement  \\
16 & include & include & $-$ & `x' to include residuals, `o' to exclude, based on $p_\Delta$ \\
17 & R & $R$ & $-$ & ratio of $A/a$, from \ref{eqn:R} \\
18 & 4ms\_pk\_cts & 4 ms peak counts & cts & measured peak counts per bin \\
19 & 4ms\_S/N & 4 ms $S/N$ & $-$ & signal-to-noise from Equation \ref{eqn:SN} \\
20 & burst\_class & burst class & $-$ & GRB class from Section \ref{sec:class} and summarized in Table \ref{tab:complete} \\
21 & p\_best & $p_{\rm best}$ & $-$ & best-fit $p-$value \\
22 & pulse\_class & pulse class & $-$ & pulse classification, described in Section \ref{sec:complex} \\
23 & S & $S$ & erg cm$^{-2}$ & energy fluence from BATSE Catalog and pulse fits \\
24 & S\_err & $\sigma_S$ & erg cm$^{-2}$ & energy fluence uncertainty \\
25 & HR & $HR$ & $-$ & energy hardness from Equation \ref{eqn:hr} \\
\enddata
\end{deluxetable*}

\begin{splitdeluxetable}{cccccccccBcccccccccBcccccccc}
\tabletypesize{\scriptsize}
\tablewidth{0pt} 
\tablenum{8}
\tablecaption{BATSE TTE GRB Pulse Catalog (Part II). 
Pulse residual fits, burst and pulse classification, and fluence and hardness characteristics. \label{tab:cat2}}
\tabletypesize{\scriptsize}
\tablehead{
\colhead{pulse\_id} & \colhead{t0} & \colhead{t0\_err} 
& \colhead{a} & \colhead{a\_err} & \colhead{omega} & \colhead{omega\_err}
& \colhead{s} & \colhead{s\_err} & \colhead{chi\string^2} & \colhead{nu} 
& \colhead{chi\string^2\_nu} & \colhead{delta\_chi\string^2} & \colhead{delta\_nu} & \colhead{p\_delta} 
& \colhead{include} & \colhead{R} & \colhead{4ms\_pk\_cts} & \colhead{4ms\_S/N} 
& \colhead{burst\_class} & \colhead{p\_best} & \colhead{pulse\_class} 
& \colhead{S} & \colhead{S\_err} & \colhead{HR} & \colhead{HR\_err} 
} 
\colnumbers
\startdata 
138 & 0.155 & 0.005 & 6.814 & 0.401 & 666.704 & 139.006 & 0.2913 & 0.0698 & 708.3 & 655 & 1.08 
& 30.7 & 4 & 3.52E-06 & x & 0.50052039 & 55 & 3.927904445 & S & 0.07314348 & Blended & 1.36E-07 & 6.41E-08 & 4.55224 & 4.55224 \\
185 & 0.113 & 0.002 & 8.339 & 1.461 & 982.423 & 170.427 & 0.9229 & 0.2057 & 80.0 & 89 & 0.90 
& 4.1 & 4 & 0.392641456 & o & 0.295932029 & 63 & 0.295932029 & S & 0.295932029 & Simple & 5.10E-08 & 6.44E-08 & 1.98 & 3.78 \\
206 & 0.13 & 0.002 & 13.035 & 0.894 & 1057.547 & 128.54 & 0.2535 & 0.0351 & 322.9 & 311 & 1.04
& 32.6 & 4 & 1.44E-06 & x & 0.377708811 & 82 & 5.545682733 & L? & 0.309297033 & Blended & 5.57E-07 & 1.42E-07 & 14.69 & 5.18 \\
218 & & & & & & & & & & & &  
& & 1 & o & & & & S & 1.42E-08 & Complex & 5.18E-07 & 1.29E-07 & 10.20 & 3.35 \\
289 & & & & & & & & & & & &  
& & 1 & o & & & & S & 0.360179945 & Simple & 9.69E-08 & 6.11E-08 & 22.28 & 17.48 \\
\enddata
\tablecomments{Table \ref{tab:cat2} is published in its entirety in the machine-readable format. A portion is shown here for guidance regarding its form and content.}
\end{splitdeluxetable}


\begin{deluxetable*}{cl}
\tablenum{9}
\tablecaption{BATSE TTE GRB Pulse Catalog (Part III). Comments. \label{tab:cat3}}
\tablewidth{0pt}
\tabletypesize{\scriptsize}
\tablehead{
\colhead{Pulse ID} & \colhead{Comments} \\
}
\startdata
138 & probably single emission episode, but could be two overlapping\\
185 & \\
206 &  \\
218 & possible Crescendo burst without Staccato pulses \\
289 &  \\
\enddata
\tablecomments{Table \ref{tab:cat3} is published in its entirety in the machine-readable format. A portion is shown here for guidance regarding its form and content.}
\end{deluxetable*}

The Comments provided in Table \ref{tab:cat3} describe a variety of characteristics that are not covered in the main catalog. These include a) instrumental reasons why pulses have been defined as TTE Partial rather than TTE Complete b) augmented descriptions of pulse characteristics, and c) burst or pulse properties measured elsewhere. In Table \ref{tab:cat3} we have delineated some of these Complex morphologies using visual descriptions rather than formal statistical ones:
\begin{itemize} 
\item {\em crown} pulses consist of a clearly defined single emission episode exhibiting many small peaks around the time of maximum emission,
\item {\em u-pulses} have u-shaped or bowl shaped double-peaked light curves characterized by short temporal spikes at the beginning and at the end of the main emission episode; sometimes they also have a spike at the center of the pulse,
\item{\em noisy double-peaked} pulses have asymmetric light curves with abnormally bright and long decay peaks,
\item {\em twin peaks} indicate single emission episodes having two closely-separated peaks overriding the main emission.
\end{itemize}
We note that some crown pulses and u-pulses have similar morphologies; it is possible that some crown pulses are merely unresolved u-pulses. 

\section{Analysis}

\subsection{Pulse Duration, Fluence, and Hardness} 

Pulse energy fluences and hardnesses have been provided in the BATSE Final Catalog ({\tt http://gammaray.msfc. nasa.gov/batse/grb/catalog/current/}) and are available for almost all of the TTE GRBs (a few have been obtained from \cite{gol13} and are identified in Table \ref{tab:cat3}). For this analysis we define energy hardness as
\begin{equation}\label{eqn:hr}
HR=(S_3+S_4)/(S_1+S_2),
\end{equation}
where $S_n$ refers to the fluence in BATSE energy channel $n$.

The BATSE Catalog publishes fluences and hardnesses for bursts, not for pulses. However, as discussed in Section \ref{sec:multi_intro}, most TTE bursts appear to contain only a single emission episode, and only a few have recognizable extended emission. Thus, pulse fluences and hardnesses are generally the same as the fluences and hardnesses of the bursts in which they are found. When bursts consist of multiple pulses, modeled pulse fits are used to extract both pulse durations and energy-dependent counts fluences of constituent pulses ({\em e.g.,} see \cite{hak11}). The counts fluences are then combined with BATSE catalog data to obtain energy fluences and hardnesses for individual pulses. This modeling approach seems to produce reasonably accurate measurements of pulse duration even if pulses contain considerable structure. In other words, large $\chi^2$ uncertainty results more from poor matches to pulse structure than from difficulties in measuring pulse boundaries. The efficacy of the approach can be seen, for example, in the fits shown in Figures \ref{fig:complexex1} through \ref{fig:complexdouble}.

\begin{figure}
\plottwo{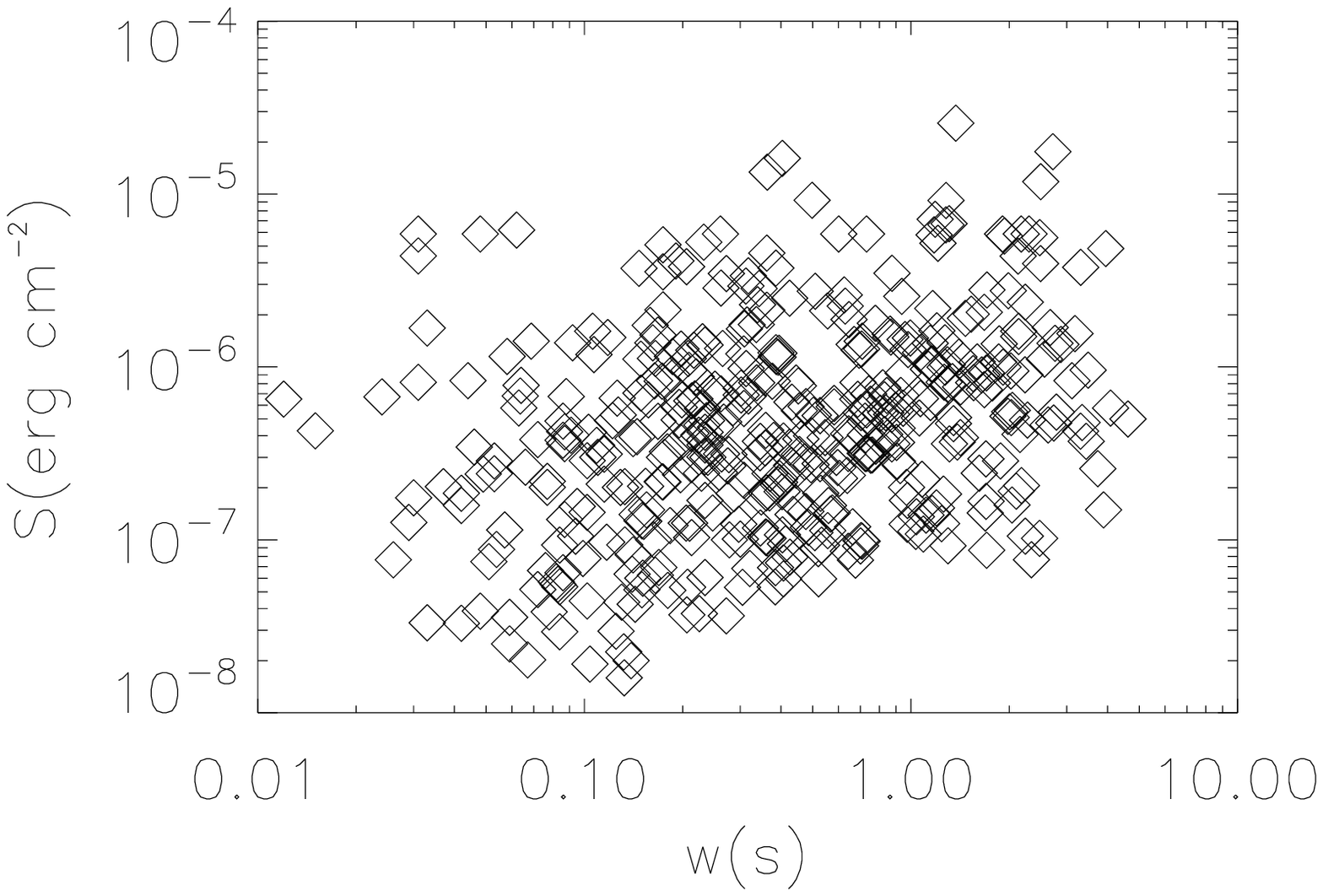}{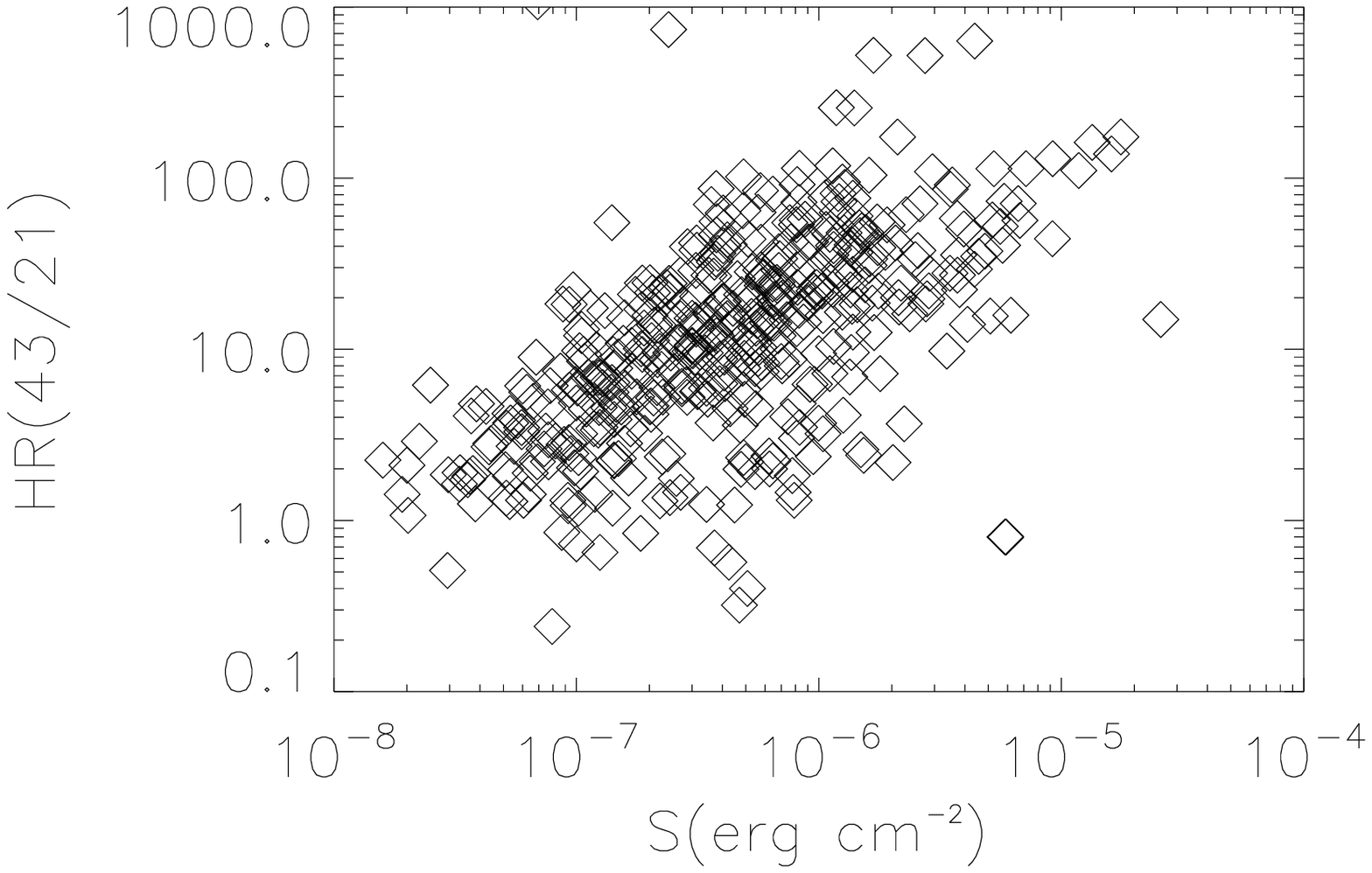}
\caption{Pulse fluence $S$ (erg cm$^{-2}$) vs. duration $w$ (left panel) and hardness $HR$ vs. fluence $S$ (erg cm$^{-2}$)(right panel). Longer pulses have greater fluences than short pulses, and harder pulses have greater fluences than soft pulses. However, no correlation is found between pulse duration and pulse hardness. \label{fig:fluence}}
\end{figure}

Using the GRB class definitions from Section \ref{sec:class}, we find that Short GRB pulse durations, fluences, and hardnesses correlate with one another in manners consistent with those described in \cite{hak11} for Long and Intermediate GRB pulses. As expected, pulses with longer durations have correspondingly larger fluences (Figure \ref{fig:fluence}, left panel). Similarly, harder pulses have correspondingly larger fluences, as high-energy photons contain significantly more energy than low-energy photons (Figure \ref{fig:fluence}, right panel). A Spearman Rank-Order Correlation test (SC) indicates that pulse fluence and duration are highly correlated (SC=$-0.31$, $p=6 \times 10^{-10}$) and that hardness and fluence are even more highly correlated (SC$=0.58$, $p=7 \times 10^{-36}$). However, hardness and duration are uncorrelated (SC$=0.46$, $p=0.37$). 

\begin{figure}
\plottwo{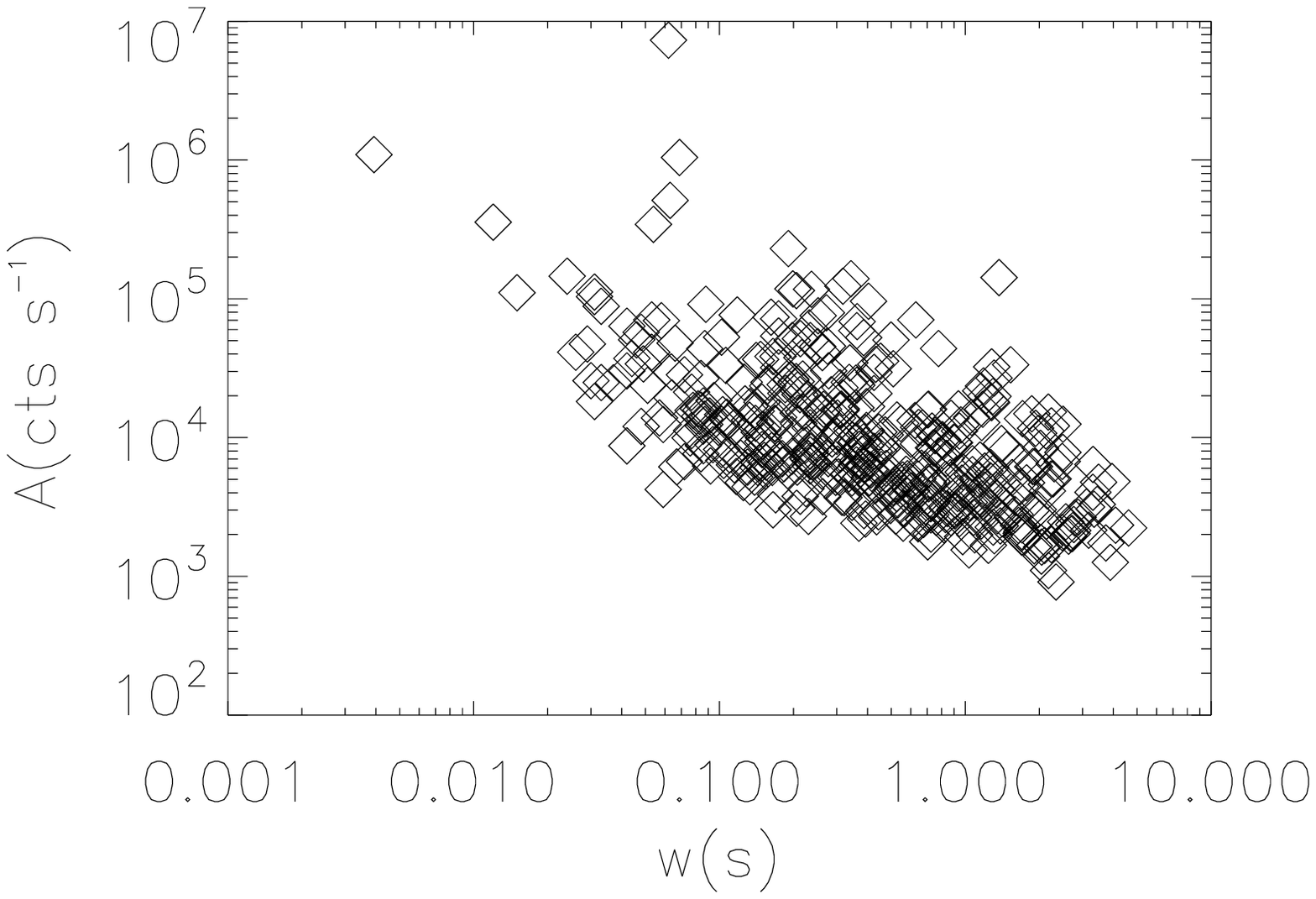}{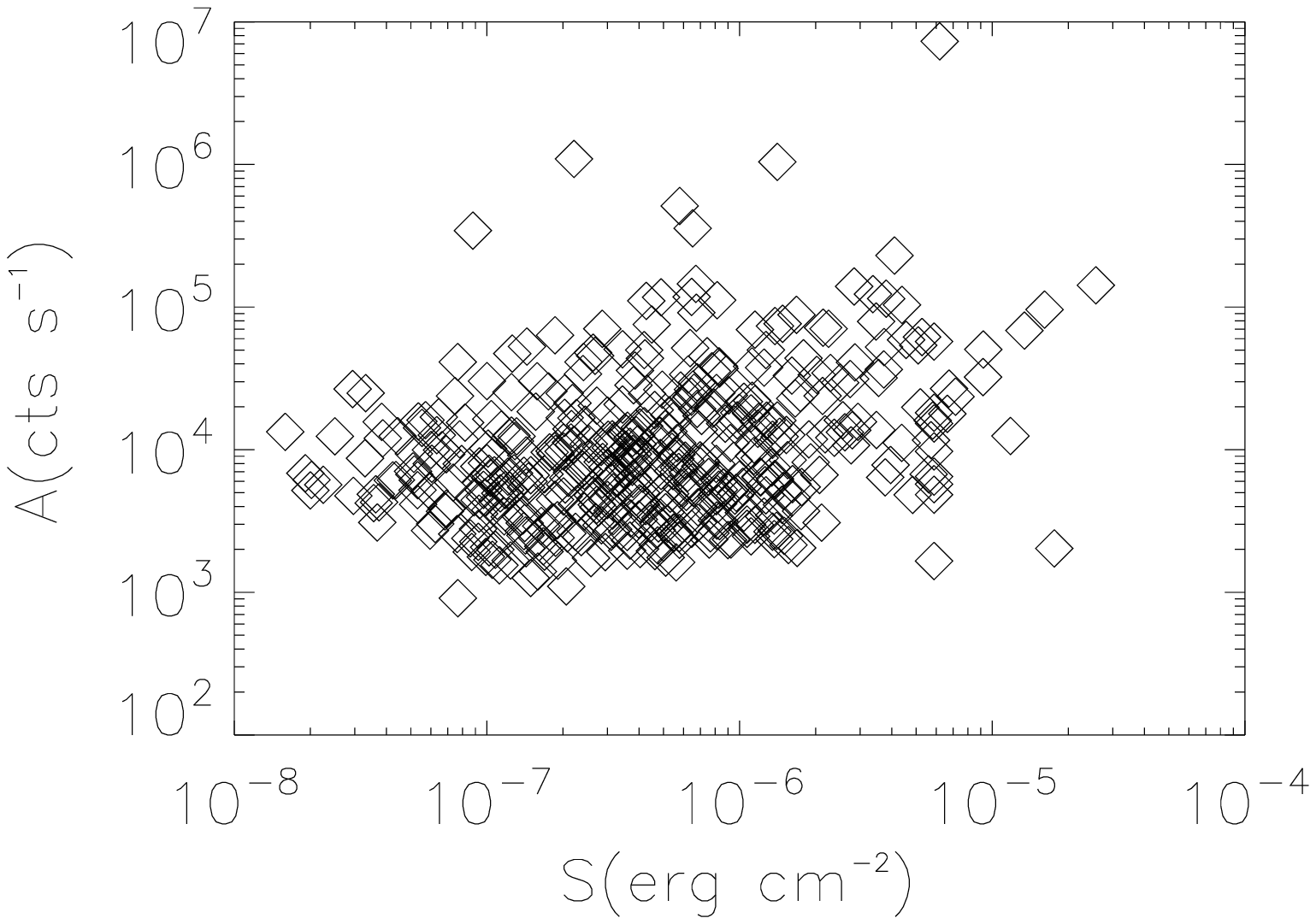}
\caption{Pulse amplitude $A$ (counts $s^{-1}$) vs. duration $w$ (left panel) and pulse amplitude (counts $s^{-1}$) vs. fluence (erg cm$^{-2}$) (right panel) for Short GRB pulses. Shorter duration pulses have larger amplitudes than longer pulses, while pulses having larger fluences also tend to have larger amplitudes. \label{fig:Avw}}
\end{figure}

\begin{figure}
\plotone{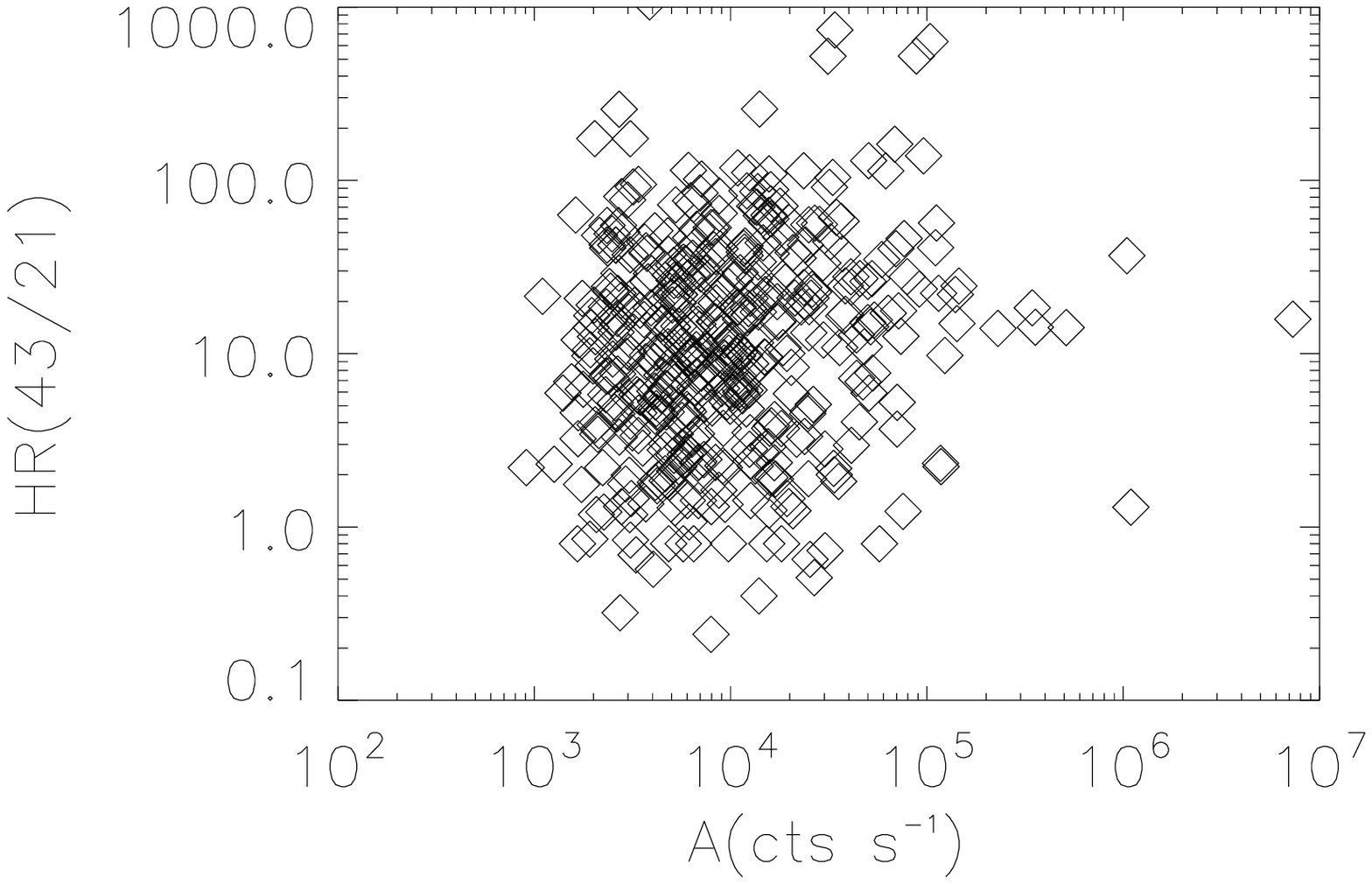}
\caption{Pulse amplitude $A$ (counts $s^{-1}$) vs. hardness $HR$. Large amplitude pulses have marginally larger spectral hardnesses than those with small amplitudes. \label{fig:AvHR}}
\end{figure}

The amplitudes of Short GRB pulses are related to their durations and fluences (Figure \ref{fig:Avw}). A Spearman Rank-Order Correlation indicates that Short GRB pulse amplitude and duration are highly anti-correlated (SC$=-0.49$, $p = 6 \times 10^{-25}$) as are amplitude and fluence (SC$=0.31$, $p = 9 \times 10^{-10}$). This relationship has been found previously for Short GRBs \citep{hak11,nor11}, and is itself an extension of the pulse amplitude vs.~duration anti-correlation found by \cite{hak08} for GRB prompt emission and known to extend to x-ray flares \citep{mar10}. Using measurements of maximum count rates divided by minimum count rates on three different timescales, and comparing these measurements with pulse durations, \cite{hak11} have demonstrated that this effect is real rather than due to a selection bias. In addition to duration and fluence, pulse hardness and amplitude (Figure \ref{fig:AvHR}) appear to be weakly correlated (SC$=0.11$, $p = 3 \times 10^{-2}$) in Short GRB pulses.

\begin{figure}
\plotone{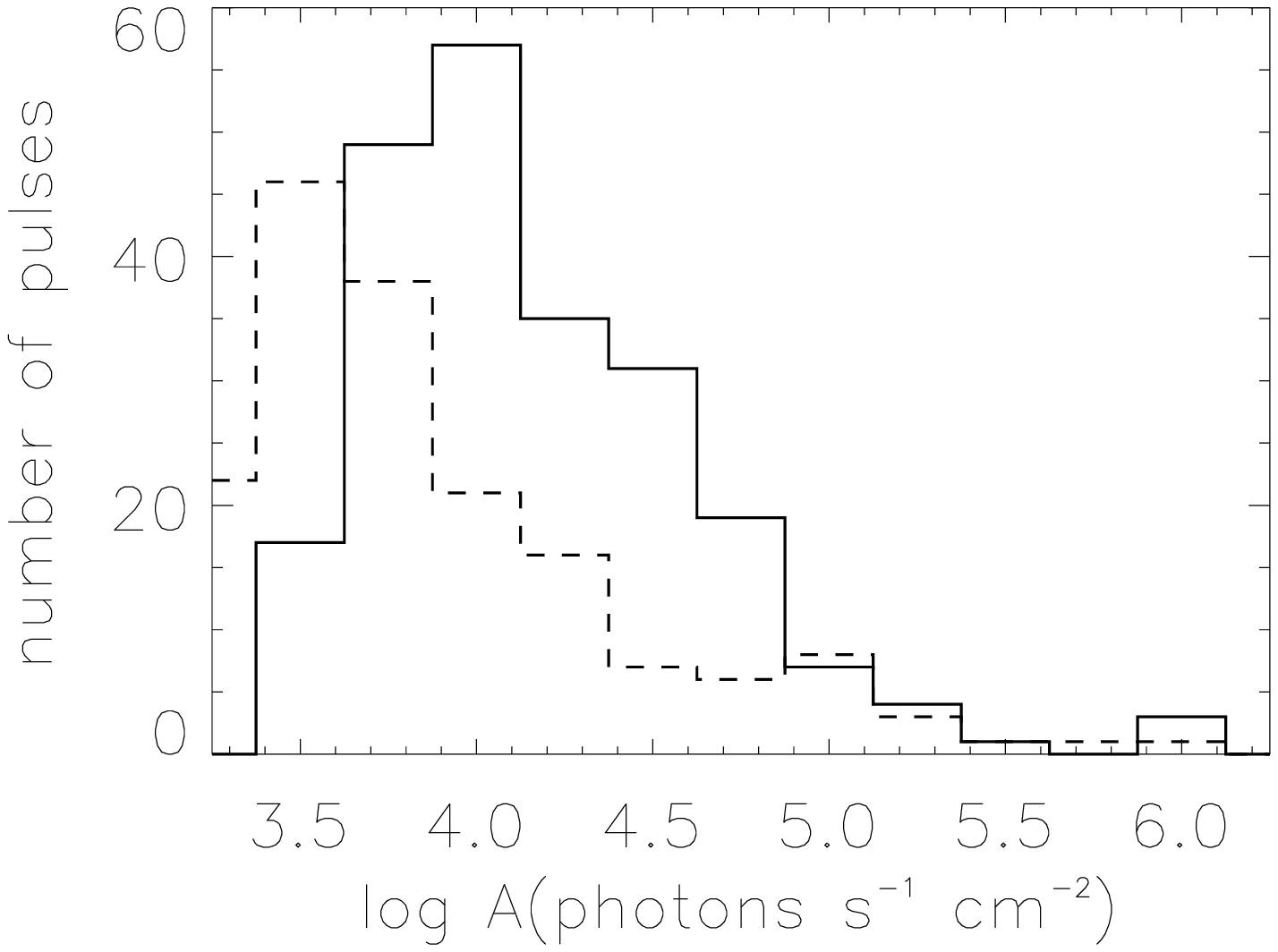}
\caption{Logarithmic pulse amplitude histograms for Short TTE GRBs in this Catalog. TTE Complete pulses (solid line) have larger pulse amplitudes than TTE Partial pulses (dashed line). \label{fig:a_hist}}
\end{figure}

The Short TTE Complete and Short TTE Partial samples are not representative of the same underlying population due to a sampling bias. Because TTE Complete pulses fit completely within the TTE window whereas TTE Partial pulses do not, TTE Complete pulses tend to have shorter durations and smaller fluences than TTE Partial bursts, as indicated by Student's T-tests (T) comparing the logarithmic distributions of duration ($T=-12.0$, $p=2 \times 10^{-28}$) and fluence ($T=-3.6$, $p=4 \times 10^{-4}$) for Short GRBs. As a result of the aforementioned anti-correlation between amplitude and duration, the difference between the TTE Complete and TTE Partial duration distributions reflects a difference between the amplitude distributions. This can be seen in Figure \ref{fig:a_hist} and in the Student's T-test results ($T=5.8$, $p=10^{-8}$).  A similar result has been previously identified by \cite{nor11} for Short Swift GRB pulses.

These results demonstrate that the overall distribution of Short GRB pulse properties cannot be described using TTE Complete pulses alone. Instead, the different properties of TTE Partial pulses must be included. Even when these have been included, the combined TTE sample is not entirely representative of the underlying Short GRB pulse distribution. Only a small fraction of bursts in this sample have been formally classified as Long/Intermediate GRBs, suggesting that the long duration end of the Short GRB sample cannot be sampled by the durations of TTE windows.

\begin{figure}
\plottwo{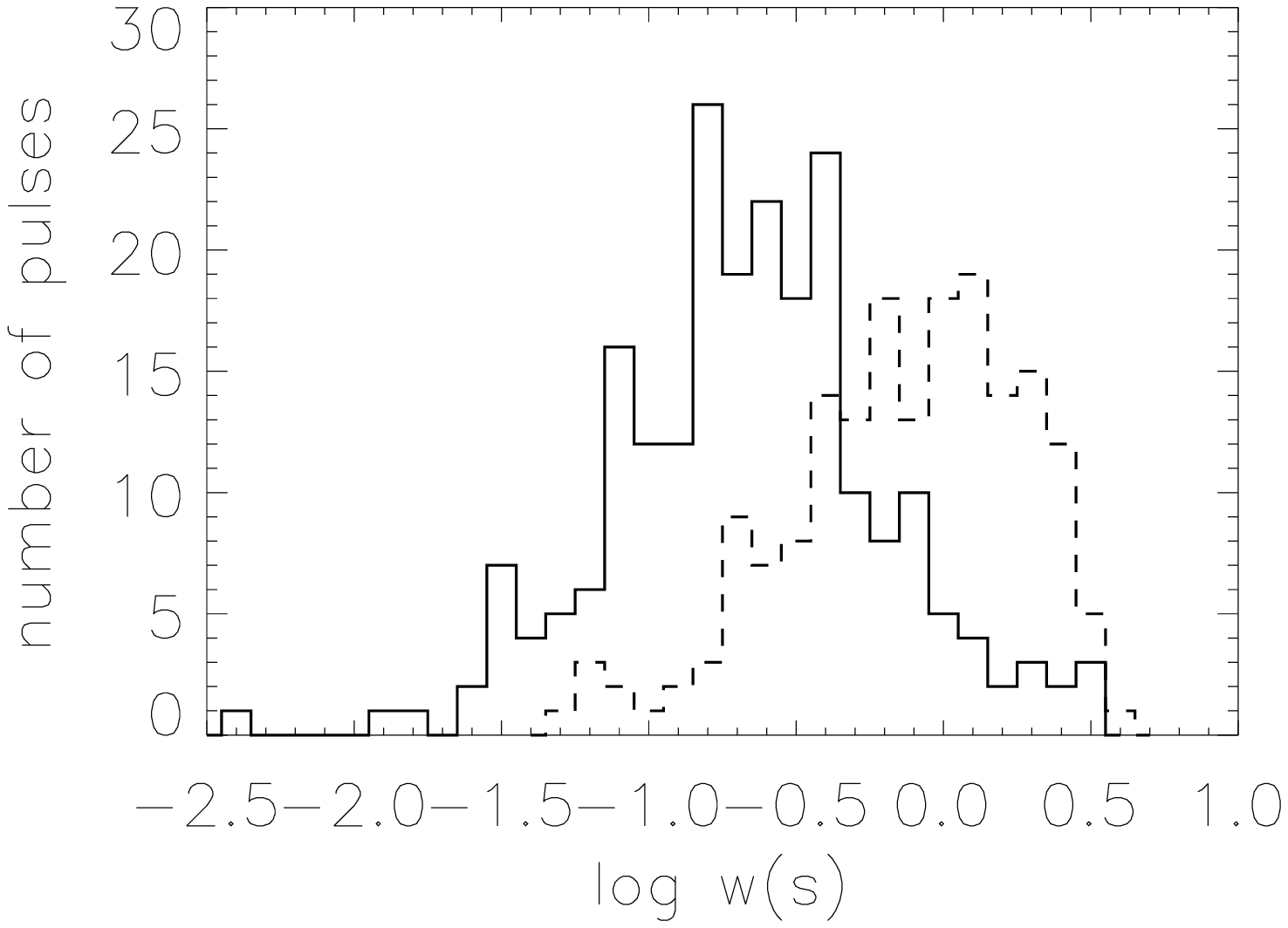}{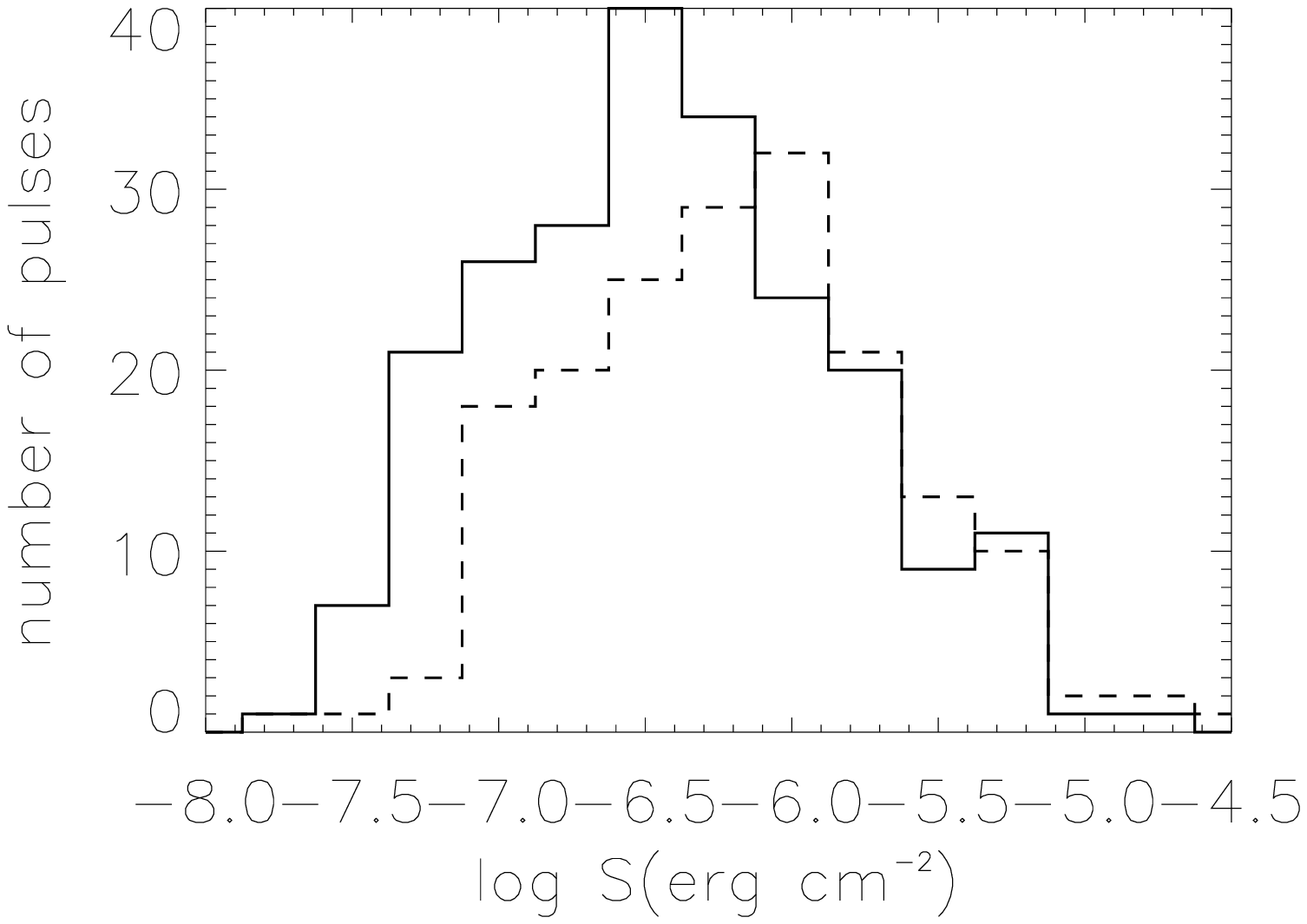}
\caption{Logarithmic pulse duration (left panel) and fluence (right panel) histograms for Short TTE GRBs in this Catalog. TTE Complete pulse distributions are indicated by solid lines and Short TTE Partial pulses are indicated by dashed lines. \label{fig:dur_hist}}
\end{figure}

\subsection{Short GRBs Exhibit a Continuum of Pulse Complexity} 

\begin{figure}[ht!]
\plotone{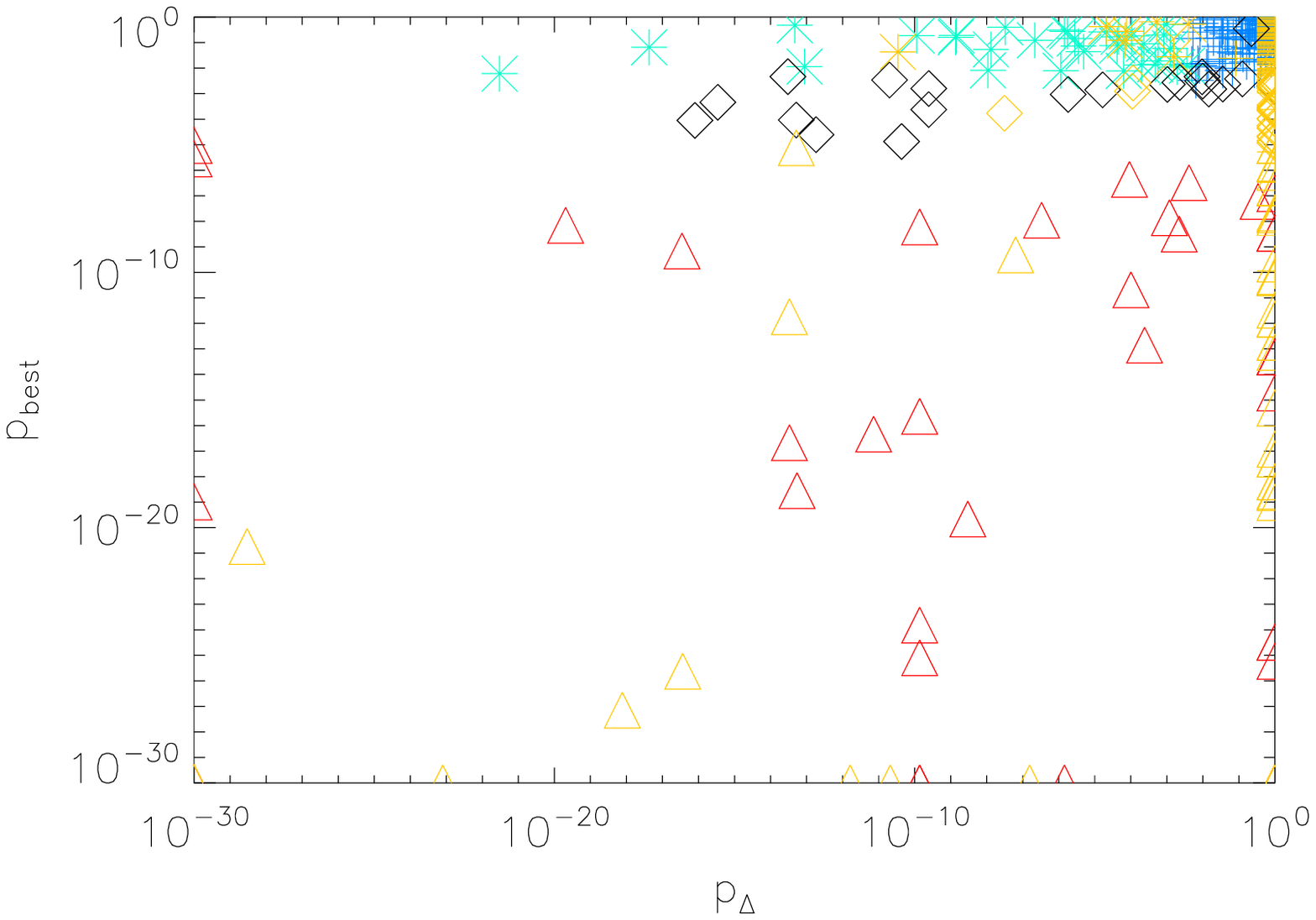}
\caption{Complexity characteristics of BATSE TTE Complete (blue, turquoise, black, and red) and TTE Partial (yellow) pulses, in terms of $\log(\Delta \chi^2)$ $p-$values (x-axis) and $\log(\chi^2_{\rm best})$ $p-$values (y-axis). Simple pulses (crosses) are characterized by good pulse fits without the residual function ({\em e.g.,} they have large $p_\Delta$ and $p_{\rm best}$ values). Blended pulses (asterisks) are significantly improved by addition of the residual function ({\em e.g.,} they have large $p_{\rm best}$ values but smaller $p_\Delta$ values). Structured pulses (diamonds) are similar to Blended pulses but have smaller $p_{\rm best}$ values. Complex pulses (triangles) are single emission episodes having poor fits to the single pulse model, even when the residual function is included. High time resolution allows a greater fraction of TTE Complete bursts to be classified as Simple and Blended pulses than TTE Partial pulses.  \label{fig:f2}}
\end{figure}

The pulse classes defined in Section \ref{sec:complex} in terms of complexity represent a continuum of characteristics; our definitions of four discrete groups is somewhat arbitrary.  Figure \ref{fig:f2} demonstrates where the Simple (crosses), Blended (asterisks), Structured (diamonds), and Complex (triangles) groups are found in terms of their fit improvement by the residual function ($p_\Delta$) and their final best fit ($p_{\rm best}$) for both the TTE Complete (blue) and TTE Partial (red) samples. The vast majority of the pulses lie in the upper right hand corner (large best fit $p-$values and large $\Delta \chi^2$ $p-$values); these Simple pulses are well-characterized using only the \cite{nor05} pulse model. Blended TTE Complete pulses along the top of the graph (large best-fit $p-$values but small $\Delta \chi^2$ $p-$values) are best fit by the \cite{nor05} model combined with the \cite{hak14} residual model. Structured TTE Complete pulses (small best fit $p-$values) are too complex to be completely characterized using combinations of the \cite{nor05} and \cite{hak14} pulse models, but there is a gradual change in complexity from Blended, to Structured, to Complex pulses. Many TTE Partial pulses light curves cannot adequately be explained by the \cite{nor05} pulse function alone, but the temporal binning of these light curves provided too few data points for any residual structure to be identified. 

\subsection{Internal Errors: Comparing Pulse Properties Measured with both 4- and 64-ms Resolution} 

Eighty-five pulses have been fitted using both 4-ms and 64-ms data. Although 64-ms data provide inadequate temporal resolution for fitting most residuals (described previously), the bulk observable \cite{nor05} pulse properties (amplitude $A$, duration $w$, and asymmetry $\kappa$) have been measured using data from both timescales. These properties can be directly compared to provide an internal check on the reliability of the pulse fitting process, and also to provide insights into the measurement uncertainties of these properties as determined by the {\tt MPFIT.PRO} nonlinear least squares routine \citep{mar09}.

\begin{figure}
\plottwo{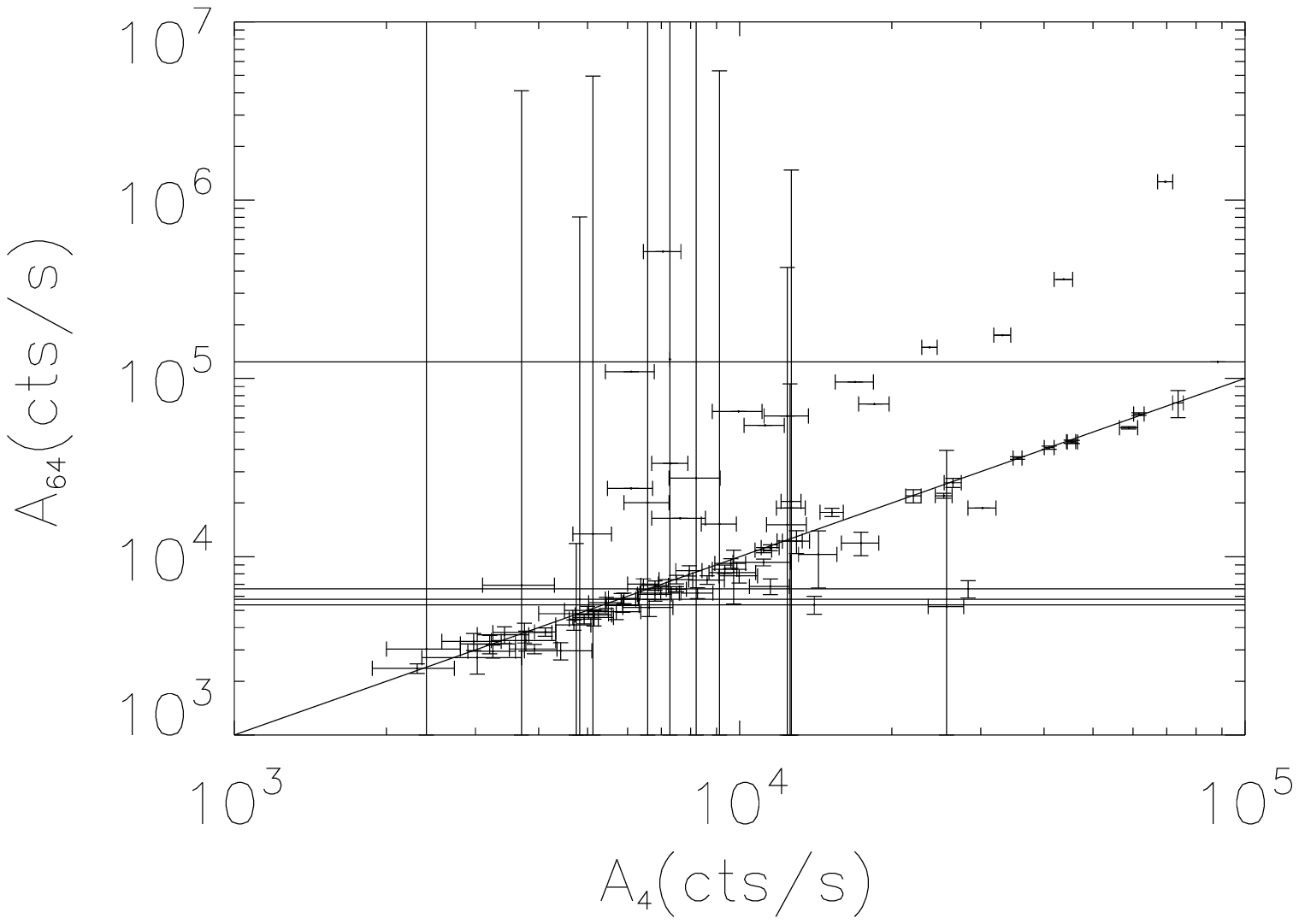}{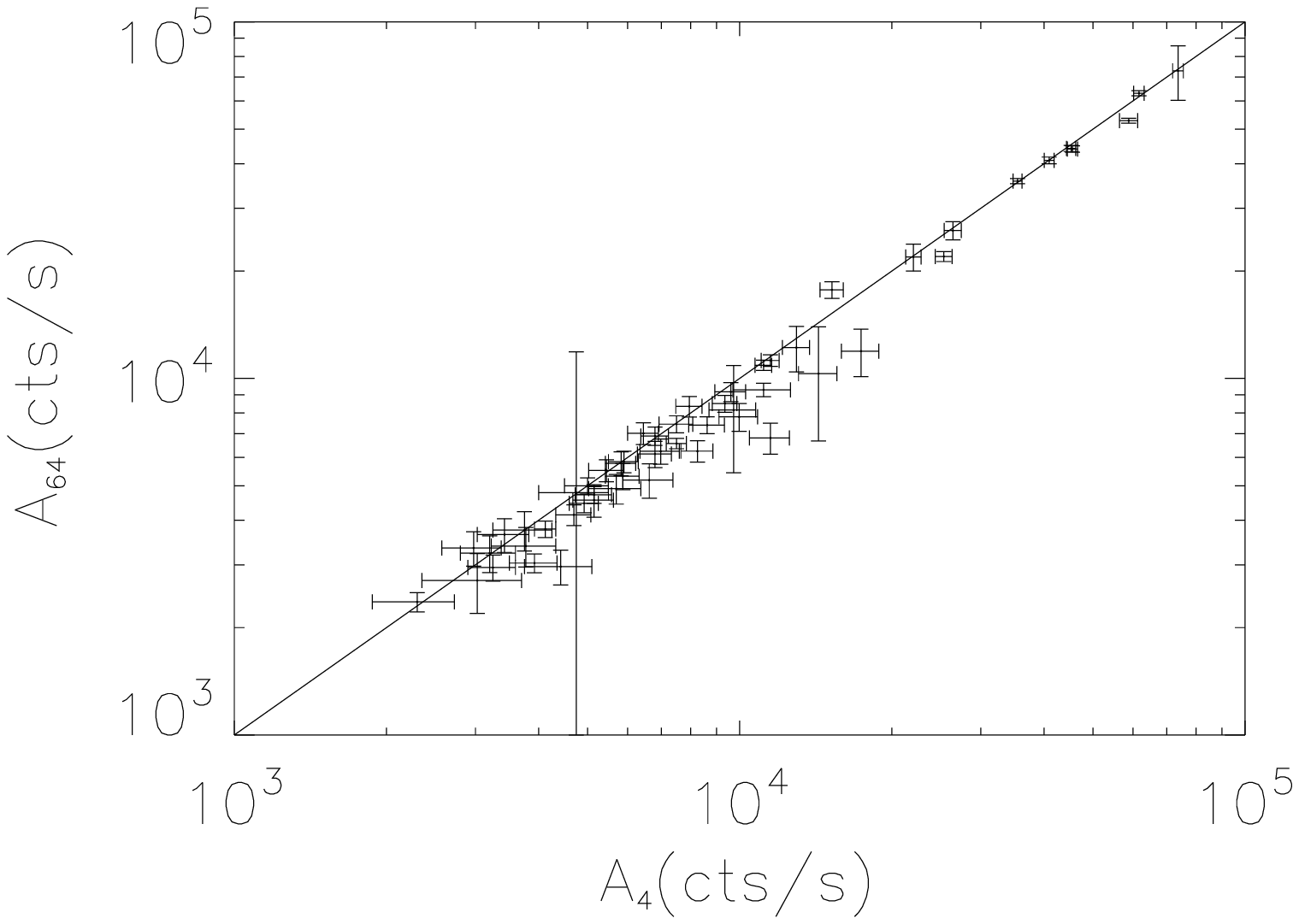}
\caption{Comparison of 64-ms vs. 4-ms pulse amplitudes ($A_{64}$ vs. $A_4$).  The left panel shows the relationship for all measurements, while the right panel shows the results when measurements with $\sigma_{A64} \ge 5 A64$, $\sigma_{A4} \ge 5 A4$, $\sigma_{A64}=0$, and $\sigma_{A4}=0$ have been removed. \label{fig:64v4A}}
\end{figure}

The 64-ms timescale pulse amplitudes ($A64$) are compared to their 4-ms pulse counterparts ($A4$) in the left panel of Figure \ref{fig:64v4A}. The majority of these amplitudes are highly correlated, and a Spearman Rank Order Correlation test indicates that these amplitude measurements are highly correlated (SC$=0.77$, $p=6\times 10^{-18}$. However, a few pulse amplitudes are found  to differ systematically by large amounts, while a few others have abnormally large statistical uncertainties. The systematically different measurements all have uncertainties of $\sigma_A=0$, and all but one of these values have been measured on the 64-ms timescale. In fact, both the large and small uncertainties are associated with large-amplitude pulses having short durations relative to the bin size. We conclude that the nonlinear least squares routine has difficulty converging at the intensity inflection point when a limited number of data points are present to describe large intensity variations. Upon excluding 32 pulses having either very large ($\sigma_{A64} \ge 5 \times A_{64}$ and $\sigma_{A4} \ge 5 \times A_4$) or small ($\sigma_{A64}=0$ and $\sigma_{A4}=0$) measurement uncertainties, the expected relationship is recovered and shown in the right panel of Figure \ref{fig:64v4A}. Although most of these measurements are consistent with unity, many 64-ms amplitudes are slightly smaller than their 4-ms counterparts. This results because the 64-ms binning washes out some of the 4-ms pulse structure.

\begin{figure}
\plottwo{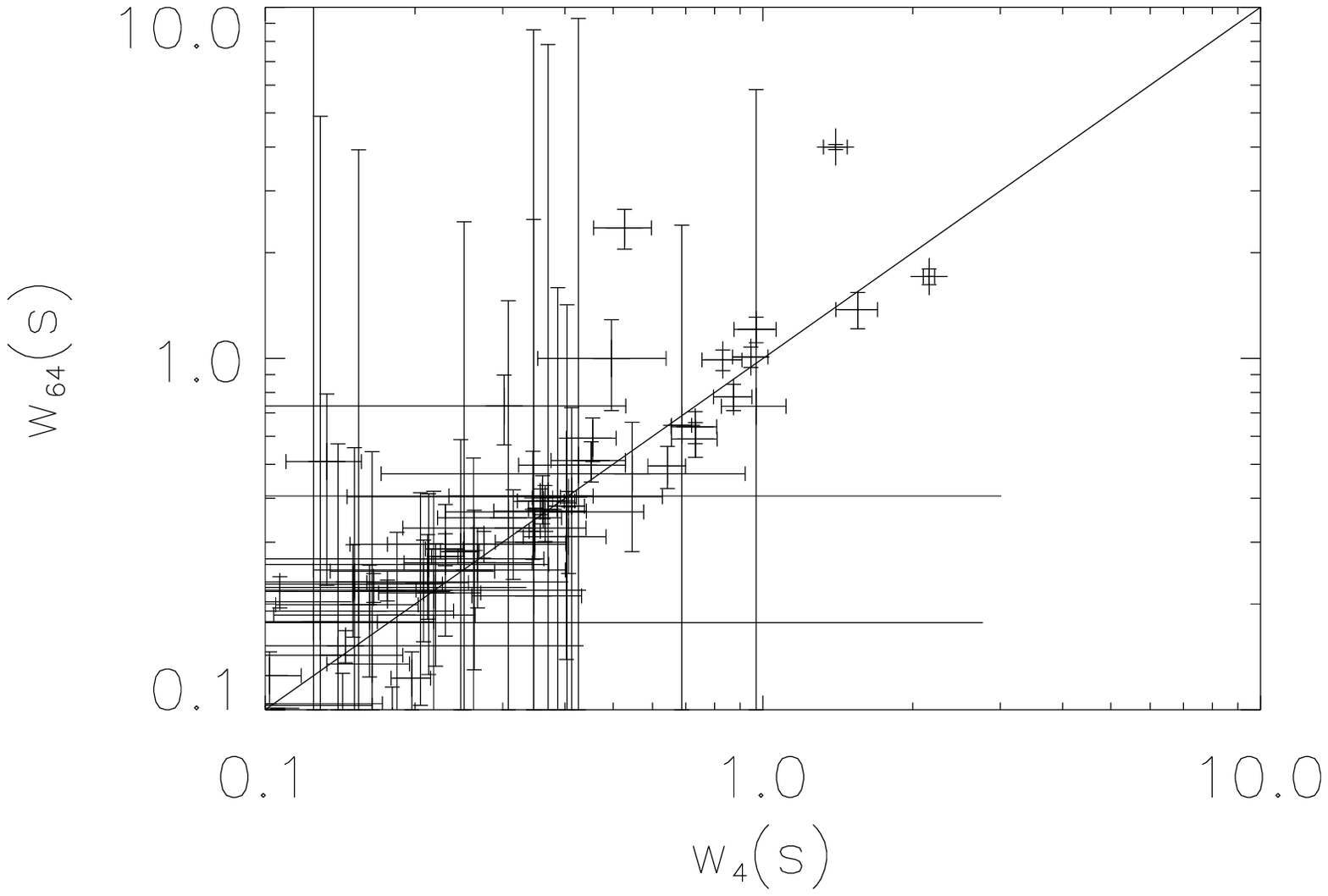}{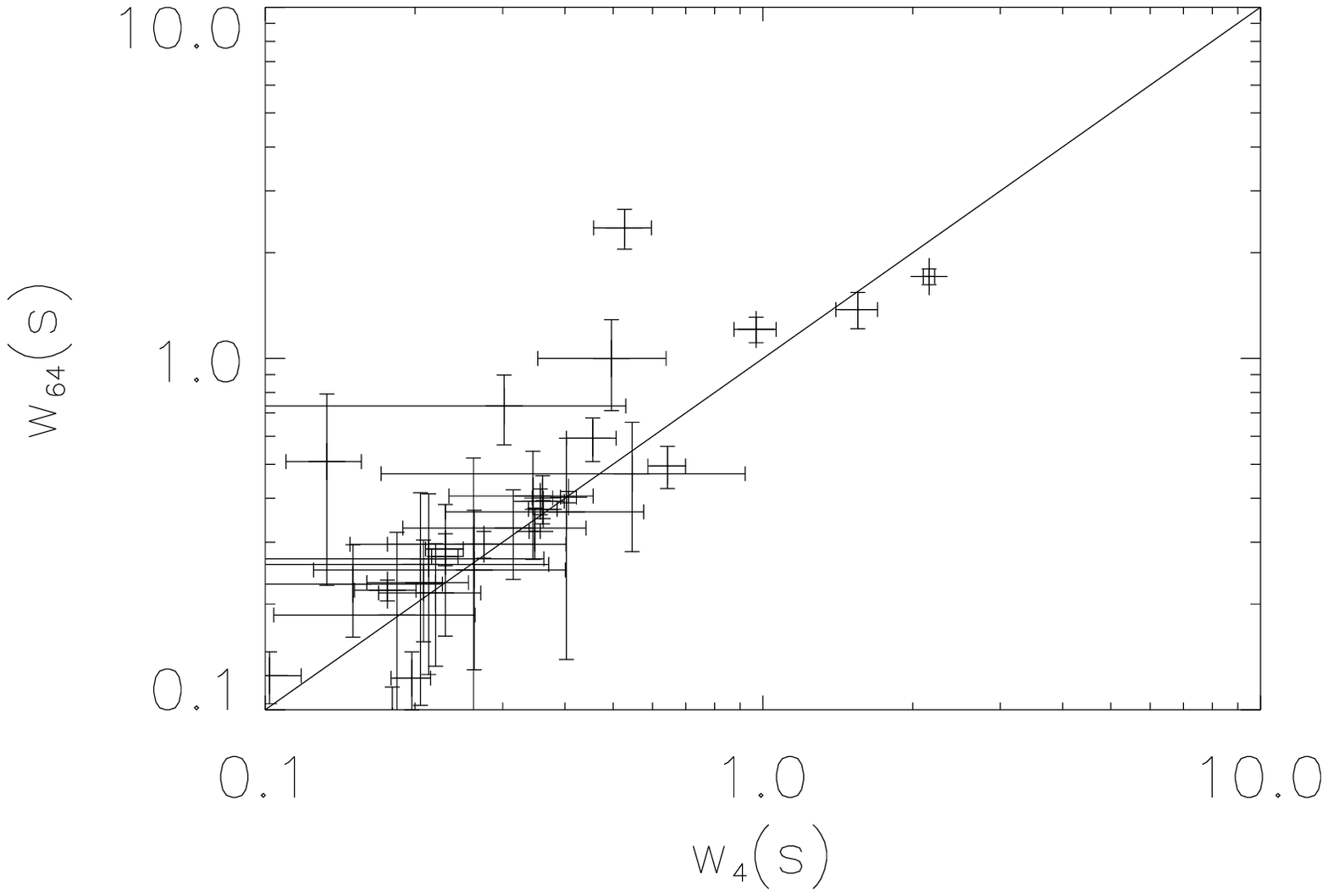}
\caption{Comparison of 64-ms vs. 4-ms pulse durations ($w_{64}$ vs. $w_4$).  The left panel shows the relationship for all measurements, while the right panel shows the results for measurements where $\sigma_{w} \le w$.\label{fig:64v4w}}
\end{figure}

The 64-ms timescale pulse durations ($w_{64}$) are compared to their 4 ms pulse counterparts ($w_4$) in the left panel of Figure \ref{fig:64v4w}. The duration measurements are highly correlated (SC$=0.92$, $p=7\times 10^{-36}$), even though the individual measurement uncertainties are large.  In other words, there do not appear to be systematic differences between durations measurements made using fits on different timescales. As expected, relative uncertainties ($\sigma_w/w$) increase as pulse durations approach the temporal resolution.  Limiting the sample to durations measured accurately on both timescales demonstrates the consistency of fitting on the two different timescales; this is shown in the right panel of Figure \ref{fig:64v4w} for 42 pulses having $\sigma_w \le w$.

\begin{figure}
\plottwo{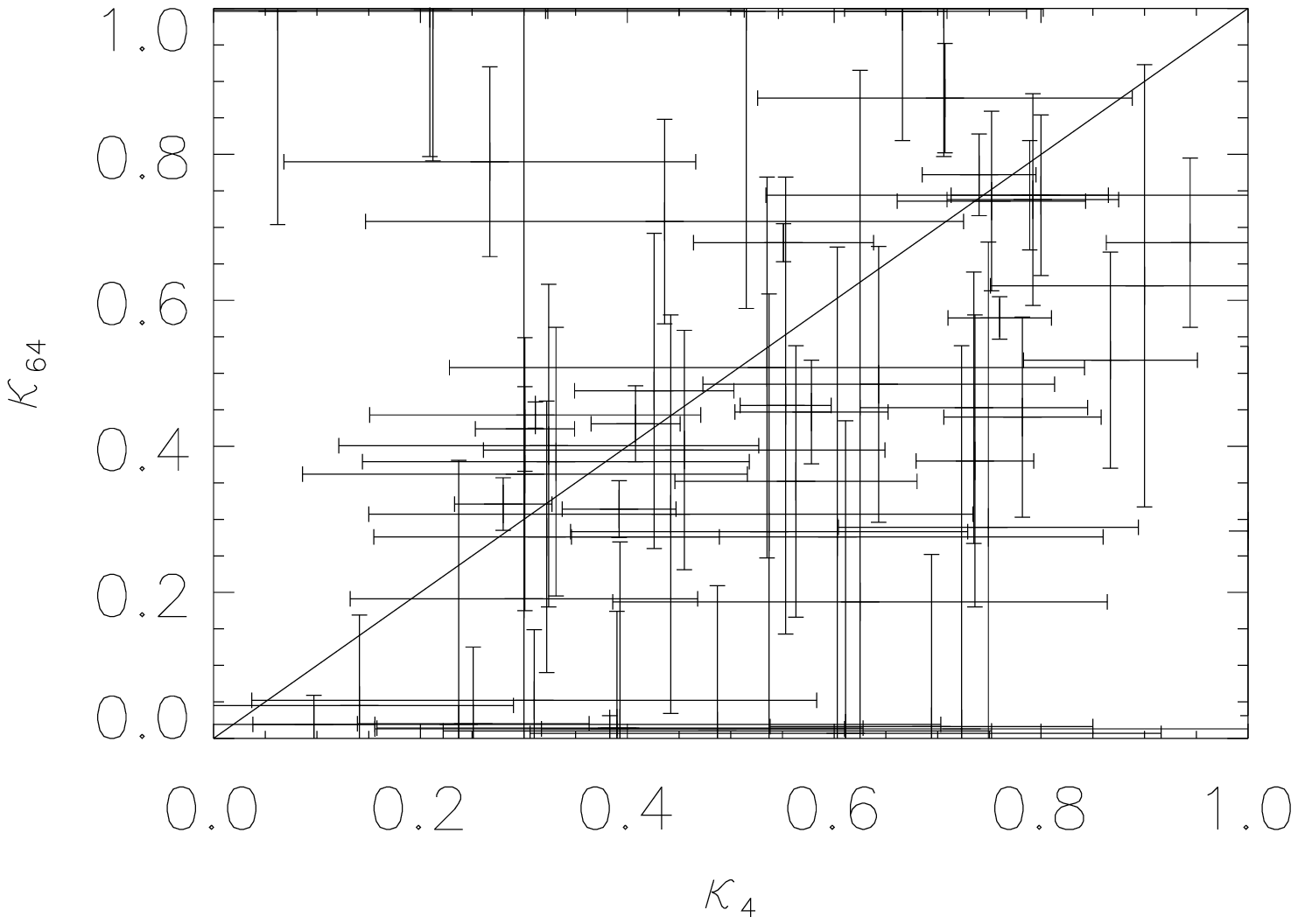}{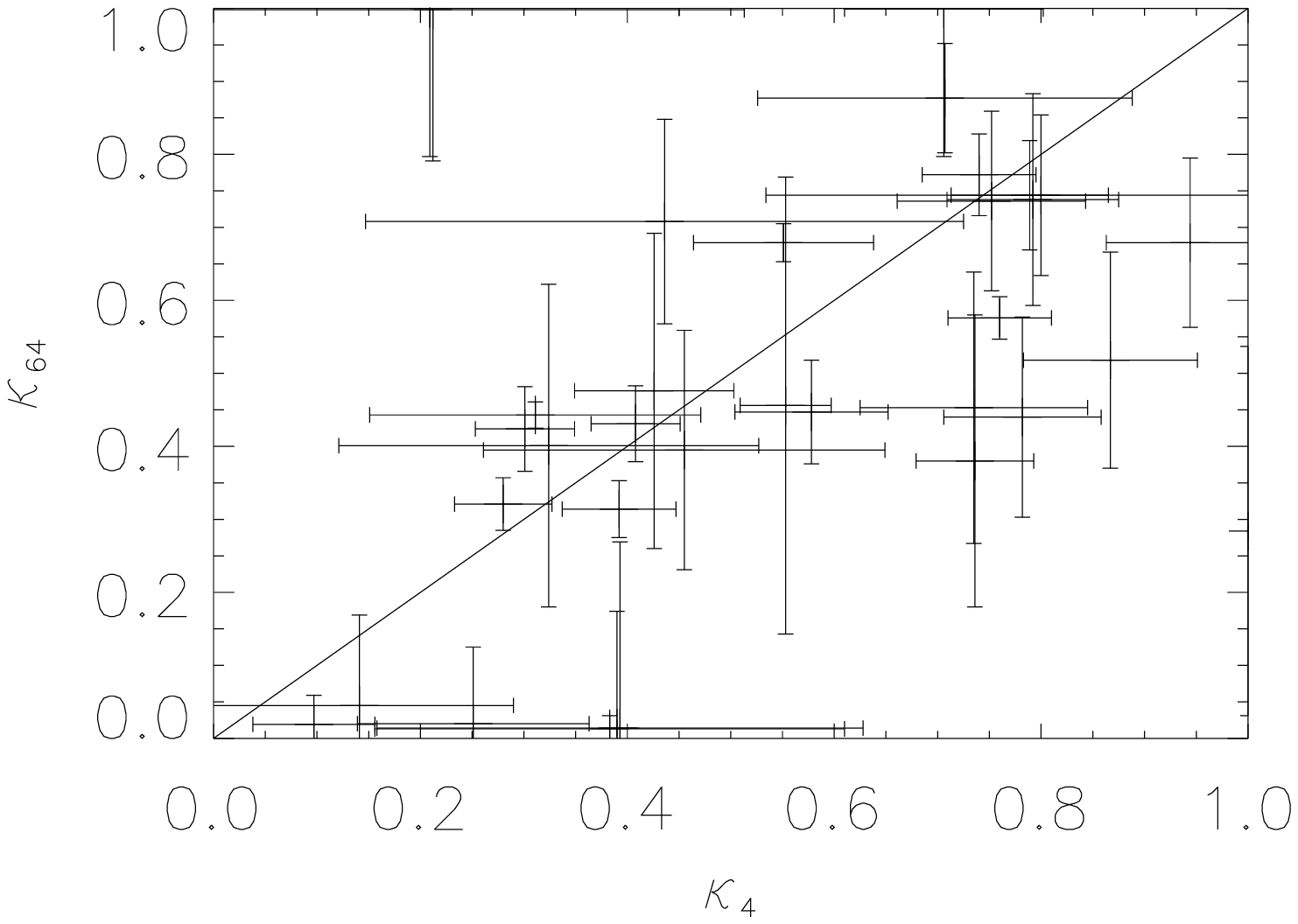}
\caption{Comparison of 64-ms vs. 4-ms pulse asymmetries ($\kappa_{64}$ vs. $\kappa_4$). The left panel shows the relationship for all measurements, while the right panel shows the results for measurements where $\sigma_\kappa \le 0.15$.\label{fig:64v4kap}}\end{figure}

Finally, the 64-ms pulse asymmetries ($\kappa_{64}$) are compared to their 4-ms counterparts ($\kappa_4$) in the left panel of Figure \ref{fig:64v4kap}. Figure \ref{fig:64v4kap} demonstrates that asymmetry measurements are difficult to make on the Short GRB timescales, as large uncertainties accompany the measurements for many of the pulses. However, a Spearman rank order correlation test indicates that the asymmetry measurements are weakly correlated (SC$=0.24$, $p=0.07$). This correlation can be clarified by limiting the sample to accurately-measured pulses. The right panel of Figure \ref{fig:64v4kap} shows that both 64-ms and 4-ms resolutions measure similar asymmetries when the sample is limited to those pulses having accurate measurements ($\sigma_\kappa \le 0.15$).

It appears that the formal fitting process has led to overestimates of many uncertainties for $\sigma_w$ and $\sigma_\kappa$. Through inspection it appears that many $\sigma_{\tau_{\rm pk}}$ measurements have also been overestimated. The uncertainties for these observables were propagated from the fitted values of $\sigma_A$, $\sigma_{ts}$, $\sigma_{tau1}$, and $\sigma_{tau2}$. Although the fitted variables are generally not observable (with the exception of $A$), the uncertainties in the measurement of many of these variables also seem to be inordinately large. 

The mathematical expression describing the \cite{nor05} intensity model (Equation \ref{eqn:function}) has several characteristics that make it difficult to fit. The largest signal exists at time $\tau_{\rm peak}$, when the pulse intensity equals the amplitude $A$ and where exponentially increasing and decreasing intensity functions involving $\tau_1$ and $\tau_2$ are joined.  The start and end of the pulse provide few additional helpful fitting constraints: $t_s$ occurs when the pulse rise intensity equals the background at the beginning of the pulse ($t_s$ prevents the intensity function from going to infinity prior to the pulse's beginning), the exponential rise $\tau_1$ determines how fast the intensity increases from $t_s$, and the exponential decay of $\tau_2$ describes the rate of intensity decrease while ensuring that this intensity will never quite reach the background. The interplay between $t_s$ and $\tau_1$ constrains the pulse rise, while the interplay between $\tau_1$, $\tau_2$, and $A$ constrains the pulse peak. These pulse parameters are harder to fit when the temporal resolution is poor, as there are fewer intensity points available with which to describe the intensity function. 

Very large and very small values of $\tau_1$ and $\tau_2$ are particularly hard to constrain.  A small $\tau_1$ value indicates a slow pulse rise while a large value indicates a rapid rise producing a correspondingly early pulse start time $t_s$. A small $\tau_2$ value indicates a rapid pulse decay while a large $\tau_2$ decay indicate a slow pulse decay. Pulse fits resulting in these large and/or small $\tau_1$ and $\tau_2$ values are often accompanied by fitting uncertainties exceeding the measured value by an order of magnitude or more. This most often happens in pulses that are short relative to the temporal resolution, as these occur where the rates of intensity rise and fall are masked by the temporal bin size.

When the $\tau_1$ rise and $\tau_2$ decay components of this function are well-behaved ($10^{-3} \lessapprox \tau_1 \lessapprox 10^3$ and $10^{-3} \lessapprox \tau_2 \lessapprox 10^3$), smoothly varying functions result and the $\tau_1$ and $\tau_2$ distributions seem to be Normally distributed. Very large or very small pulse rise and decay values produce uncertainty distributions that appear to be asymmetric. 

Increased temporal resolution improves the quality of both the pulse fits and measured pulse parameters.  For this reason, the formal pulse-fitting parameters obtained from the 4-ms TTE Complete sample have smaller formal uncertainties than their 64-ms TTE Partial counterparts. However, constraints are present for all pulses in the TTE pulse catalog as a result of poor counting statistics: the higher TTE resolution results in fewer counts per bin, which provides its own limits on pulse property measurement. 

The formal duration errors obtained from {\tt MPFIT} can be compared to the internal error distribution taken from the differences between $w_4$ and $w_{64}$. In other words,
\begin{equation}
\sigma_{w-{\rm internal}}=|w_4-w_{64}|/\sqrt{2}.
\end{equation}
We find that the internal and external duration error distributions are consistent with one another such that
\begin{equation}
\sigma_{w-{\rm internal}}^2 \approx \sigma_{w4}^2+\sigma_{w64}^2
\end{equation}
upon excluding pulses with poorly measured durations ($\sigma_{w4} \ge 10$ s and $\sigma_{w64} \ge 10$ s). It should also be noted that duration uncertainties increase for faint pulses (as measured both by fluence and by peak flux). This is not surprising, as the duration definition (given in Equation \ref{eqn:duration}) is dependent on intensity.

\subsection{Pulse Complexity as a Function of Signal-to-Noise} 

Some pulse complexity appears to result from a selection bias stemming from inadequate temporal resolution; this can be found by examining the different numbers of events in each of the pulse complexity classes (see Table \ref{tab:complex}). Far more TTE Complete pulses can be characterized by the \cite{nor05} pulse function plus the \cite{hak14} residual function than TTE Incomplete pulses. In other words, poor temporal resolution appears to have created false pulse structures by rebinning and smearing out the known triple-peaked pulse characteristics.
 
\begin{figure}[ht!]
\plotone{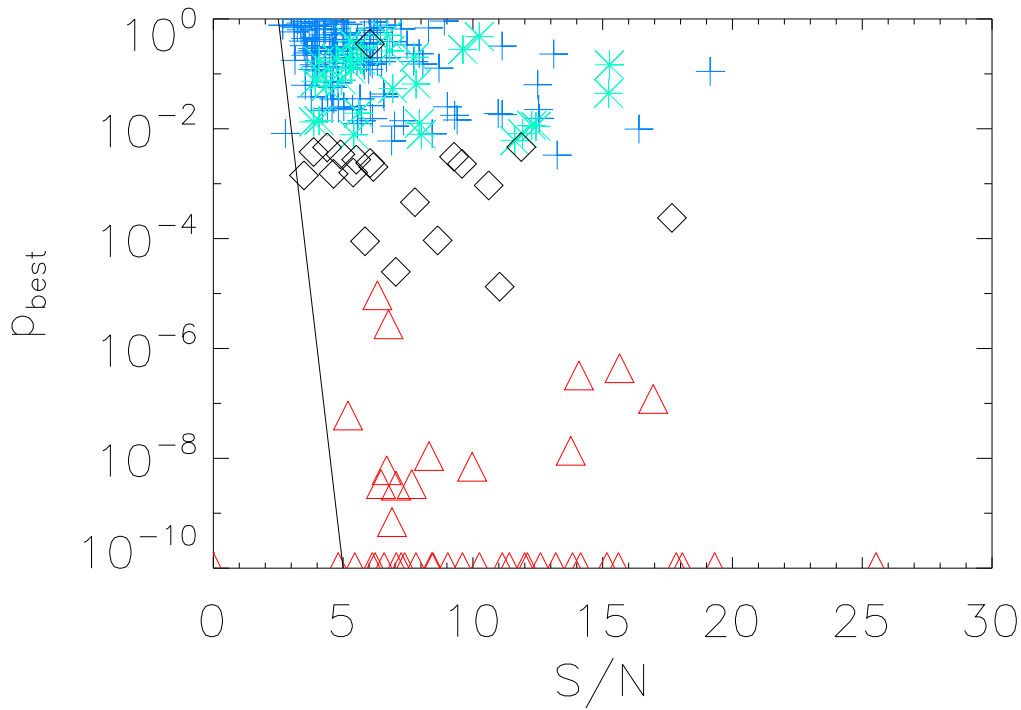}
\caption{Complexity of TTE Complete pulses ($p_{\rm best}$) as a function of 4-ms $S/N$. Here, blue crosses indicate Simple pulses, green asterisks indicate Blended pulses, black diamonds indicate Structured pulses, and red triangles indicate Complex pulses. Pulses with $p_{\rm best} < 10^{-10}$ have been plotted as having $p_{\rm best}=10^{-10}$. The simplest pulses are found at low $S/N$ while the most complex pulses are found at high $S/N$. \label{fig:complexvsn}}
\end{figure}

Once we exclude TTE Partial pulses from our sample we find that bright GRB pulses tend to have more complex structures than faint pulses, in agreement with previous results obtained for Long/Intermediate GRB pulses (\cite{hak14,hak15}). We characterize the TTE Complete sample by a 4-ms definition of $S/N$ (see Equation \ref{eqn:SN}). Figure \ref{fig:complexvsn} demonstrates that pulse complexity (characterized by the best-fit $p-$value $p_{\rm best}$) increases as $S/N$ increases; The correlation between these characteristics is significant (a Spearman Rank-Order Correlation analysis finds SC$=0.48$, $p=6 \times 10^{-15}$). Figure \ref{fig:complexvsn} shows that Simple and Blended pulses are the faintest, Structured pulses are brighter, and Complex pulses are the brightest. We draw several conclusions from Figure \ref{fig:complexvsn}:
\begin{itemize}
\item Short GRB pulses, like their Long and Intermediate burst counterparts, exhibit a triple-peaked structure.
\item A smaller percentage of Short GRB pulses seem to exhibit measurable residual structure (Blended and Structured) compared to Long/Intermediate GRBs, although it is difficult to imagine what a complete sample of Long/Intermediate burst pulses should look like based on existing analyses.
\item The triple-peaked structure is less pronounced for low $S/N$ GRB pulses (see Figure \ref{fig:complexvsn}), suggesting a sampling bias by which structure might be present in most or all pulses but cannot be resolved with low photon counts.
\item More pronounced pulse structures are observed at high $S/N$ (as denoted by the relative number of Structured and Complex pulses), suggesting that most or all GRB pulses contain complex structures, but these also are washed out at low $S/N$.
\end{itemize}

\begin{figure}[ht!]
\plotone{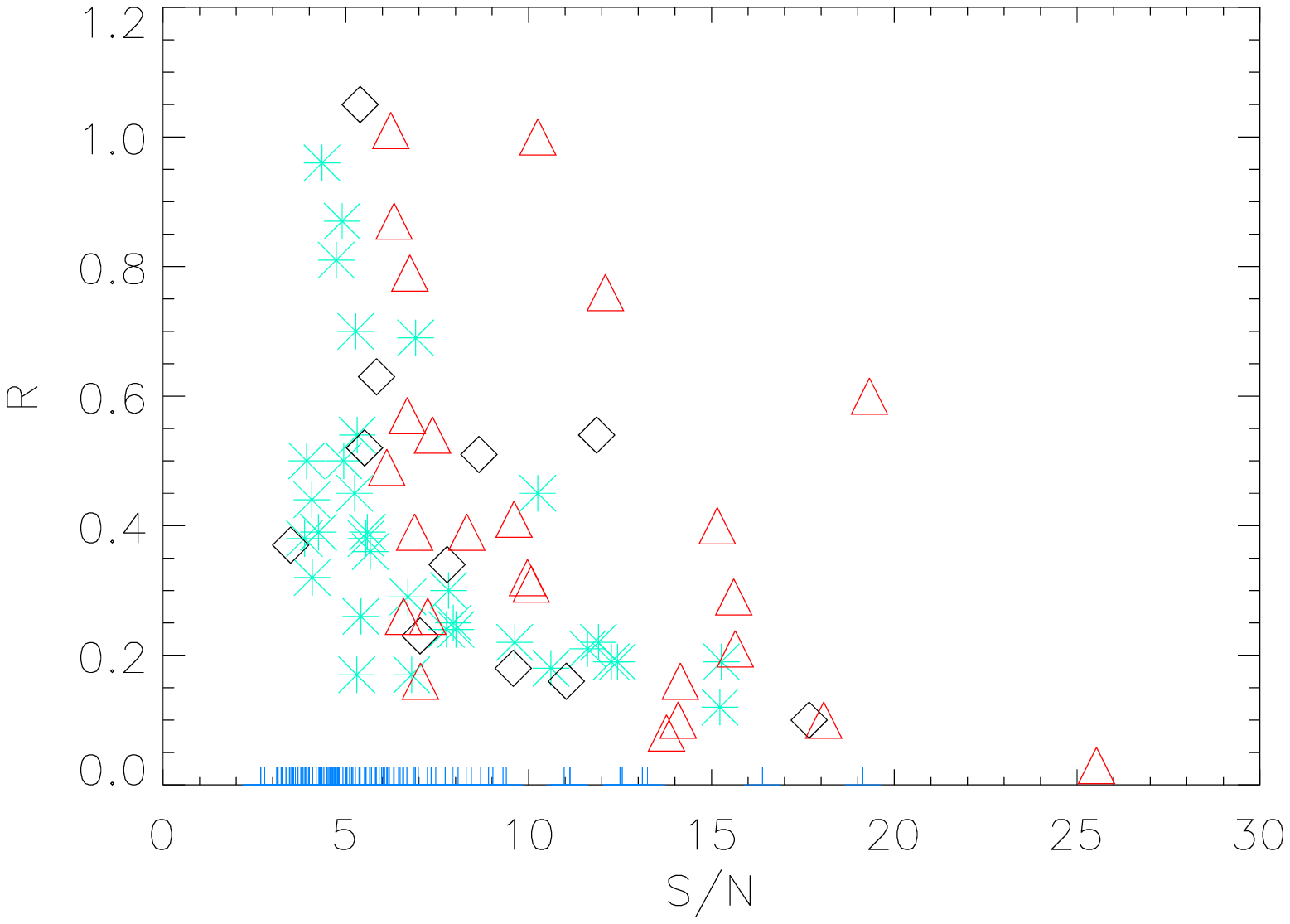}
\caption{Relative amplitudes of the fitted residual structures to the fitted pulses ($R$) as a function of $S/N$. Blue crosses indicate Simple pulses, green asterisks indicate Blended pulses, black diamonds identify Structured pulses, and red triangles identify Complex pulses. \label{fig:RvSN}}
\end{figure}

GRB pulses are difficult to resolve and to fit at low signal-to-noise, resulting in less-certain measurements of their properties relative to bright pulses. This can have the undesired effect of altering pulse properties near the $S/N$ threshold. Figure \ref{fig:RvSN} demonstrates that faint TTE pulse properties do indeed differ from those of bright pulses, as measured by $R$ (the ratio of the residual fit amplitude to the pulse fit amplitude; see Equation \ref{eqn:R}) relative to $S/N$. We find that:

\begin{itemize}
\item Pulses characterized by large complexities (Structured and Complex) are observed at larger $S/N$ than those having simpler structures (Simple and Blended). See Table \ref{tab:SN}.
\item Pulses with large residual structures ($R > 0.8$) are primarily found near the minimum $S/N$ threshold. See Figure \ref{fig:RvSN}.
\item Pulses observed at the largest $S/N$ have the smallest measured $R$ values. See Figure \ref{fig:RvSN}.
\end{itemize}

The light curves of bright TTE pulses exhibit more pronounced structural complexity than the smooth light curves of fainter pulses. Some of this can be explained by the simple observation that noise is capable of washing out pre-existing pulse structures and making pulse light curves look smoother. However, the large $S/N$ range spanned by pulses suggests that there might also be an intrinsic effect such that bright pulses exhibit larger temporal variabilities than faint ones. This can only be explained if bright pulses are also more luminous than faint ones. Such a conclusion is consistent if a pulse lag vs.~pulse luminosity relationship exists for Short GRBs that is analogous to the relationship identified previously for Long ones ({\em e.g} \cite{hak08}).

\begin{deluxetable*}{ccc}
\tablenum{10}
\tablecaption{TTE Signal-to-noise of TTE Pulses by Complexity\label{tab:SN}}
\tablewidth{0pt}
\tablehead{
\colhead{Pulse Complexity} & \colhead{$\langle S/N \rangle $} & \colhead{$\sigma_{S/N}$}  \\
}
\startdata
Simple & 5.8  & 2.7 \\
Blended & 7.3 & 3.3 \\
Structured & 8.5 & 4.0 \\
Complex & 10.9 & 5.1 \\
\enddata
\end{deluxetable*}

\subsection{Complexity in Blended and Structured Pulses: Characterizing the Residual Function} 

The addition of the residual function improves many of the Short GRB pulse fits. 
The wavelike form of the residual function, that can be described by a
modified Bessel function attached to a compressed mirror image
of itself, produces a rippled or multi-peaked shape to the 
otherwise monotonic underlying pulse. 
The multi-peaked shape is common among the
isolated pulses in Long/Intermediate bursts detected by BATSE,
Swift, and Fermi GBM \citep{hak14, hak15}, and the 
characteristics of the residual function correlate with
a number of other pulse properties.

\begin{figure}
\plottwo{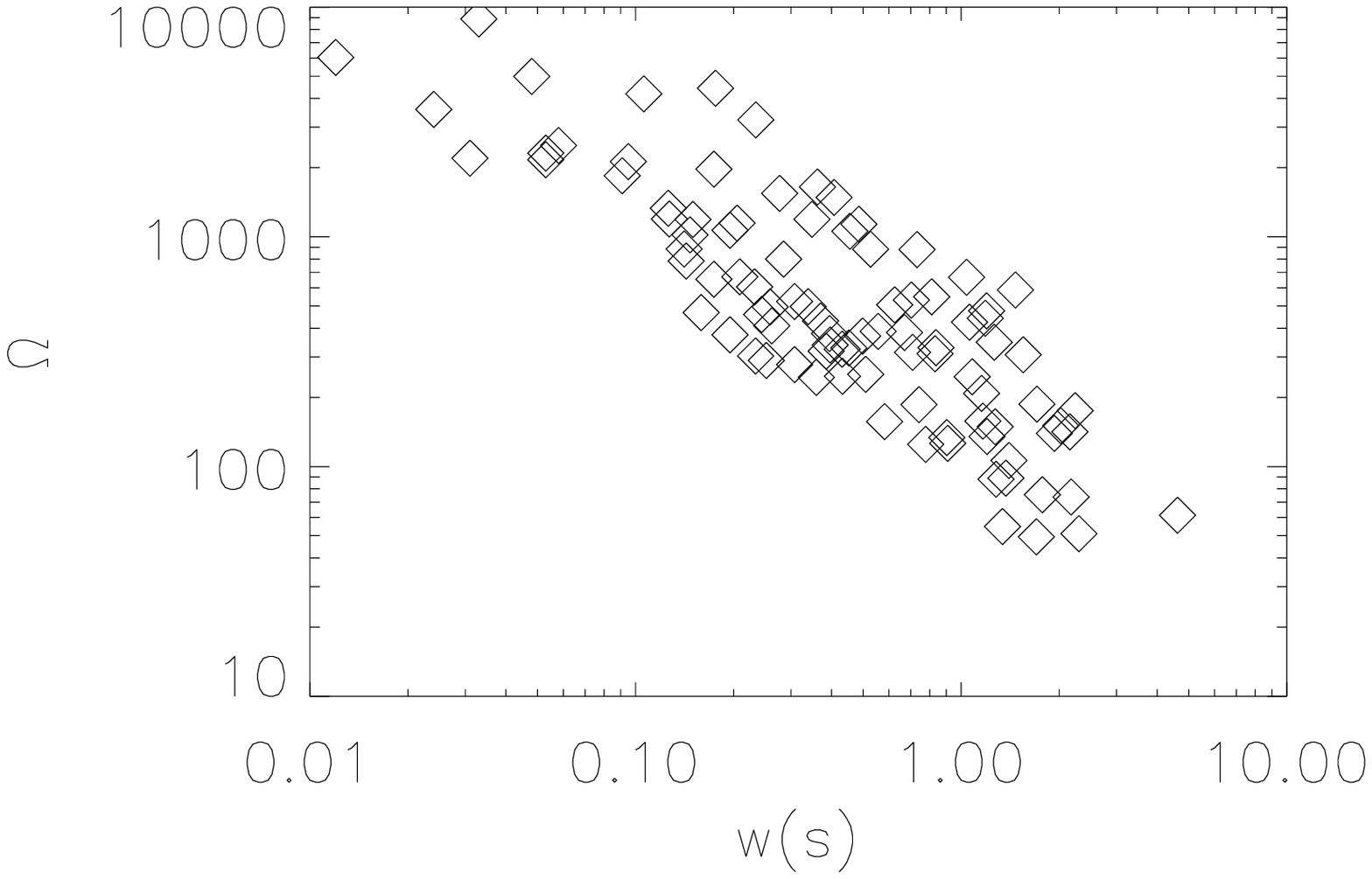}{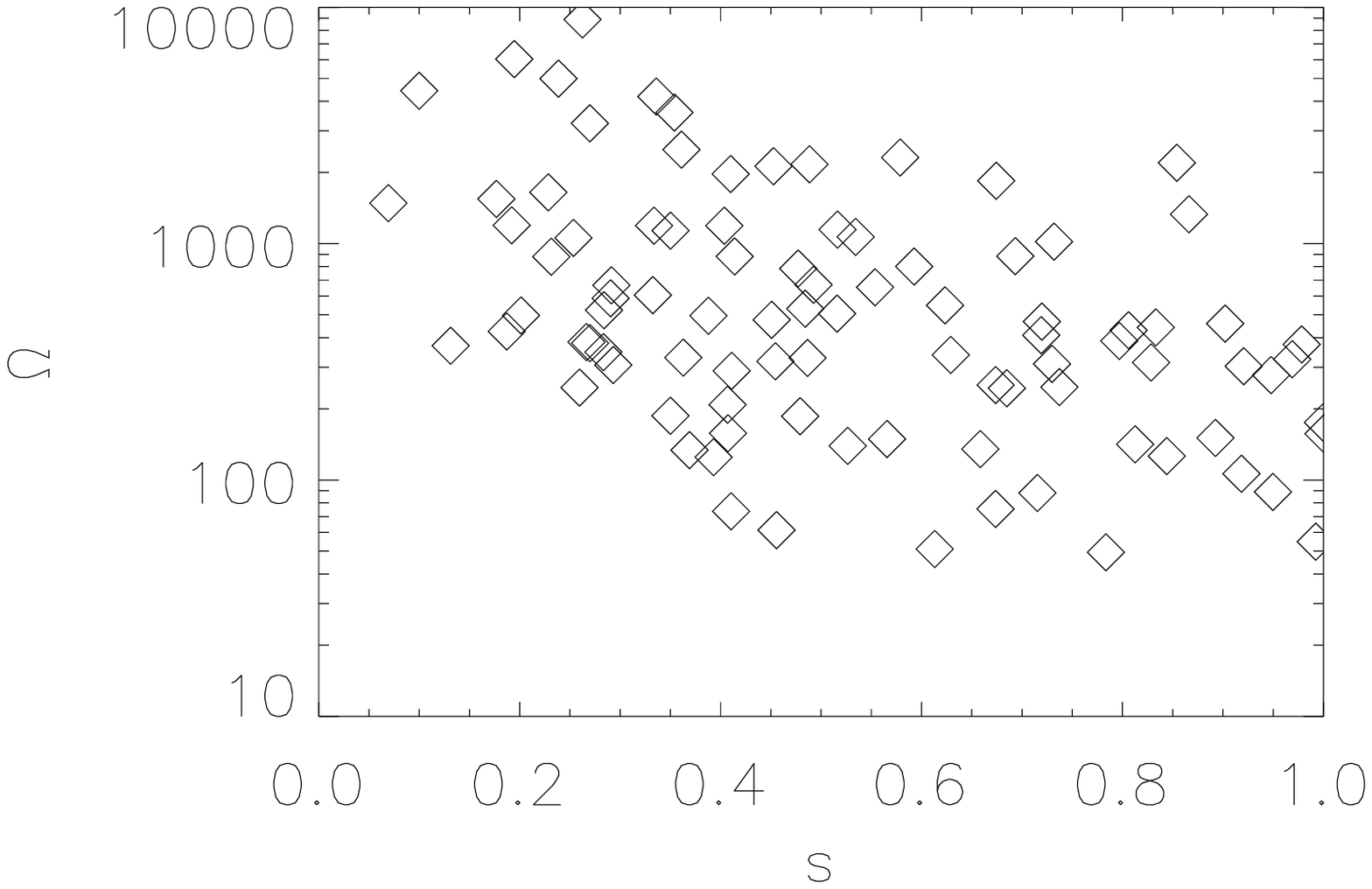}
\caption{Left panel: Bessel function frequency ($\Omega$) vs. pulse duration ($w$). Shorter duration pulses
have larger values of $\Omega$, indicating that shorter pulses have correspondingly shorter
residual functions. Right panel: Bessel function frequency ($\Omega$) vs. pulse residual stretching parameter ($s$). Pulses
with greater stretching, indicating asymmetry in their residual function, have larger values of $\Omega$. \label{fig:omegavW}}
\end{figure}

The residual function is generally confined to the temporal 
interval occupied by the underlying pulse: the duration of the
residual function (characterized by the Bessel frequency $\Omega$)
correlates with the pulse duration ($w$), which is similar to results
found for Long/Intermediate GRB pulses \citep{hak14, hak15}.
The left panel of Figure \ref{fig:omegavW} demonstrates 
this correlation (SC$=-0.80$, $p=6 \times 10^{-20}$).

For Long/Intermediate GRB pulses, the inherent asymmetry 
of the residual function (characterized by $s$)
anti-correlates with the pulse asymmetry ($\kappa$), indicating that the
residual structure is aligned with the underlying pulse shape.
Unfortunately, a similar correlation cannot be verified 
for Short GRB pulses (SC$=-0.35$, $p=2 \times10^{-2}$ is found), as the low 
$S/N$ environment in which these pulses are found make 
accurate $\kappa$ measurements difficult.

An anti-correlation is found between $\Omega$
and $s$ (demonstrated in the right panel of 
Figure \ref{fig:omegavW}, with a Spearman Rank-Order 
Correlation of SC$=0.46$, $p=8 \times10^{-6}$). 
This is surprising because this correlation
suggests that duration ($w$) and asymmetry ($\kappa$) are related,
whereas no correlation is found ($p=0.82$).  
We suspect that this correlation is not entirely real;
it might result from the low $S/N$ environment in which 
Short GRB pulses are found, the potentially interdependent
ways in which $\Omega$ and $s$ contribute to the 
residual function in Equation \ref{eqn:residual}, 
and the fact that our initial estimates of $\Omega$ 
and $s$ are based on $\kappa$.

\begin{figure}
\plotone{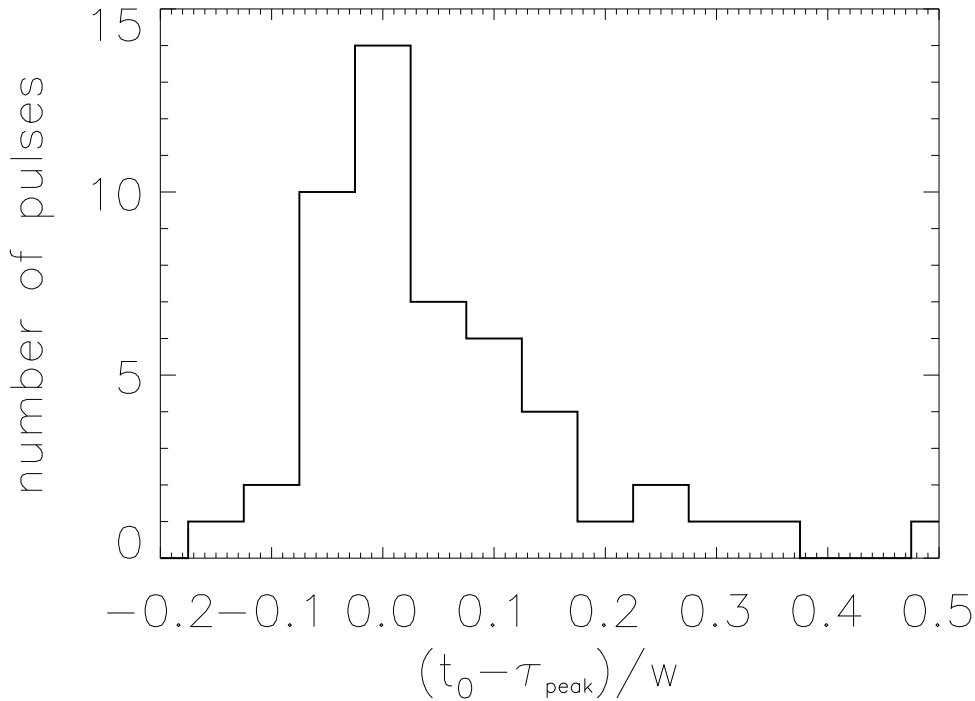}
\caption{Histogram showing the offset between the peak of the residual function ($t_0$)
and the peak of the underlying pulse ($\tau_{\rm peak}$), normalized to the pulse duration ($w$). 
The pulse and residual peaks are generally
not aligned, although the offset can be either positive or negative.
This sample has been limited to TTE Complete pulses having the most easily-measured
residual functions ({\em e.g.,} Blended and Structured). \label{fig:t0-taupeak}}
\end{figure}

The peak time of the residual function $t_0$ is found to not always
align with the peak time of the underlying pulse $\tau_{\rm peak}$. This
is demonstrated in Figure \ref{fig:t0-taupeak}, where the difference
$t_0-\tau_{\rm peak}$ has been normalized to a standard time by dividing
it by the pulse duration $w$. Although this offset appears to be real,
the reason for the offset (which is positive for some bursts and
negative for others) is still not understood, because it implies that the
pulse and the residual function are somewhat independent of one another. 

The amplitude of the 
residual function ($a$) varies from near zero to roughly the
pulse amplitude ($A$); the ratio of these amplitudes is 
characterized by $R$. However, we find no obvious correlation between
the normalized difference and the alignments of $t_0$ and $\tau_{\rm peak}$ 
with other pulse parameters ({\em e.g.,} $R$, $HR$, $S$).

\subsection{Pulse Spectral Evolution} 

Long/Intermediate GRB
pulse light curves evolve from hard to soft, with re-hardening 
occurring at or just prior to each of the three pulse peaks \citep{hak15}. 
Asymmetric pulses are hard overall and have
pronounced hard-to-soft evolution; these contrast with symmetric pulses 
that are softer and have weak hard-to-soft evolution. 
This weak evolution can result in softer precursor peaks
than central peaks, giving pulses the appearance of having
intensity tracking behaviors.

It is interesting to see if Short GRB pulses undergo similar spectral evolutions
as Long/Intermediate GRB pulses. Finding that they do would independently
validate our initial assumption that Short GRB emission episodes are
indeed individual pulses, because we made no spectrally-dependent
assumptions about spectral evolution in our pulse definition (see
\ref{sec:method}).

The TTE pulse light curves have been collected in the four energy channels:
described previously.
Although the count rates in each of these four channels are low, 
they provide some information that can be used to infer pulse
spectral evolution. We define the counts hardness ($hr$) in each
time bin $i$ as:
\begin{equation}
hr_i= \frac{C_{3i}+C_{4i}}{C_{1i}+C_{2i}}
\end{equation}
where $C_{1i}, C_{2i}, C_{3i}$, and $C_{4i}$ are the counts/bin in
channels 1, 2, 3, and 4, respectively. We track 4 ms pulse spectral 
evolution by measuring $hr_i$ in each
bin between $t_{\rm start}$ and $t_{\rm end}$, in a manner similar
to that done in \cite{hak15} for BATSE and Swift data.

\begin{figure}
\plottwo{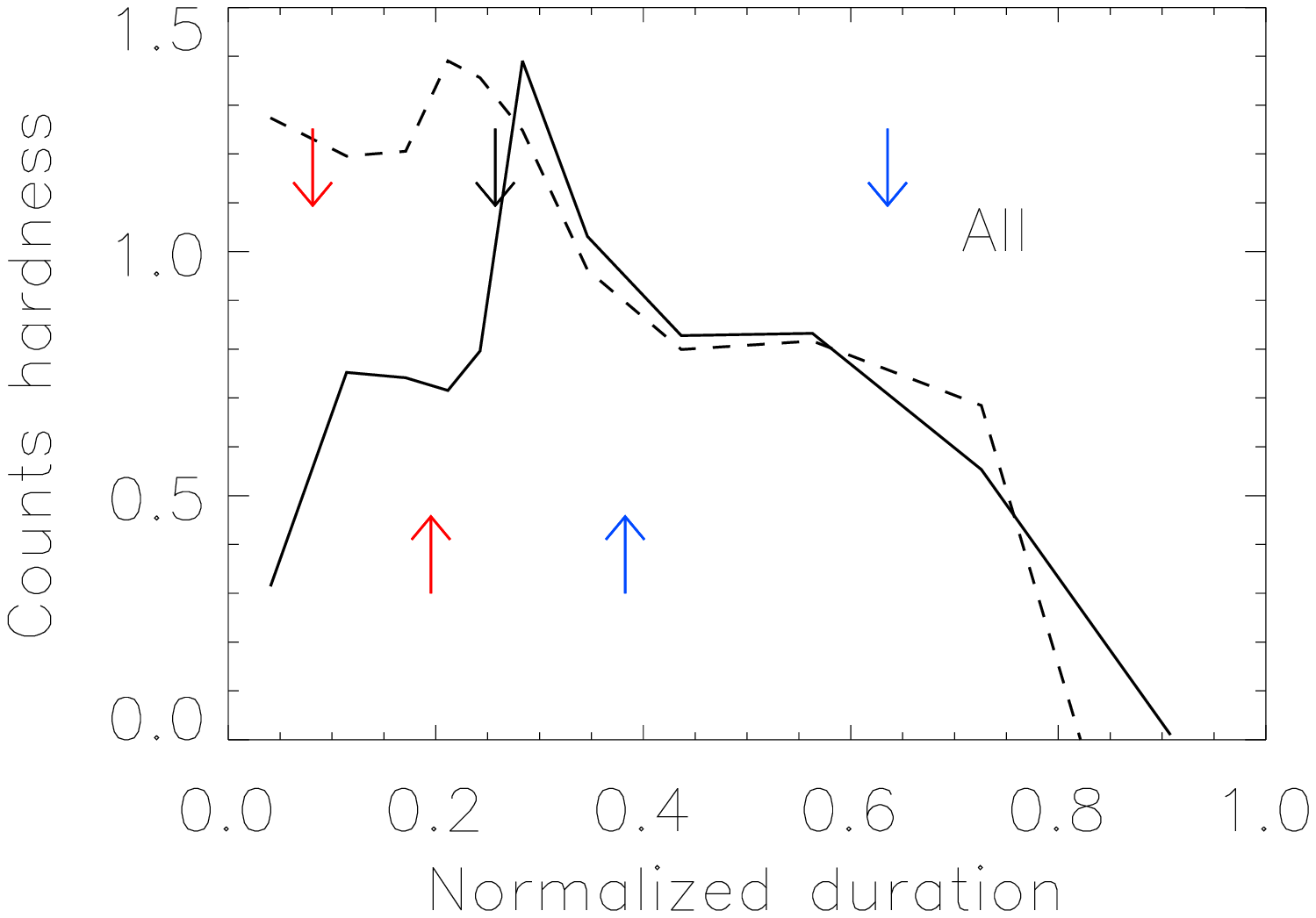}{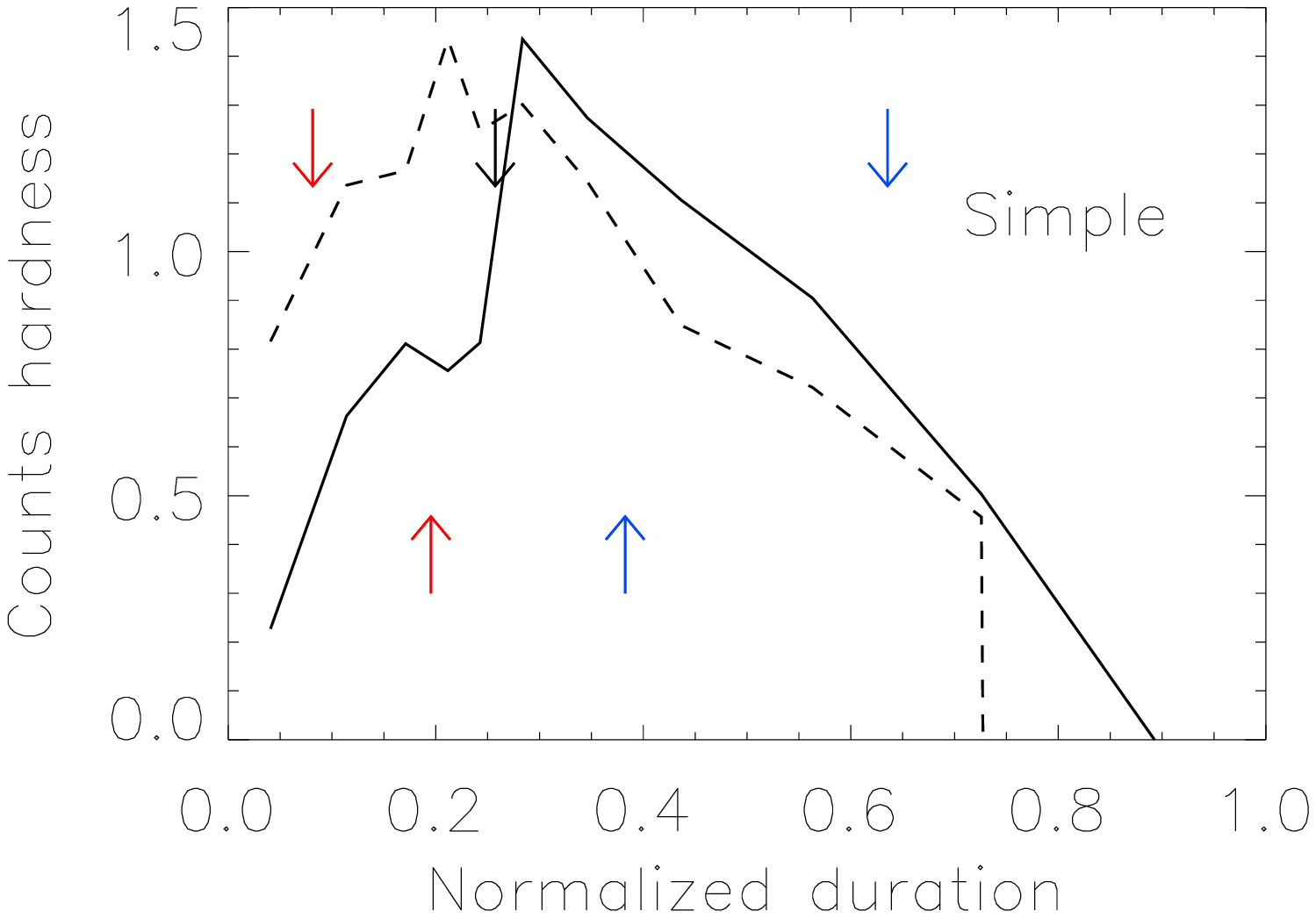}
\caption{Normalized mean light curve (solid line) and counts hardness ($hr$) 
evolution (dashed line) of 159 Short GRB pulses (left panel)
and 102 Simple pulses (left panel). In this and subsequent
figures, downward facing
arrows indicate the approximate times of the precursor peak (red),
central peak (black), and decay peak (blue). Upward facing arrows
indicate the time of the valley separating the precursor peak (red)
and decay peak (blue) from the central peak.\label{fig:spec_1}}
\end{figure}

\begin{figure}
\plottwo{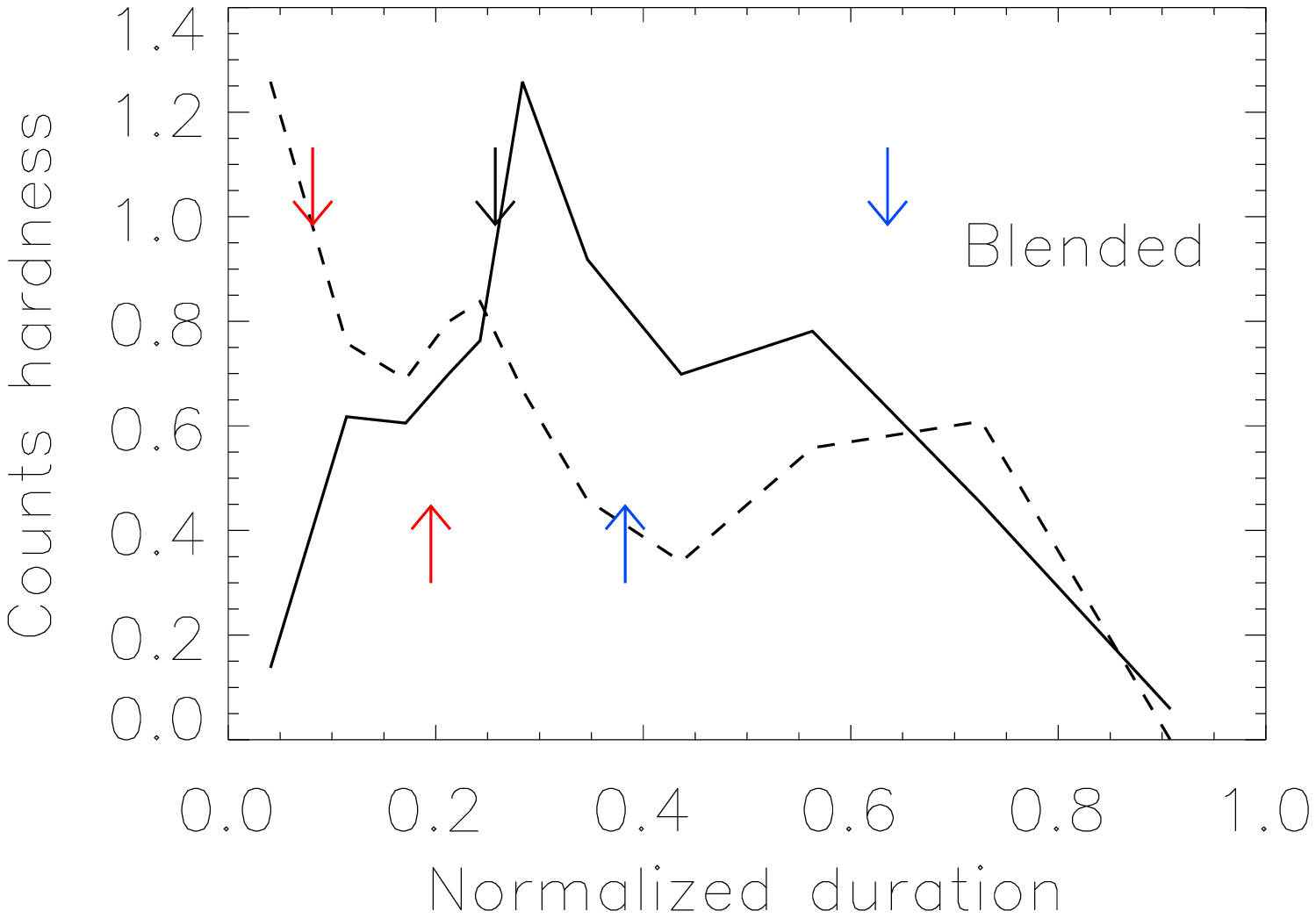}{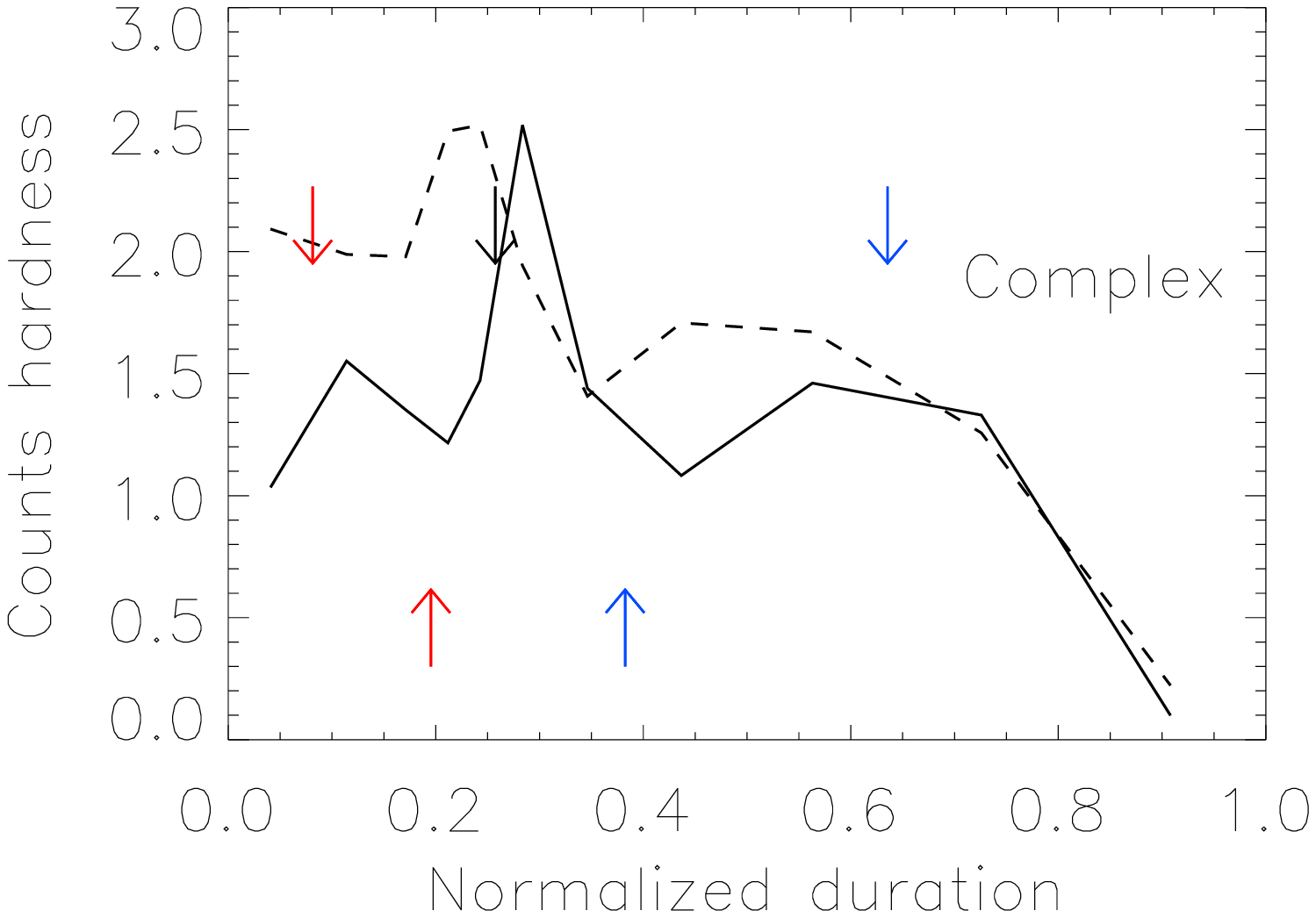}
\caption{Normalized mean light curves (solid line) and counts hardness ($hr$) 
evolution (dashed line) of 25 Blended pulses (left panel) and 32 
Structured and Complex pulses (right panel). \label{fig:spec_2}}
\end{figure}

\begin{figure}
\plottwo{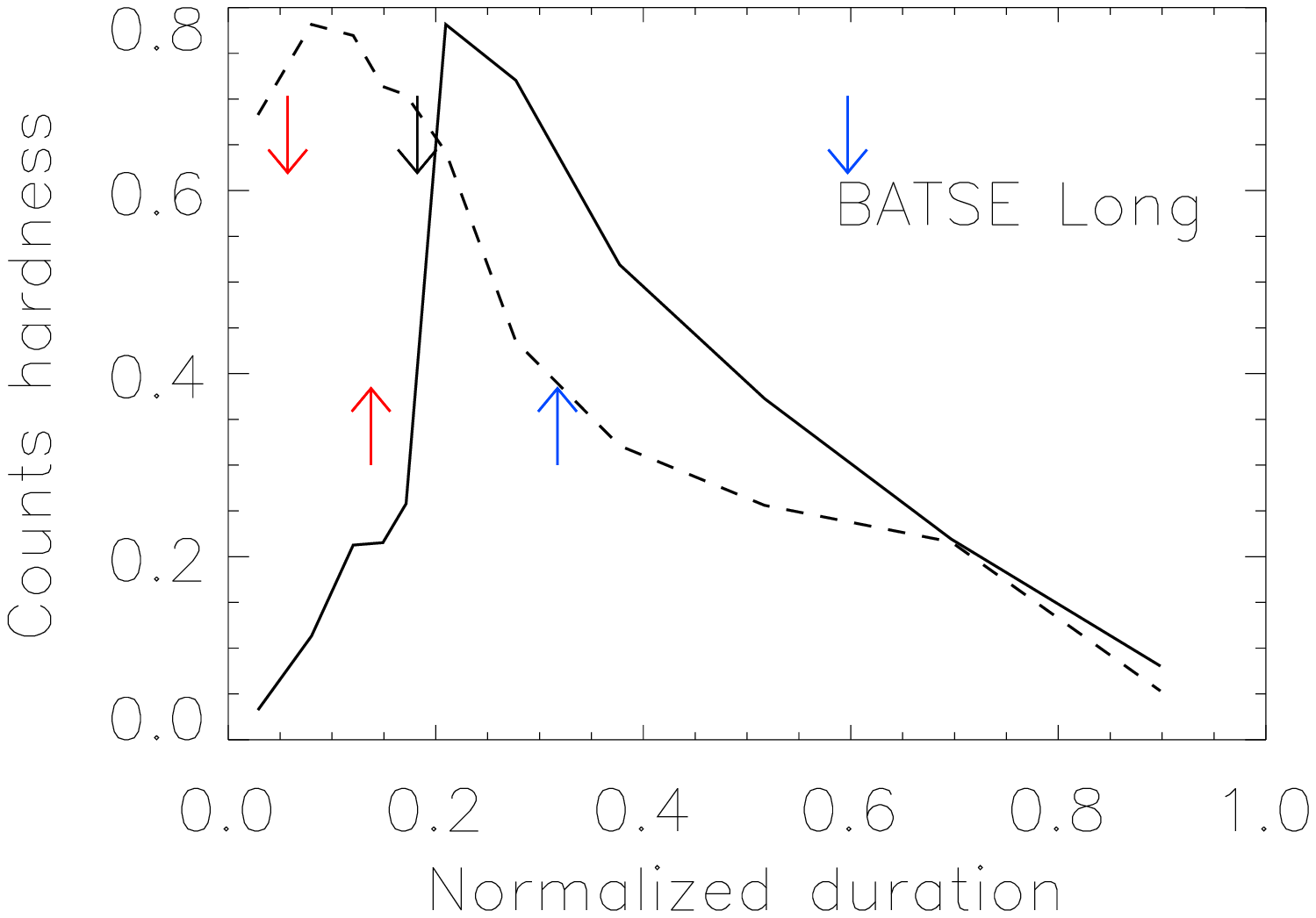}{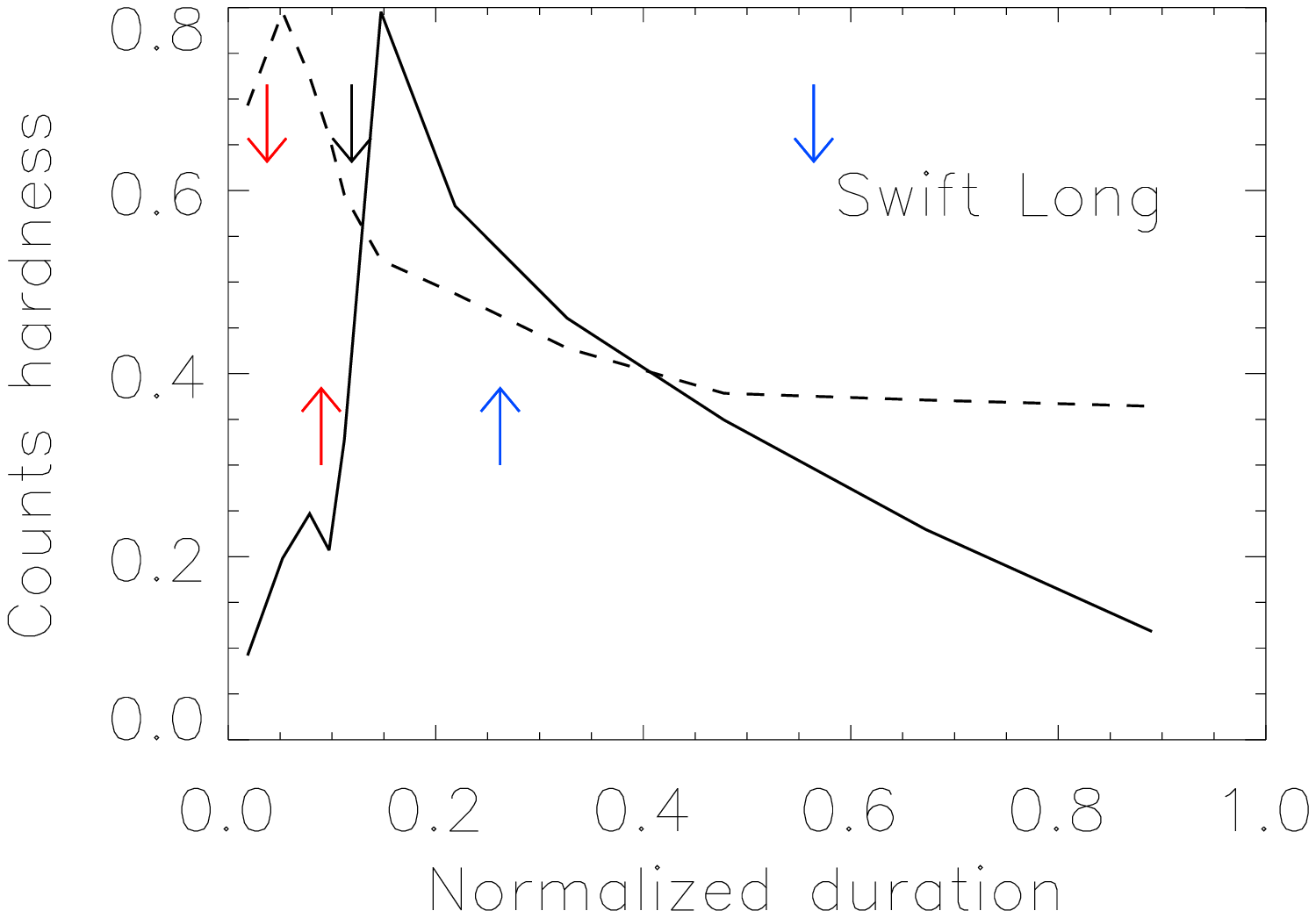}
\caption{Normalized mean light curves (solid line) and counts hardness ($hr$) 
evolution (dashed line) of Long/Intermediate
pulses observed by BATSE (left panel) and Swift (right panel) \label{fig:spec_3}}
\end{figure}

We sum the counts from many pulses to get summed light curves and
spectral evolutionary averages; this approach overcomes
limits imposed by small number counting statistics and
allows us to examine spectral evolution as a function of
pulse structure. (Note: we have excluded pulses with
negative total hardness ratios, as well as pulses with trigger numbers
between 3282 and 3940 having incorrectly transcribed Channel 1 counts data). 
Figure \ref{fig:spec_1} shows the normalized mean light curve
(solid line) and $hr$ evolution (dashed line) for all 159 BATSE 
TTE Complete pulses (left panel). Also shown are normalized mean light curves
(solid lines) and counts hardness evolutions (dashed lines) of 102 Simple pulses (right panel). 
Figure \ref{fig:spec_2} contrasts these
with the normalized mean light curves (solid lines) and counts hardness evolution (dashed lines)
of 25 Blended pulses (left panel) 
and 32 Structured/Complex pulses (right panel). Figure \ref{fig:spec_3} 
shows the normalized mean light curves (solid lines) and counts hardness evolutions
(dashed lines) of Long/Intermediate BATSE pulses (left panel) and 
Long/Intermediate Swift pulses (right panel).

It is not surprising that the summed light curves exhibit the triple-peaked structure, 
as the light curves have been co-added using
this structure as a temporal template. However, this co-adding should not
produce the observed hard-to-soft pulse evolution, with spectral re-hardening 
occurring at or just before each of the three peaks. This behavior is similar
to that seen in Long/Intermediate GRB pulses, independently 
demonstrating that these Short GRB emission episodes are individual pulses.

The normalized mean light curves verify the hypothesis that each emission episode 
contains but a single pulse. This appears to be true even for Structured and
Complex pulses, where highly variable light curves are co-added to produce
smooth light curves exhibiting only the triple-peaked structure.
The rapidly-varying component does not alter the underlying hard-to-soft
spectral evolution, which is similar to that found in Simple and Blended pulses. 
However, Structured/Complex pulses appear to be harder than smoother pulse types,
suggesting that the highly-variable component is responsible for this.
This result leads us to two important conclusions: 1) despite their highly 
variable structures, complex emission episodes are also {\em single pulses}, and
2) the highly-variable component found in Structured/Complex pulses contains
higher energy photons than what is found in the smoothly evolving component.

This verification leads us to draw an additional important conclusion:
{\em structure and complexity} beyond the triple-peaked pulse shape
represents an additional, randomly-distributed emission component that is 
not present in all Short GRB pulses. Summing together a large number
of Structured and Complex pulses should itself produce a complex
light curve rather than the triple-peaked structure seen in the right
panel of Figure \ref{fig:spec_2}. As described in the previous paragraphs,
this additional emission component is bright, hard, and variable.

\subsection{Multi-Pulsed Short GRBs} 

Although multi-pulsed TTE bursts are uncommon (making up only $10\%$
of the population), their light curves are interesting because they contain
{\em interpulse separations} as well as pulse durations. 
For multi-pulsed GRBs we define the interpulse separation ($w_{\rm sep}$) as
\begin{equation}
w_{\rm sep}=\tau_{\rm peak2}-\tau_{\rm peak1}
\end{equation}
where $\tau_{\rm peak1}$ and $\tau_{\rm peak2}$ are the times of maximum
amplitude for pulses 1 and 2, respectively.
The $T_{90}$ duration of a two-pulsed GRB is thus
\begin{equation}
T_{90} \approx  \tau_{1; \rm rise}+w_{\rm sep}+\tau_{2; \rm decay}
\end{equation}
where $\tau_{1; \rm rise}$ is the rise time of the first pulse
and $\tau_{2; \rm decay}$ is the decay time of the second pulse. 
Since $w_{\rm sep}$
is generally larger than the durations of either pulse, the $T_{90}$
duration of a GRB is generally dominated by the interpulse
separation ({\em e.g.,} see \cite{hak03}). Measurements of $w_{\rm sep}$
allow us to explore relationships between emission times of pulses
in multi-pulsed GRBs as well as the pulsed emission itself.

\begin{figure}
\plottwo{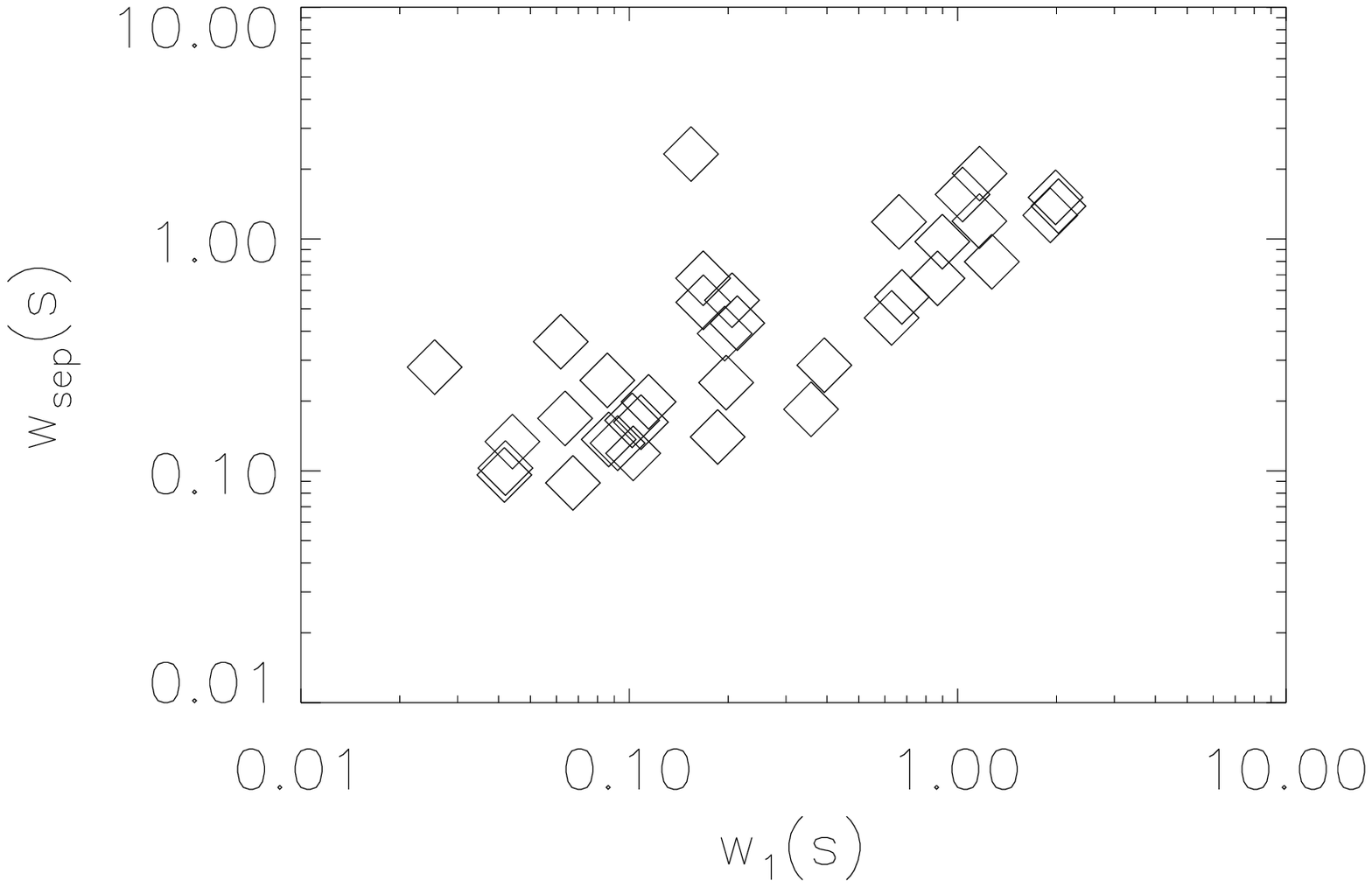}{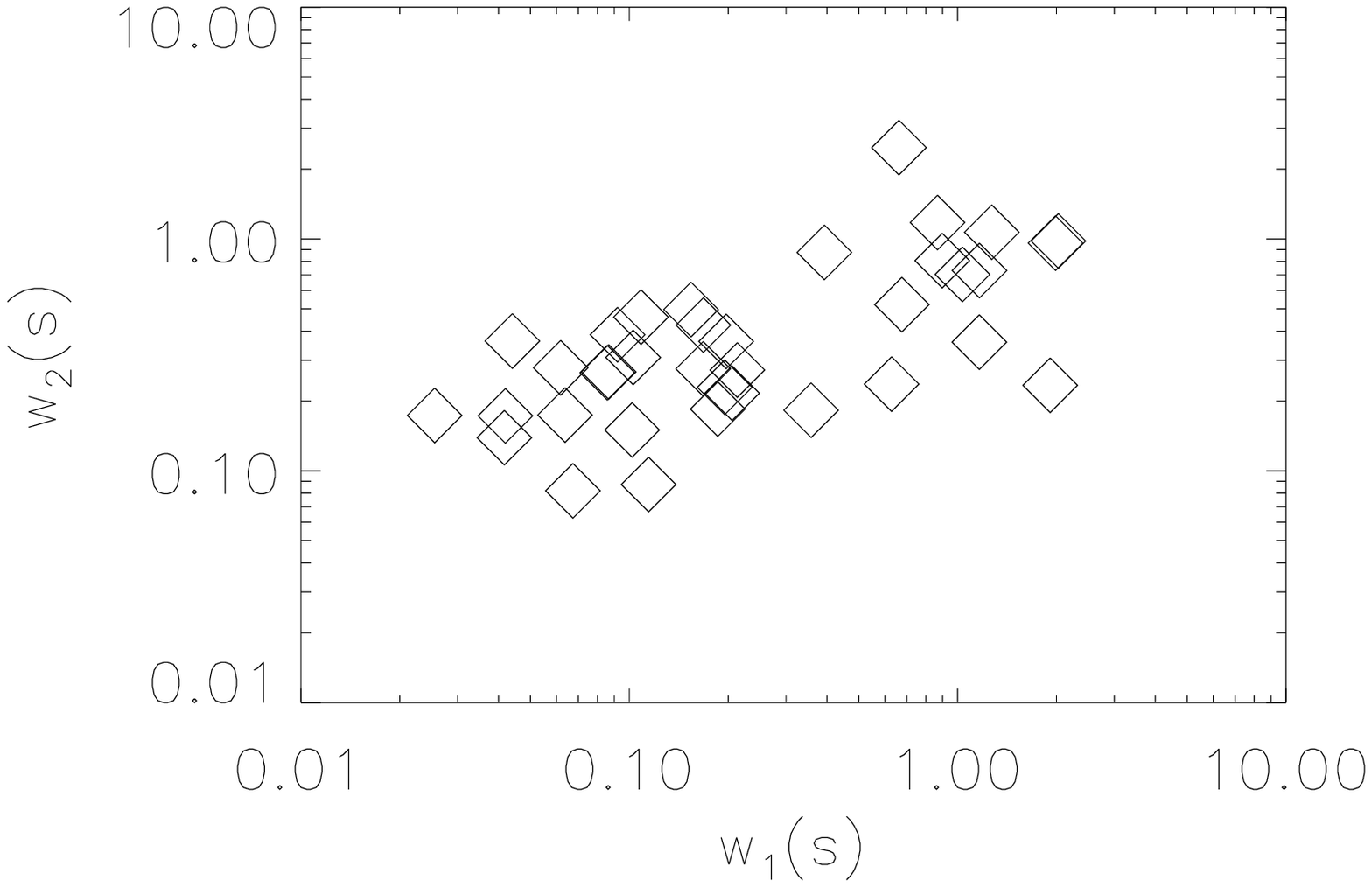}
\caption{The separation between pulses ($w_{\rm sep}$; left panel) and the duration of
the second pulse ($w_2$; right panel) in multi-pulsed GRBs both increase 
as the duration of the first pulse $w_1$ increases.\label{fig:double}}\end{figure}

\begin{figure}
\plotone{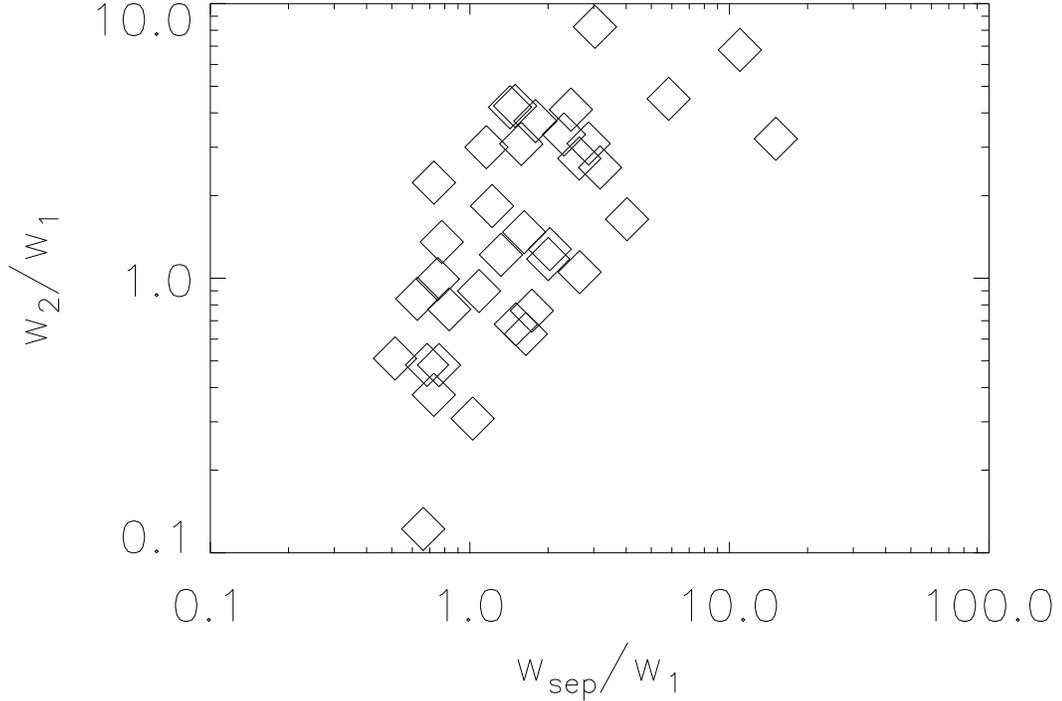}
\caption{Redshift-independent characteristics of double-pulsed Short GRBs. 
A Spearman Rank-Order correlation value of $p=5 \times 10^{-5}$) demonstrates
a lengthening of the observed emission episodes coupled
with a lengthening of the waiting time between these episodes. \label{fig:spread}}
\end{figure}

Strong correlations exist between the 
times of the emission episodes and the intervals separating them.
The left panel of Figure \ref{fig:double} demonstrates
that interpulse separations strongly correlate 
with first pulse durations; a Spearman Rank-Order test finds 
SC$=0.78$ and $p=6 \times 10^{-8}$. This 
correlation indicates that longer 
energy release times in the first pulse introduce 
correspondingly longer wait times until energy is released in the next pulse.
The right panel of Figure \ref{fig:spread} shows that
the second pulse's duration is also longer when the 
first pulse's duration is long; a Spearman Rank-Order 
correlation test finds SC$=0.60$ and $p=2 \times 10^{-4}$. This
correlation indicates that the energy release time
of the second pulse lasts longer when energy release time
of the first pulse is also long.

Even though few GRB redshifts were available 
during the BATSE era, the three durations
we have measured independently in each burst ($w_1, 
w_2,$ and $w_{\rm sep}$) can provide us with sufficient information
to develop two redshift-independent parameters. 
We define these parameters by
dividing the second pulse's duration and the burst's inter-pulse separation by the
corresponding first pulse's duration. Since all three parameters are time dilated
by the same factor $1+z$ (where $z$ is the redshift), the ratios $w_2/w_1$ and
$w_{\rm sep}/w_1$ are redshift-independent.
Figure \ref{fig:spread} demonstrates the intrinsic correlation between
$w_2/w_1$ and $w_{\rm sep}/w_1$ (SC$=0.67$, $p=2 \times 10^{-5}$). This correlation demonstrates
a lengthening of the observed emission episodes coupled
with a lengthening of the waiting time between these episodes.

Pulse durations and interpulse separations
are not independent quantities: later pulses have
memories of at least some properties of the initial pulses,
as well as of the gaps separating the pulses. If pulses
represent structures undergoing kinematic motion,
then a long pulse duration indicates that the pulsed emission
occurs over a large distance.  Similarly large interpulse
gaps indicate either a large distance between locations
where a pulse occurs, or a deceleration in the
bulk flow velocity. One interpretation of the increase
in duration between the second pulse and the first pulse
would then be that the emitting material has slowed
and/or lost energy. This could be the result of jet
expansion and/or slowing of the bulk flow.

\subsection{Crescendo GRBs}\label{sec:crescendo}

\begin{figure}
\plottwo{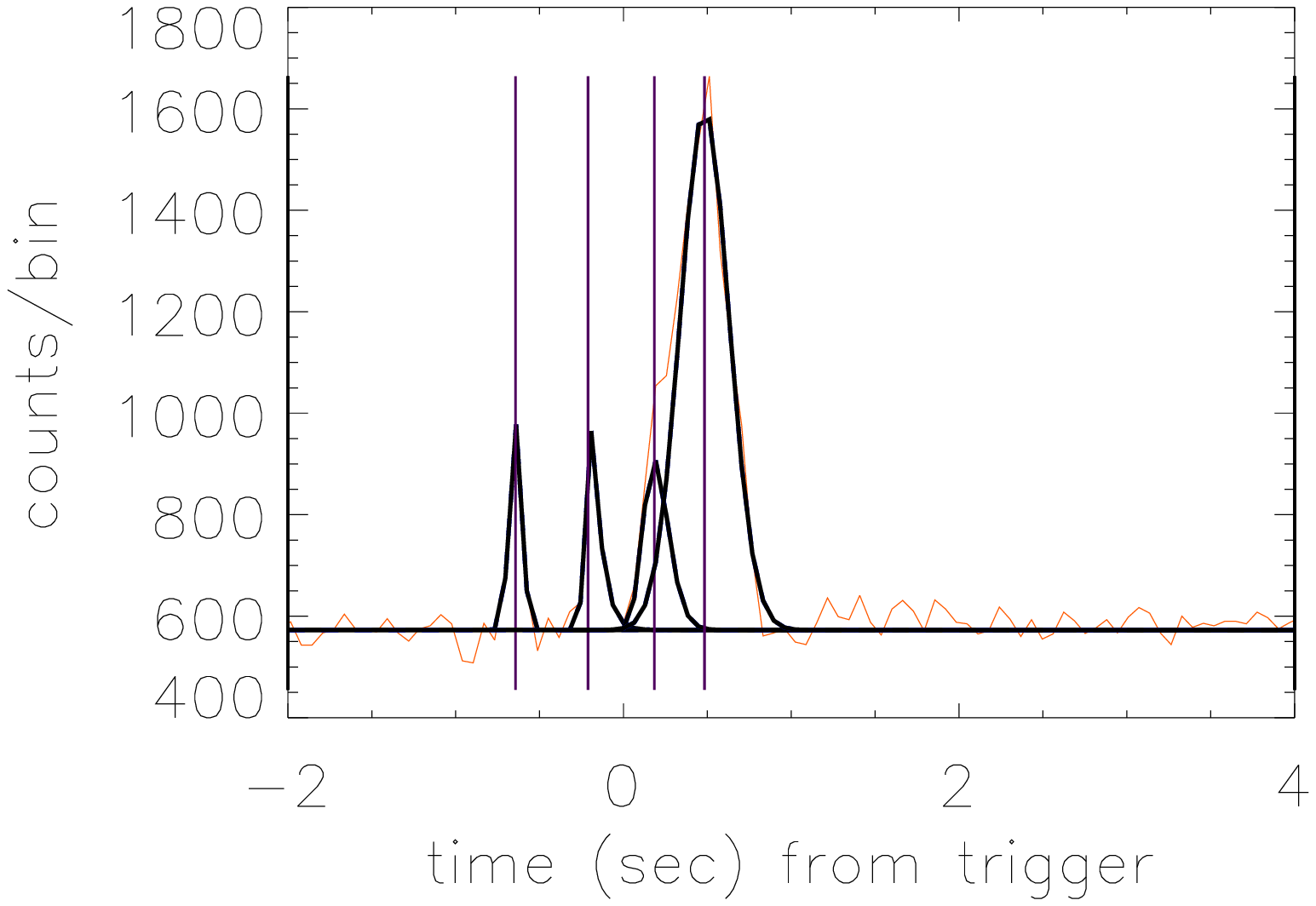}{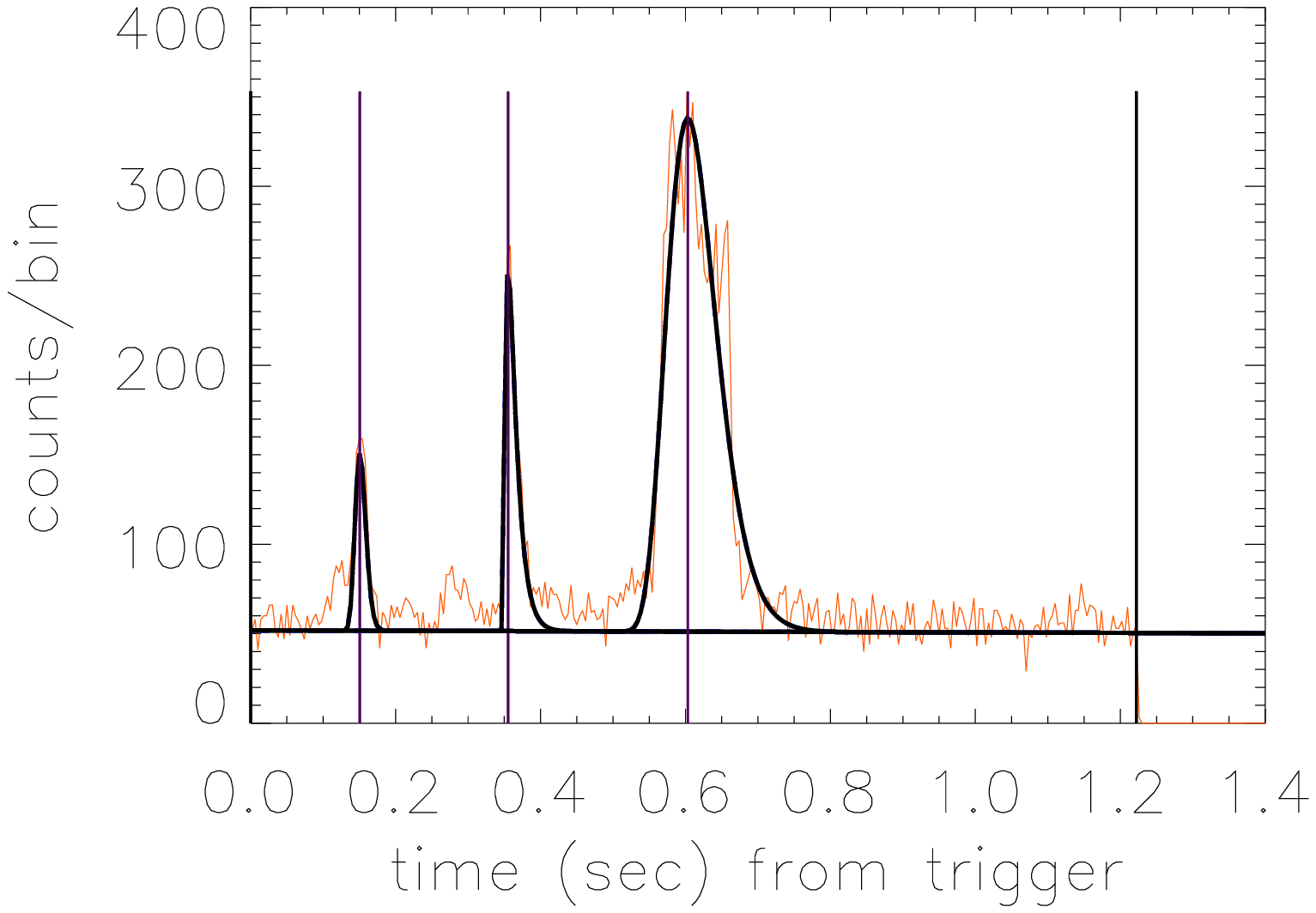}
\caption{Crescendo GRBs 3735 (left panel) and 5439 (right panel). These bursts contain Staccato pulses. \label{fig:stac1}}
\end{figure}

\begin{figure}
\plottwo{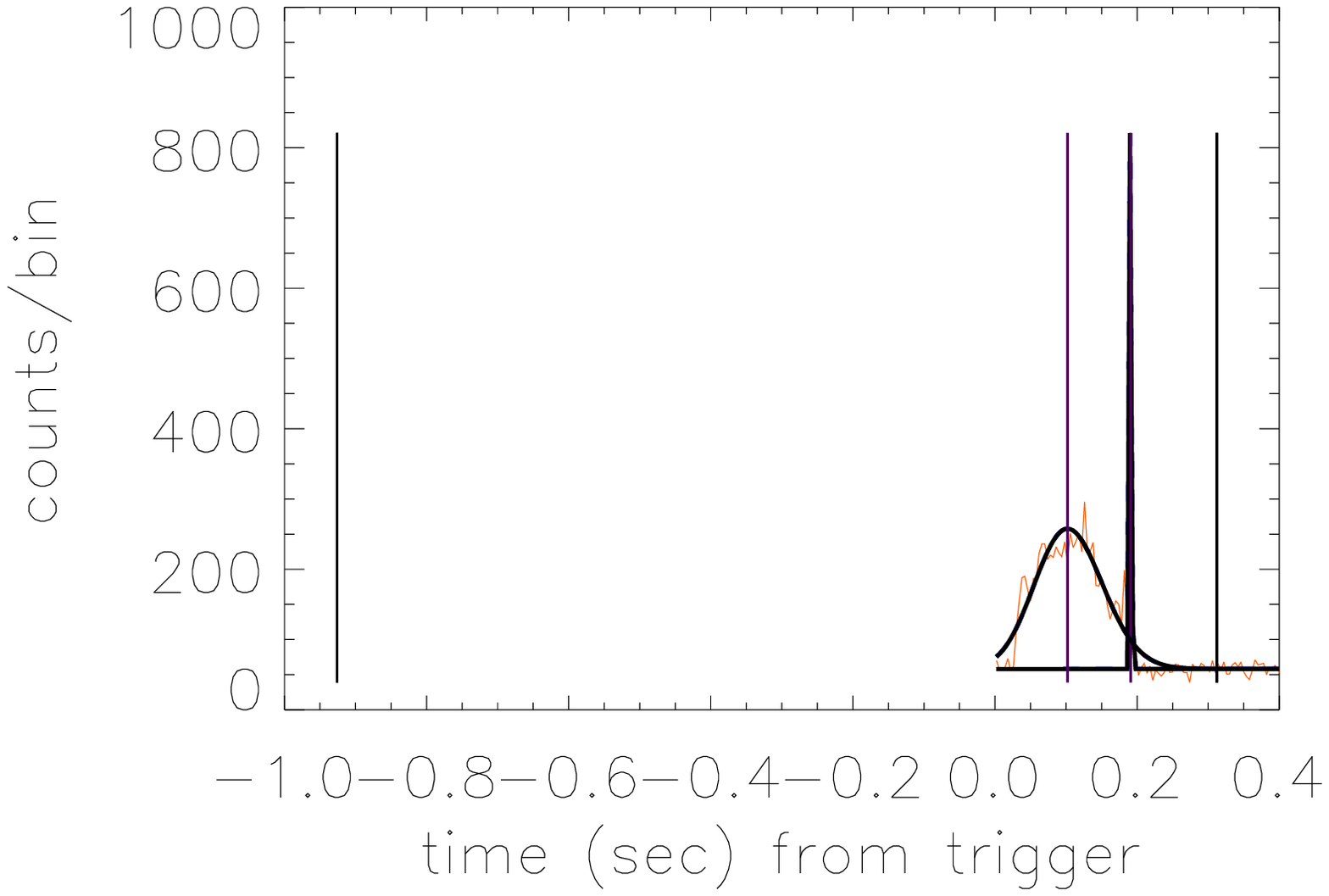}{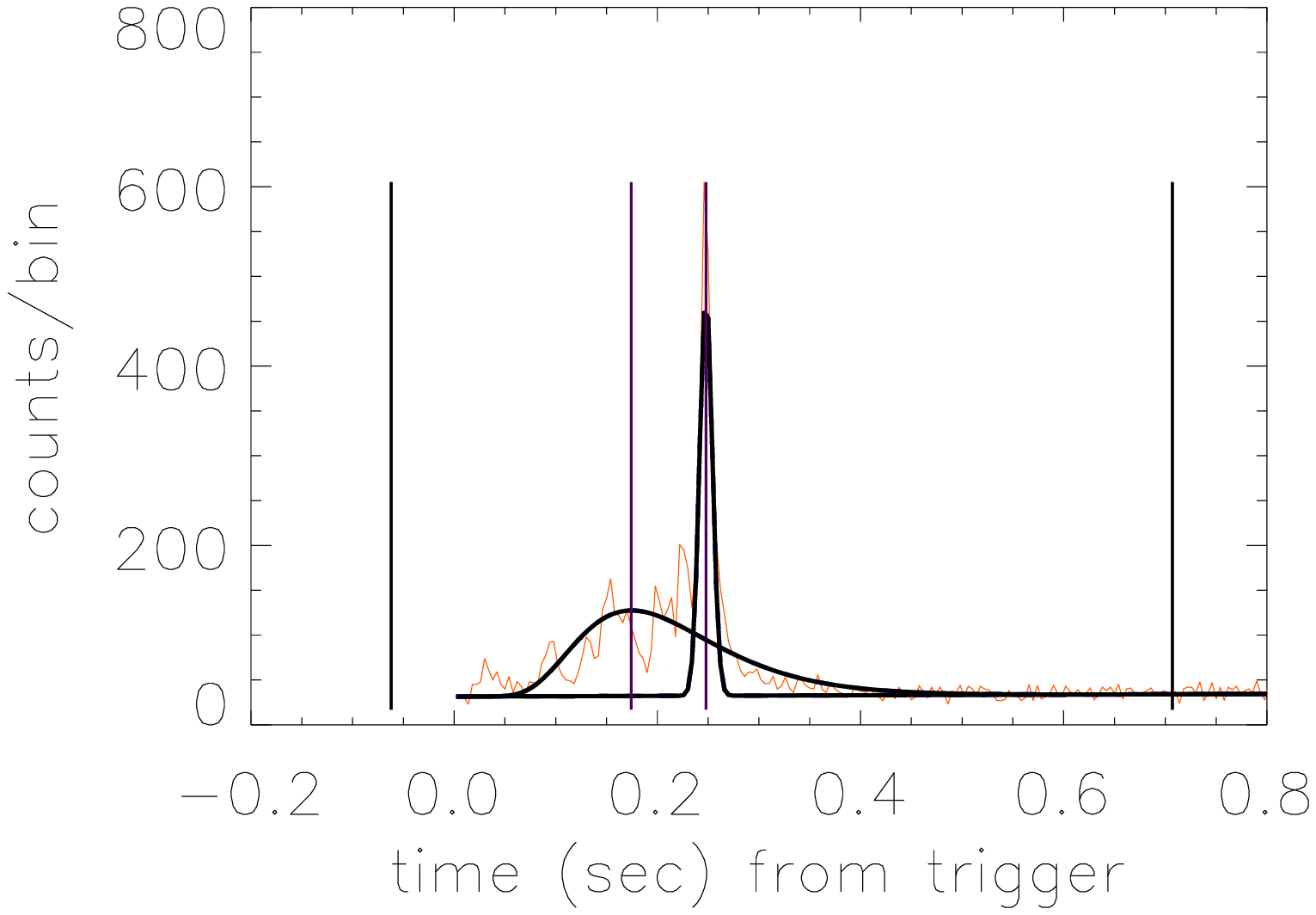}
\caption{Crescendo GRBs 1453 (left panel) and 3173 (right panel). Temporal resolution makes it difficult to tell if these bursts contain Staccato pulses. \label{fig:stac2}}
\end{figure}

As described in Table \ref{tab:complex}, four of the TTE bursts have pulse structures that are inconsistent with the standard GRB pulse paradigm. These bursts are instead characterized as asymmetric structures that increase gradually in intensity, then end with an abrupt crescendo (see Table \ref{tab:staccato}). The individual pulse structures leading to the crescendo are clearly visible for triggers 3735 and 5439 (Figure \ref{fig:stac1}), whereas they are unresolved for triggers 1453 and 3173 (Figure \ref{fig:stac2}) and 7375 (Figure \ref{fig:stac3}). Because the bursts all increase in intensity with time, we refer to these gamma-ray transients as {\em Crescendo bursts}, and the rapid-fire pulses as {\em Staccato} pulses. Our limited temporal resolution, coupled with the fact that trigger 5439 is an Intermediate GRB, prohibits us from determining if there is more than one category of Crescendo bursts.

In reevaluating the pulses in the BATSE TTE Pulse Catalog with this new definition in mind, we notice that the pulses associated with Triggers 218 and 7753 also exhibit possible Crescendo behavior. We have identified these pulses as possible Crescendo bursts in the Comments column of the catalog (Table \ref{tab:cat3}).

Although GRB pulse structure provides minimal evidence that Short and Long GRBs have different progenitors, Crescendo GRBs with Staccato pulses exhibit signatures of emission predicted from neutron star $-$ black hole mergers. Tidal disruption of the neutron star is expected in a coalescing system of this type, forming a torus around the black hole. The black hole spin should cause the torus to precess via Lense-Thirring torques \citep{sto13}, resulting in a signal consisting of a small number of quasi-periodic events with interpulse separations of around 30 to 100 ms. The predicted precession period $T_p$ should increase as $T_p \propto t^{4/3}$, leading to a corresponding increase in the interpulse separation. The separations between the Staccato pulses in BATSE triggers 3735 and 5439 exceed the expected 30 to 100 ms window, and these separations do not increase as $t^{4/3}$, so it seems unlikely that these Crescendo bursts are consistent with the neutron star $-$ black hole merger model. However, the variable emission in Crescendo bursts 1453, 3173, and 7375 is of a shorter timescale, and may be consistent with the model, although this is undetermined due to the unresolved temporal binning. Regardless, the rarity of bursts having non-pulsed emission of the type predicted by \cite{sto13} is in agreement with the results of \cite{dic13}, who find that events having these predicted properties do not dominate the Short GRB population. 

Not all Crescendo bursts necessarily belong to the Short GRB class. At least one Long GRB (BATSE trigger 1425; Figure \ref{fig:stac3}) appears to exhibit Crescendo behavior along with Staccato pulses. However, it should be noted the pulses in this burst appear to have asymmetric shapes consistent with the \cite{nor05} pulse model, unlike the pulses in the Crescendo GRBs 3735 and 7375.

\begin{deluxetable*}{lcl}
\tablenum{11}
\tablecaption{BATSE Crescendo bursts\label{tab:staccato}}
\tablewidth{0pt}
\tablehead{
\colhead{TTE Crescendo Bursts} & \colhead{Description} 
}
\startdata
1453 & 4-5 overlapping peaks increasing in intensity to a short bright final pulse \\
3173 & 4-5 overlapping peaks increasing in intensity to a long, bright final pulse \\
3735 & 3-4 symmetric short Staccato constant-intensity pulses followed by a longer, final bright pulse \\
3904 & TTE Partial bursts: overlapping peaks increasing in intensity to a short bright final pulse \\
5439 & 2 symmetric short Staccato pulses followed by a bright symmetric pulse \\
7375	 & 7-8 overlapping peaks increasing in intensity \\
\tableline
Example of a possible Long Crescendo Burst? \\
1425 & 5-6 symmetric overlapping Staccato pulses increasing in intensity  \\
\enddata
\end{deluxetable*}

\begin{figure}
\plottwo{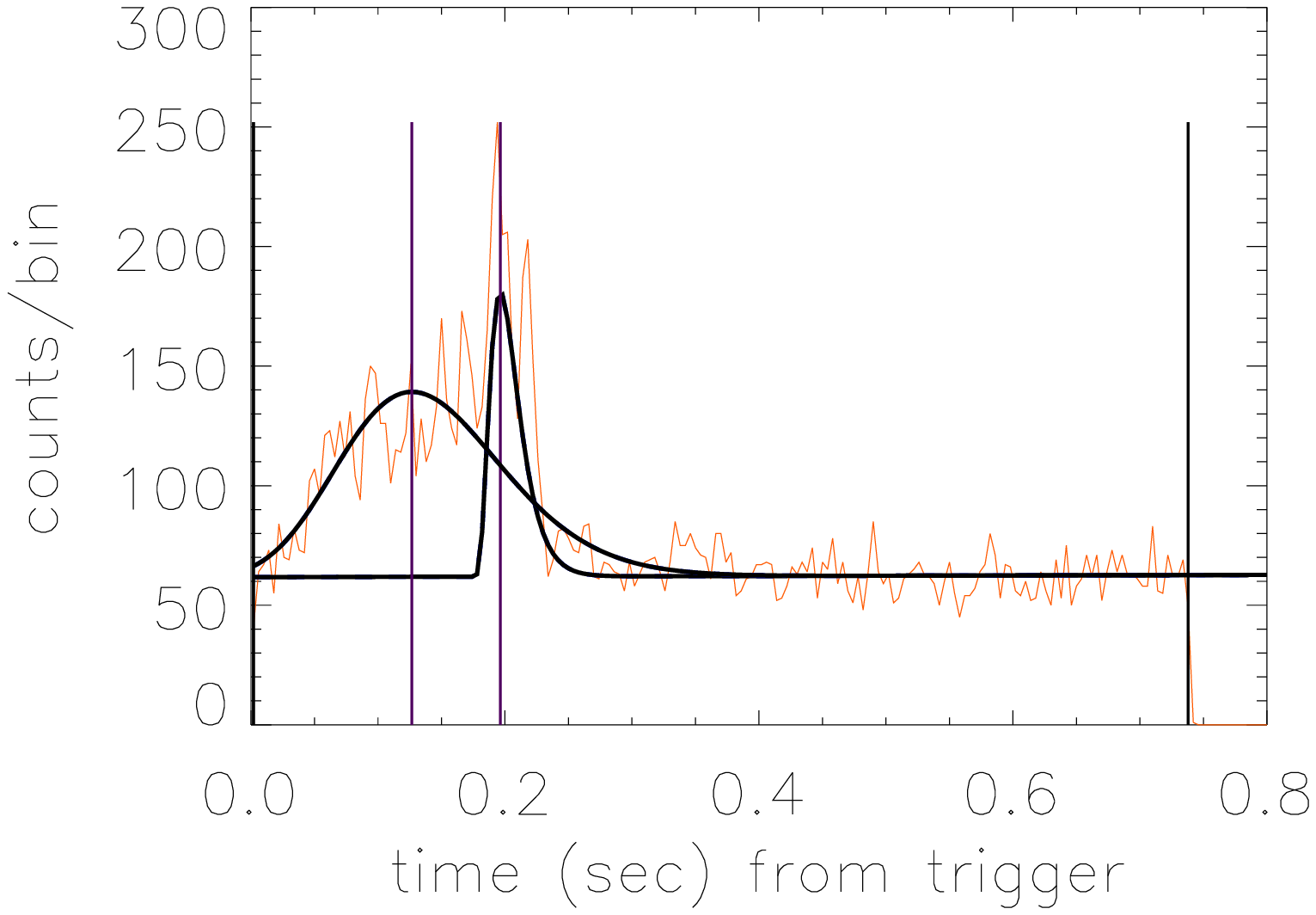}{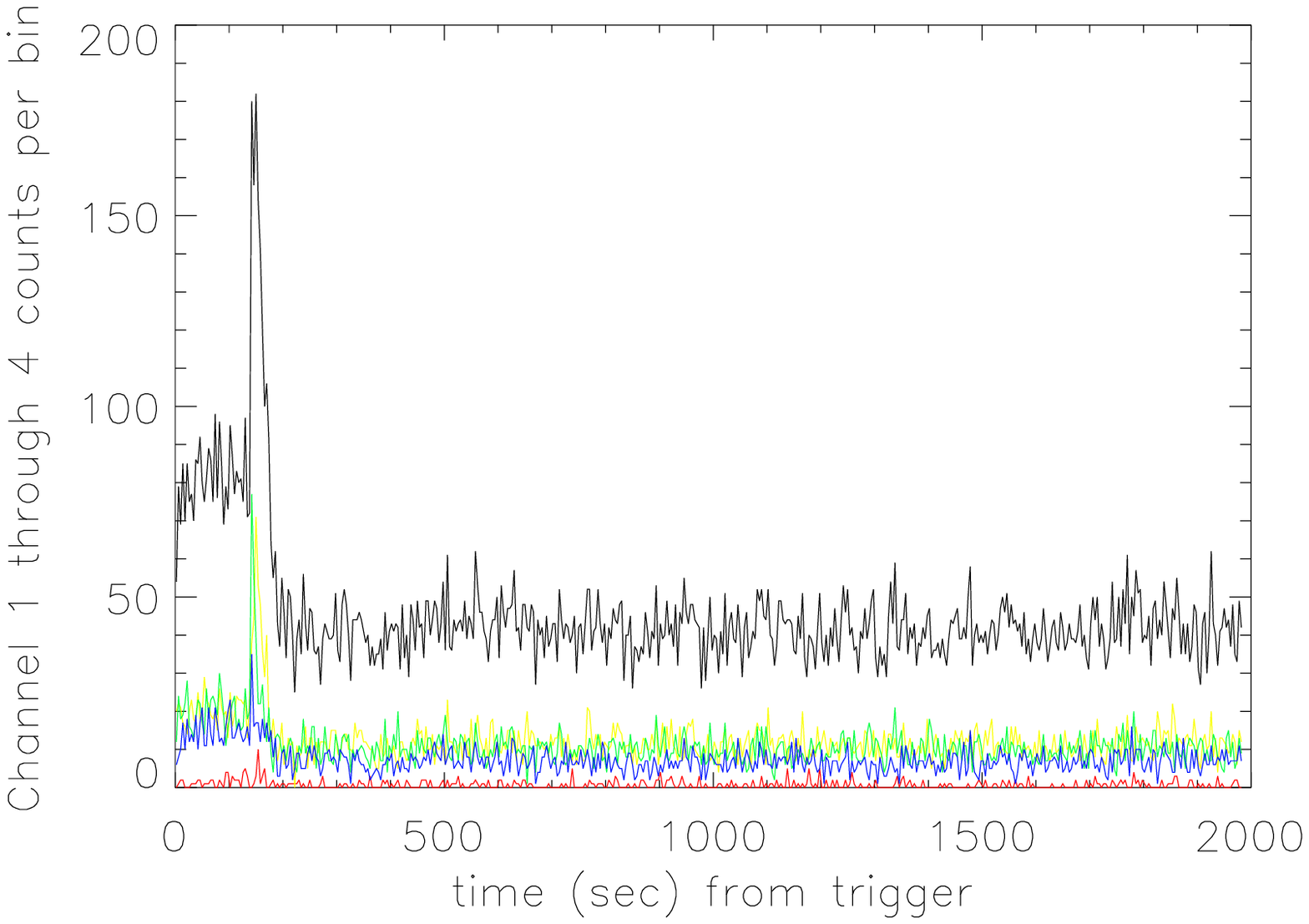}
\caption{Crescendo GRB 7375 (left panel) and TTE Partial Crescendo GRB 3904 (right panel). The Crescendo structure of 3904 cannot be seen in 64-ms data, and is only clearly seen in 4-ms data when comparing energy-dependent light curves. Light curves in BATSE energy channels are identified by different colors: channel 1 (red; $25 - 50$ keV), channel 2 (yellow; $50 - 100$ keV), channel 3 (green; $100 - 300$ keV), and channel 4 (blue; 300 keV$ - $1 MeV). \label{fig:stac3}}
\end{figure}

\begin{figure}
\plotone{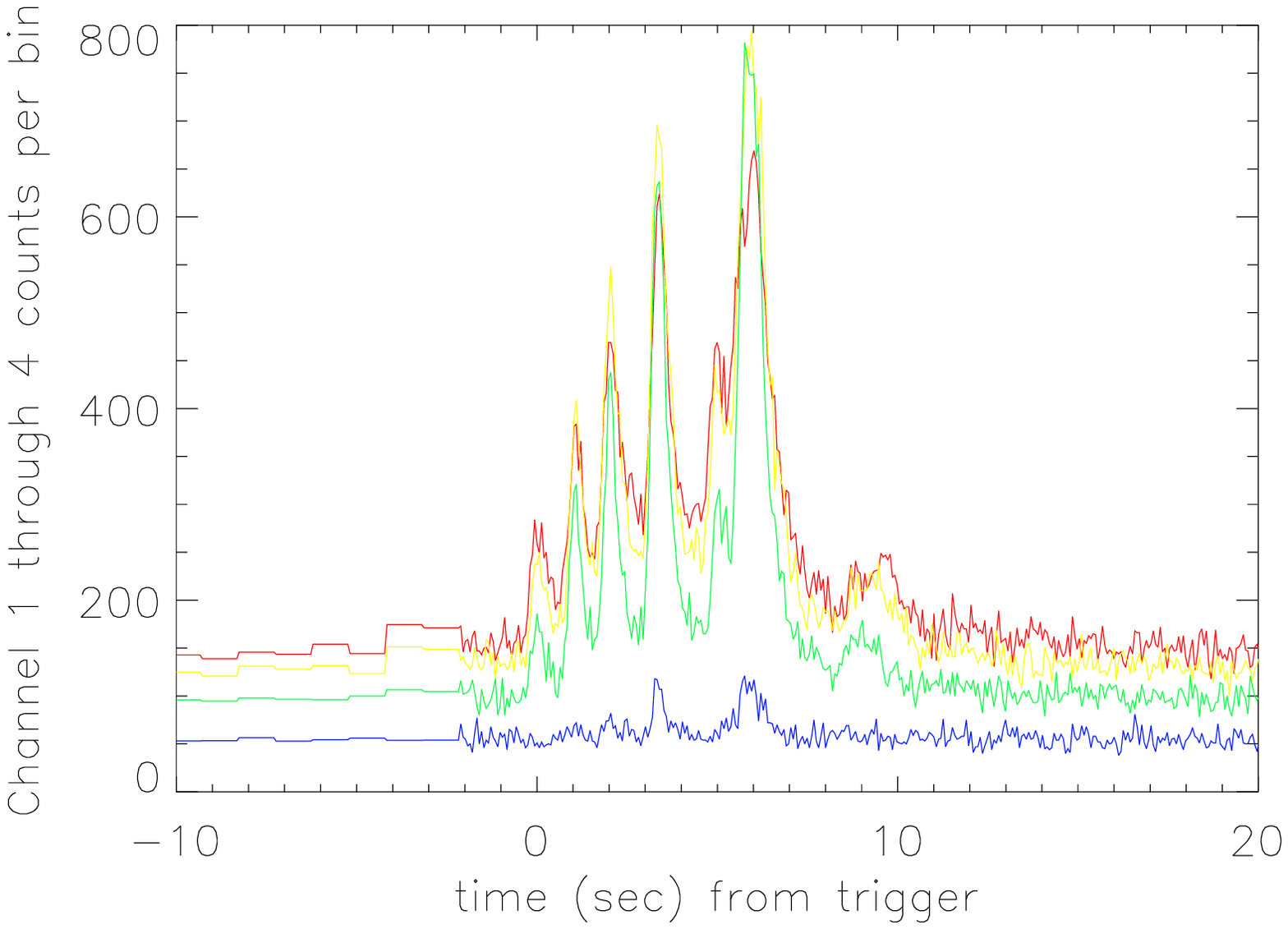}
\caption{Long GRB 1425. This possible Long Crescendo GRB contains Staccato pulses that overlap, providing a potential intermediate case between Crescendo bursts with and without Staccato pulses. Light curves in BATSE energy channels are identified by different colors: channel 1 (red; $25 - 50$ keV), channel 2 (yellow; $50 - 100$ keV), channel 3 (green; $100 - 300$ keV), and channel 4 (blue; 300 keV$ - $1 MeV). \label{fig:stac3}}
\end{figure}

\section{Conclusions}

Pulses are the dominant structures in Short GRB
light curves, as they are in Long and Intermediate GRBs.
We have verified this by producing a catalog of BATSE TTE GRB
pulses and their properties; the vast majority of bursts in this
catalog belong to the Short GRB class.
The catalog has been compiled
under the assumption that most GRB emission structures
can be explained to first order by the \cite{nor05} empirical pulse model.
The catalog contains 434 pulses in 387 GRBs, 
characterized by those fit at 4-ms resolution and
those fit at 64-ms resolution.

Statistical and machine learning tools form the basis of the approach
used to construct the pulse catalog.
The identification of Short GRBs is based on statistical clustering
methods and has been extended to this dataset using supervised classification, 
rather than from using the common but more arbitrary $T_{90} \le 2$ s rule. 
Pulse light curves and pulse residuals are fitted to empirical models with 
flexible parameters using a non-linear least squares modeling approach. 
An iterative, heuristic statistical approach is used both to characterize
pulses that fit the model as well as to classify pulses that exhibit complexities 
suggestive of an augmented model.

We note that two important selection biases, {\em temporal resolution} and
{\em signal-to-noise}, must be considered carefully in the pulse-
fitting process and when treating complexity as a departure from 
non-stochastic light curve variations.
Our definitions of complexity are interwoven with both of these biases: 
our pulse- and residual-fitting models identify when emission episodes
are likely to be pulses, and they help us to characterize complexity. 
We have binned flux data in 
order to apply pulse and residual models, and we have
demonstrated that the temporal resolution of our binned data 
affects both our ability to identify pulses and to characterize 
their complexities. Signal-to-noise plays a similar role to binning
in washing out existing structure. Both of these effects must also
be considered when comparing pulses observed by gamma-ray detectors
having different sensitivities, spectral responses, and temporal resolutions
({\em e.g.,} Swift, Fermi GBM, Suzaku).

Most Short GRB pulses exhibit correlated behaviors 
suggesting that they are produced by mechanisms
governed by only a few free parameters.
These processes, whatever they are,
seem to be responsible for producing not just Short pulses,
but also pulses found within all GRB classes. Among these correlated
properties: shorter duration pulses have higher amplitudes
(peak fluxes) than longer duration pulses, larger fluence pulses 
also have harder spectra than faint ones, and larger fluence pulses 
have higher amplitudes than lower fluence ones.
{\em These correlated properties are common among 
Short, Intermediate, and Long GRBs,
thus linking all three burst classes and suggesting similar
emission mechanisms.}
Unlike in Long and Intermediate 
burst pulse evolution \citep{hak14, hak15}, 
the role of asymmetry in Short GRB pulses
is difficult to determine because asymmetry is difficult to measure
given the small number of photons detected.

The triple-peaked behavior seen in Long/Intermediate GRB pulses
is also present in Short GRB light curves, which exhibit a continuum of
structural complexity. The simplest form can be modeled 
by a monotonically increasing and decreasing pulse structure \citep{nor05}.  
A slightly more complex pulse shape is non-monotonic
but still smooth; we represent this with the \cite{nor05}
pulse model augmented by the \cite{hak14} residual structure.
Additional structural complexity appears to be added on top
of the \cite{nor05} pulse model plus \cite{hak14} residual model;
these pulses have excess complex emission overlaying 
a recognizable triple-peaked structure. The most complex
pulses are dominated by complex and chaotic structures;
they are only recognizable as pulses because
their chaotic structure is found within a single emission episode.
Not all of this complex structure is chaotic; many complex
pulses exhibit what appear to be recognizable and repeated behaviors
that suggest the existence of complex pulse subclasses.
{\em However, composite light curves made by summing the fluxes of many complex
pulses show only the smooth triple-peaked structure, validating our
hypothesis that complexity represents a randomly-distributed augmentation
of the light curve.}

The triple-peaked pulse behavior is supportive of 
emission from a shocked medium \citep{hak14}, with the 
mirroring effect seen in the precursor and decay peaks suggesting 
forward and reverse shock behavior. The hard-to-soft
evolution observed in GRB pulses also indicates
that the time of maximum energy release is at the beginning
of the pulse, when the light curve intensity is still
increasing. The additional structure seen in the light curves
of Structured and Complex pulses may indicate 
GRBs in which additional, more chaotic radiation processes 
are also involved. These chaotic patterns are only
present in conjunction with pre-existing pulse light curves, 
further supporting the idea that {\em pulses} are the underlying, 
foundational units of GRB emission. The additional complex
structures might represent some more localized
behavior, such as microjets or electromagnetic 
fluctuations of some sort.

Double- and triple-pulsed Short GRBs are uncommon, but 
they exist. These bursts provide valuable insights
into the processes by which GRBs release energy.
The interpulse separations in these multiple-pulsed bursts
correlate with the duration of the initial pulse, suggesting
that first pulse duration is a predictor for the time
that will pass before the next pulse is emitted. Similarly,
the duration of the second pulse correlates with both the
duration of the first pulse and the interpulse separation, 
indicating that there is memory within the burst of the
energy released from the first pulse. If the pulse emission
timescale indicates the kinematics of relativistically
jetted material, then these correlations suggest energy
loss as the jet moves outward. We have shown that these results are
redshift-independent, and therefore intrinsic, as
expected from models involving external shocks.
However, this interpretation is inconsistent
with results obtained previously for Long GRBs \citep{rrm00},
as Long GRBs do not show either increasing interpulse separations
or increasing pulse durations. 

The original basis for the Short GRB class 
was BATSE's duration bimodality \citep{kou93}:
this discovery led to the idea that the Short GRB emission timescale
necessitated compact merger models rather than 
those involving hyper- and supernovae. 
Afterglows, host galaxies, and a wide
range of evidence provided from non-prompt emission
support the idea that Short and Long/Intermediate GRBs 
originate in different environments, produced by different hosts.
The recent discovery of a gravitational wave ``chirp" 
associated with Short (or possibly Intermediate)
GRB 170817A \citep{LIGO17} is consistent with the
neutron star-neutron star merger model of Short GRBs.

Most theoretical models
explain GRB emission as originating
from emitting regions located far from
the progenitor. 
These models assume that the progenitors contribute 
indirectly to the pulse properties via 
the amount of material they eject, the 
relativistic velocity of this ejected material,
and the angular characteristics of the beamed jets produced.
The physics of pulsed GRB emission can thus be 
similar for different GRB classes even if their progenitors are very different. 
Jet models involve relativistic material
moving away from the progenitor and towards
the observer at extremely high velocities (Lorentz
factor $100 \le \Gamma \le 1000$); these
can lead to significantly time-compressed
durations ($w_{\rm observed} = w / (1+\Gamma^2)$)
for any emission that is produced in the jet frame.
Such extreme time compression should
produce pulses too short to be consistent with observed pulse
durations. For example, the duration of GRB 170817A's pulsed emission
is too long to have undergone time-compression of order $10^4$,
thus the emission could not have been created in the frame of the expanding
shell. One solution to this problem is to have the emitting region relatively 
stationary with respect to the observer. In other words,
the pulsed emission needs to be produced
in a stationary external medium
rather than internal to the expanding jet.
Furthermore, the {\em burst} duration must
reflect the activity time of the central engine
rather than episodic emission from within
the moving jet, because if it did time compression
would smear out the duration bimodality 
and observed class boundaries.

The similarities
between pulse properties observed across GRB classes
suggests that the prompt emission in Long, Short,
and Intermediate bursts alike originates from 
a similar physical mechanism, even if multi-pulsed 
Long/Intermediate GRBs do not exhibit pulse
lengthening associated with external shocks.
This inconsistency might be resolved if
Short GRB pulses represent sequential episodes, 
moving outward from a single event, while Long/Intermediate GRB 
pulses are independent (unlinked) episodes, corresponding
to different events occurring within the line-of-sight. 



Despite their many similarities,
Short and Long/Intermediate GRBs exhibit several different
prompt emission characteristics that can be used to help classify them:
\begin{itemize}
\item More Short GRBs appear to be single-pulsed ($90\%$) 
than Long/Intermediate GRBs ($25-40\%$).
\item Multi-pulsed Short GRBs exhibit correlated pulse durations and interpulse separations,
whereas multi-pulsed Long/Intermediate GRBs do not.
\item Durations of Short GRB pulses are shorter than those of Long/Intermediate GRB pulses. 
\item Short GRB pulses are spectrally harder than Long/Intermediate
GRB pulses and undergo greater hard-to-soft evolution. 
\item The light curves of Short GRBs generally exhibit more 
pronounced precursor and decay peaks than 
Long/Intermediate GRB pulses.
\end{itemize}

Finally, our catalog development approach has led to the
discovery of a new type of gamma-ray transient. 
Crescendo GRBs have longer rise times than decay times, 
and cannot be adequately modeled by asymmetry in
the \cite{nor05} pulse model. Some Crescendo GRBs are
characterized by a series of rapid-fire Staccato pulses
leading up to the crescendo, while others have a crescendo
that is preceded by a complex pulse that may be composed
of unresolved Staccato pulses or may be composed of a
complex emission episode that is similar to the
extended emission found in some Short GRBs. 
Crescendo GRBs might be a subset of GRBs (representing,
for example, neutron star $-$ black hole mergers), but 
they might also represent a completely 
different type of gamma-ray transient. We have found
at least one example of what appears to be a Long Crescendo 
GRB, suggesting that Crescendo characteristics, like triple-peaked
pulse structures, do not belong only in the realm of 
Long or Short GRB classification.

We have demonstrated that the prompt pulses from 
Short GRBs share much in common with pulses from 
Long/Intermediate GRBs, even as they exhibit important differences.
The authors hope that this BATSE TTE GRB Pulse Catalog
helps invite new patterns of inquiry on Short gamma-ray bursts,
as well as on potential common emission mechanisms.


\acknowledgments

This work was supported in part by NASA EPSCoR grant NNX13AD28A
and  OTKA grant NN111016. We 
gratefully acknowledge conversations with Robert D. Preece concerning
the relationship between the catalog results and theoretical models, and 
we thank the referee and statistical expert for their helpful comments.
We acknowledge helpful and valuable discussions during the
development of this project with Stanley McAfee, Rebecca Brnich, 
Corrine Taylor, Bailey Williamson, Thomas Cannon, and Alex Greene.

\end{document}